\documentclass[11pt]{article}

\usepackage{graphicx} 
\usepackage{amsmath}
\usepackage{amsfonts}
\usepackage{amssymb}

\newcommand{\mypicturewidth}{0.45\textwidth}

\newcommand{\be}{\begin{equation}}
\newcommand{\ee}{\end{equation}}
\newcommand{\bea}{\begin{eqnarray}}
\newcommand{\eea}{\end{eqnarray}}
\newcommand{\mbb}{\mathbb}
\newcommand{\ti}{\times}

\newcommand{\mc}{\mathcal}
\newcommand{\K}{\mc{K}}

\def\lsim{\mathrel{\rlap{\lower4pt\hbox{\hskip1pt$\sim$}}
    \raise1pt\hbox{$<$}}}                
\def\gsim{\mathrel{\rlap{\lower4pt\hbox{\hskip1pt$\sim$}}
    \raise1pt\hbox{$>$}}}                

\begin{document}

\title{\bf Roulette Inflation with K\"ahler Moduli and their Axions}
\author{J.~Richard~Bond$^{\rm a}$, Lev Kofman$^{\rm a}$, Sergey Prokushkin$^{\rm a}$, 
and Pascal M.~Vaudrevange$^{\rm a, b}$\\
\\
$^{\rm a}${\it Canadian Institute for Theoretical Astrophysics}\\
{\it University of Toronto, 60 St. George St., Toronto, ON M5S 3H8, Canada}
\\
$^{\rm b}${\it Department of Physics}\\
{\it University of Toronto, 60 St. George St., Toronto, ON M5S 1A7, Canada}}

\date{\today}

\maketitle

\abstract{We study 2-field inflation models based on the
``large-volume'' flux compactification of type IIB string theory.  The
role of the inflaton is played by a K\"ahler modulus $\tau$
corresponding to a 4-cycle volume and its axionic partner
$\theta$. The freedom associated with the choice of Calabi-Yau
manifold and the non-perturbative effects defining the potential
$V(\tau, \theta)$ and kinetic parameters of the moduli brings an
unavoidable statistical element to theory prior probabilities within
the low energy landscape. The further randomness of $(\tau, \theta)$
initial conditions allows for a large ensemble of trajectories.
Features in the ensemble of histories include ``roulette trajectories'',
with long-lasting inflations in the direction of the rolling axion,
enhanced in number of e-foldings over those restricted to lie in the
$\tau$-trough. Asymptotic flatness of the potential makes possible an
eternal stochastic self-reproducing inflation. A wide variety of
potentials and inflaton trajectories agree with the cosmic microwave
background and large scale structure data. In particular, the observed
scalar tilt with weak or no running can be achieved in spite of a
nearly critical de Sitter deceleration parameter and consequently a
low gravity wave power relative to the scalar curvature power. }

\section{Introduction} \label{sec:intro}

The ``top-down'' approach to inflation seeks to determine cosmological
consequences beginning with inflation scenarios motivated by
ever-evolving fundamental theory.  Most recent attention has been
given to top-down models that realize inflation with string theory.
This involves the construction of a stable six-dimensional
compactification and a four-dimensional extended de Sitter (dS) vacuum
which corresponds to the present-day (late-time) universe {\it e.g.},
the KKLT prescription \cite{KKLT}. Given this, there is a
time-dependent, transient non-equilibrium inflationary flow in four
dimensions towards the stable state, possibly involving dynamics in
both sectors.

Currently, attempts to embed inflation in string theory are far from
unique, and indeed somewhat confused, with many possibilities
suggested to engineer inflation, using different axionic and moduli
fields \cite{Blanco-Pillado:2004ns, Blanco-Pillado:2006he}, branes 
in warped geometry \cite{KKLMMT}, D3-D7 models \cite{Dasgupta:2002ew},
{\it etc.} \cite{Buchel:2003qj, em}. These pictures are increasingly being 
considered within a 
string theory landscape populated locally by many scalar fields.

Different realizations of stringy inflation may not be mutually
incompatible, but rather may arise in different regions of the
landscape, leading to a complex statistical phase space of solutions.
Indeed inflation driven by one mechanism can turn into inflation
driven by another, {\it e.g.}, \cite{Kofman:1985aw}, thereby increasing the
probability of inflation over a single mechanism scenario.

So far all known string inflation models require significant
fine-tuning.  There are two classes that are generally discussed
involving moduli. One is where the inflaton is identified with brane
inter-distances. Often the effective mass is too large (above the
Hubble parameter) to allow acceleration for enough e-folds, if at all.
To realize slow-roll inflation, the effective inflaton mass should be
smaller than the Hubble parameter during inflation, $m^2 <
H^2$. Scalar fields $\phi$ which are not minimally but conformally coupled to gravity
acquire effective mass terms $\frac{1}{12} R \phi^2$ which prevents 
slow-roll. An example of this problem is warped brane inflation where the
inflaton is conformally coupled to the four-dimensional gravity
\cite{KKLMMT}. A similar problem also arises in supergravity.  The case has
been constructed that has masses below the Hubble parameter $H$ which
avoids this $\eta$-problem at the price of severe fine-tuning
\cite{KKLMMT}.  Another class is geometrical moduli such as K\"ahler
moduli associated with 4-cycle volumes in a compactifying Calabi-Yau
manifold as in \cite{Blanco-Pillado:2004ns, Blanco-Pillado:2006he},
which has been recently explored in \cite{CQ} and which we extend here
to illustrate the statistical nature of possible inflation histories.

Different models of inflation predict different spectra for scalar and
tensor cosmological fluctuations. From cosmic microwave background and
other large scale structure experiments one can hope to reconstruct
the underlying theory that gave rise to them, over the albeit limited
observable range.  Introduction of a multiple-field phase space
leading to many possible inflationary trajectories necessarily brings
a statistical element prior to the constraints imposed by data. That
is, a theory of inflation embedded in the landscape will lead to a
broad theory ``prior'' probability that will be updated and sharpened
into a ``posteriori'' probability through the action of the data, as
expressed by the likelihood, which is a conditional probability of the
inflationary trajectories given the data. All we can hope to
reconstruct is not a unique underlying acceleration history with
data-determined error bars, but an ensemble-averaged acceleration
history with data-plus-theory error bars \cite{BCKP}.

The results will obviously be very dependent upon the theory prior. In
general all that is required of the theory prior is that inflation
occurs over enough e-foldings to satisfy our homogeneity and isotropy
constraints and that the universe preheats (and that life of some sort
forms) --- and indeed those too
are data constraints rather than a priori theory
constraints. Everything else at this stage is theoretical prejudice. A
general approach in which equal a priori theory priors for
acceleration histories are scanned by Markov Chain Monte Carlo methods
which pass the derived scalar and tensor power spectra though cosmic
microwave background anisotropy data and large scale clustering data
is described in \cite{BCKP}. But since many allowed trajectories would
require highly baroque theories to give rise to them, it is essential
to explore priors informed by theory, in our case string-motivated
priors.

The old top-down view was that the theory prior would be a
delta-function of {\it the} correct one and only theory. The new view is
that the theory prior is a probability distribution on an energy
landscape whose features are at best only glimpsed, with huge number
of potential minima, and inflation being the late stage flow in the
low energy structure toward these minima. 

\begin{figure}
  \begin{center}
    \includegraphics[width=10cm]{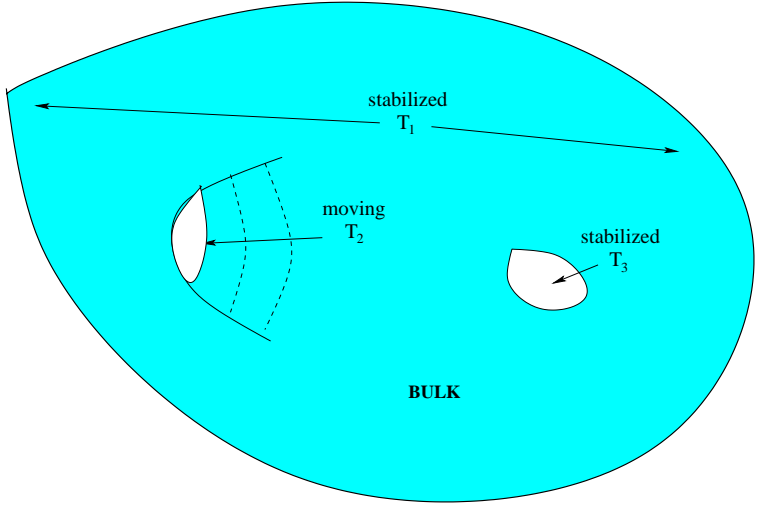}
  \end{center}
  \caption{Schematic illustration of the ingredients in K\"ahler moduli
    inflation. The four-cycles of the CY are the K\"ahler moduli $T_i$
    which govern the sizes of different holes in the manifold. We
    assume $T_3$ and the overall scale $T_1$ are already stabilized,
    while the last modulus to stabilize, $T_2$, drives inflation while
    settling down to its minimum. The imaginary parts of $T_i$ have to
    be left to the imagination.  The outer $3+1$ observable dimensions
    are also not shown.}
  \label{fig:ill}
\end{figure}

In the picture we adopt for this paper, the flow is of collective
geometrical coordinates associated with the settling down of the
compactification of extra dimensions. The observed inflaton would be
the last (complex) K\"ahler modulus to settle down. We shall call this
$T_2$. The settling of other K\"ahler moduli associated with 4-cycle
volumes, $T_3, T_4, ...$ and the overall volume modulus, $T_1$, as
well as ``complex structure'' moduli and the dilaton and its axionic
partner, would have occurred earlier, associated with higher energy
dynamics, possibly inflations, that became stabilized at their
effective minima. The model is illustrated by the cartoon
Fig.~\ref{fig:ill}. We work within the ``large volume'' moduli
stabilization model suggested in \cite{BB, BBCQ,CQS} in which the
effective potential has a stable minimum at a large value of the
compactified internal volume, $\mc{V} \sim 10^5 - 10^{20}$ in string
units.  An advantage of this model is that the minimum exists for
generic values of parameters, {\it e.g.}, of the flux contribution to
the superpotential $W_0$. (This is in contrast to the related KKLT
stabilization scheme in which the tree-level $W_0$ is fine-tuned at
$\sim 10^{-4}$ in stringy units in order for the $\mc{V}$ minimum to
exist.)

In this paper, we often express quantities in the relatively small
``stringy units'' $m_s\propto M_P/\sqrt{\mc{V}}$, related to the
(reduced) Planck mass \be
\label{eq:MP}
M_P = 1/ \sqrt{8 \pi G} = 2.4 \ti 10^{18} \textrm{GeV} \, , 
\ee 
where $G$ is Newton's constant. 

In this picture, the theory prior would itself be a Bayesian product
of a number of conditional probabilities: (1) of manifold
configuration defining the moduli; (2) of parameters defining the
effective potential and the non-canonical kinetic energy of the
moduli, given the manifold structure; (3) of the initial conditions
for the moduli and their field momenta given the potentials. The
latter will depend upon exactly how the ``rain down'' from higher
energies occurs to populate $T_2$ initial conditions. An effective
complication occurs because of the so-called eternal inflation regime,
when the stochastic kicks that the inflaton feels in an e-folding can
be as large as the classical drift. This $T_2$-model is in fact another
example of stringy inflation with self-reproduction. (See 
\cite{Blanco-Pillado:2004ns} for another case.) If other
higher-energy moduli are frozen out, most inflationary trajectories
would emerge from this quantum domain. However we expect other quantum
domains for the higher-energy moduli to also feed the $T_2$ initial
conditions, so we treat these as arbitrary.

The K\"ahler moduli are flat directions at the stringy tree level. The
reason this picture works is that the leading non-perturbative
(instanton) and perturbative ($\alpha'$) corrections introduce only
an exponentially flat dependence on the K\"ahler moduli, avoiding the
$\eta$-problem.  Conlon and Quevedo \cite{CQ} focused on the real part
of $T_2$ as the inflaton and showed that slow-roll inflation with
enough e-foldings was possible. A modification \cite{Simon:2006du} of
the model  considered inflation in a new $\mc{V}$ direction
but with a negative result. 

The fields $T_i$ are complex, $T_i=\tau_i +i\theta_i$.  In this paper
we extend the model of \cite{CQ} to include the axionic direction
$\theta_2$. There is essentially only one trajectory if $\theta_2$ is
forced to be fixed at its trough, as in \cite{CQ}. The terrain in the
scalar potential $V(\tau_2, \theta_2)$ has hills and valleys in the
$\theta_2$ direction which results in an ensemble of trajectories
depending upon the initial values of $\tau_2, \theta_2$. The field
momenta may also be arbitrary but their values quickly relax to
attractor values. The paper \cite{Holman:2006ek} considered
inflation only along the $\theta$ direction while the dynamics in the
$\tau$ direction were artificially frozen. We find motion in $\tau$ 
always accompanies motion in $\theta$.

In K\"ahler moduli models, there is an issue of higher order perturbative
corrections. Even a tiny quadratic term would break the exponential
flatness of the inflaton potential and could make the $\eta$-problem
reappear. However, the higher order terms which depend on the inflaton
only through the overall volume $\mc{V}$ of the Calabi-Yau manifold
will not introduce any mass terms for the inflaton.  Although these
corrections may give rise to a mass term for the inflaton, it might
have a limited effect on the crucial last sixty e-folds.

In \S~\ref{sec:model} we describe the model in the context of type IIB
string theory.  In \S~\ref{sec:corr} we address whether higher
(sub-leading) perturbative corrections introduce a dangerous mass term
for the inflaton. In \S~\ref{sec:pot} we discuss the effective
potential for the volume, K\"ahler moduli and axion fields, showing
with 3 moduli that stabilization of two of them can be sustained even
as the inflaton $T_2$ evolves. Therefore in \S~\ref{sec:inf} we
restrict ourselves to $V(\tau_2, \theta_2)$ with the other moduli
stabilized at their minima. \S~\ref{sec:plots} explores inflationary
trajectories generated with that potential, for various choices of
potential parameters and initial conditions.  In \S~\ref{sec:stoch} we
investigate the diffusion/drift boundary and the possibility of
self-reproduction.  In \S~\ref{sec:conc} we summarize our results and
outline issues requiring further consideration, such as the
complication in power spectra computation that follows from the
$\tau_2, \theta_2$ freedom.

\section{The Type IIB String Theory Model}\label{sec:model}

Our inflationary model is based on the ``large-volume" moduli
stabilization mechanism of \cite{BB,BBCQ,CQS}. This mechanism relies
upon the fixing of the K\"ahler moduli in IIB flux compactifications
on Calabi-Yau (CY) manifolds $M$ by non-perturbative as well as
perturbative effects. As argued in \cite{BB,BBCQ,CQS}, a minimum of
the moduli potential in the effective $4d$ theory exists for a large
class of models.  The only restriction is that there should be more
complex structure moduli in the compactification than K\"ahler moduli,
i.e.  $h^{1,2}>h^{1,1}>1$, where $h^{1,2}, h^{1,1}$ are the Hodge
numbers of the CY. (The number of complex structure moduli is 
$h^{1,2}$ and the number of K\"ahler moduli is $h^{1,1}$. 
Other Hodge numbers are fixed for a CY threefold.) 
The ``large-volume" moduli stabilization mechanism is
an alternative to the KKLT one, although it shares some features with
KKLT.  The purpose of this section is to briefly explain the model of
\cite{BB,BBCQ,CQS}.

An effective $4d$ $\mc{N} = 1$ supergravity is completely specified by
a K\"ahler potential, superpotential and gauge kinetic function.  In
the scalar field sector of the theory the action is 
\be S_{\mc{N}=1} =
\int d^4 x \sqrt{-g} \left[ \frac{M_P^2}{2} \mc{R} - \K_{,i\bar{j}}
D_\mu \phi^{i} D^\mu \bar{\phi}^j - V(\phi, \bar{\phi})
\right], 
\ee where 
\be
\label{ScalP}
V(\phi, \bar{\phi}) = e^{\K/M_P^2} \left(\K^{i \bar{j}} D_i
\hat{W} D_{\bar{j}} \bar{\hat{W}} - \frac{3}{M_P^2} \hat{W}
\bar{\hat{W}} \right) + \textrm{ D-terms}.  
\ee 
Here $\K$ and $\hat{W}$ are the K\"ahler potential and the
superpotential respectively, $M_P$ is the reduced Planck mass
eq.(\ref{eq:MP}), and $\phi^i$ represent all scalar moduli.  (We closely
follow the notations of \cite{CQS} and keep $M_P$ and other numerical
factors explicit.)

The $\alpha'^3$-corrected K\"ahler potential \cite{BBHL} is
\be
\label{Kahler}
\frac{\mc{K}}{M_P^2} = - 2 \ln \left(\mc{V} + \frac{ \xi
  g_s^{\frac{3}{2}}}{2 e^{\frac{3 \phi}{2}}} \right) - \ln(S +
  \bar{S}) - \ln \left(-i \int_{M} \Omega \wedge \bar{\Omega}\right).
\ee 

Here $\mc{V}$ is the volume of the CY manifold $M$ in units of
the string length $l_s = 2 \pi \sqrt{\alpha'}$, $\mc{V}=\mc{V}_s
l_s^6$ and we set $\alpha'=1$.  The second term in the logarithm
represents the $\alpha'$-corrections with $\xi = -\frac{\zeta(3)
\chi(M)}{2 (2 \pi)^3}$ proportional to the Euler characteristic
$\chi(M)$ of the manifold $M$. $S = -iC_0 + e^{-\phi}$ is the IIB
axio-dilaton with $\phi$ the dilaton component and $C_0$ the
Ramond-Ramond 0-form. $\Omega$ is the holomorphic 3-form of $M$.
The superpotential depends explicitly upon the K\"ahler moduli $T_i$
when non-perturbative corrections are included \be
\label{Super}
\hat{W}   =  \frac{g_s^{\frac{3}{2}} M_P^3}{\sqrt{4 \pi}} 
\left(W_0 + \sum_{i=1}^{h^{1,1}} A_i e^{-a_i T_i} \right), 
\quad W_0 = \frac{1}{l_s^2}\int_{M} G_3 \wedge \Omega .
\ee
Here, $W_0$ is the tree level flux-induced superpotential which is
related to the IIB flux 3-form $G_3 = F_3 - i S H_3$ as shown.  The
exponential terms $A_i e^{-a_i T_i}$ are from non-perturbative
(instanton) effects. (For simplicity, we ignore higher instanton
corrections. This should be valid as long as we restrict ourselves to
$a_i \tau_i \gg 1$, which we do.) The K\"ahler moduli are complex, 
\be
\label{T}
T_i = \tau_i + i \theta_i \,, 
\ee 
with $\tau_i$ the 4-cycle volume and $\theta_i$ its axionic partner,
arising from the Ramond-Ramond 4-form $C_4$.  The $A_i$
encode threshold corrections. In general they are functions of the complex 
structure moduli and are independent of the
K\"ahler moduli. This follows from the requirement that $W$ is a
holomorphic function of complex scalar fields and therefore can depend
on $\tau_i$ only via the combination $T_i = \tau_i + i \theta_i$.  On
the other hand, $W$ should respect the axion shift symmetry $\theta_i
\to \theta_i + \frac{2\pi}{a_i}$ and thus cannot be a polynomial
function of $T_i$. (See \cite{W} for discussion.)

The critical parameters $a_i$ in the potential are constants which
depend upon the specific nature of the dominant non-perturbative
mechanism. For example, $a_i=\frac{2\pi}{g_s}$ for Euclidean D3-brane
instantons and $a_i=\frac{2\pi}{g_s N}$ for the gaugino condensate on
the D7 brane world-volume. We vary them freely in our exploration of
trajectories in different potentials.

It is known that both the dilaton and the complex structure moduli can
be stabilized in a model with a tree level superpotential $W_0$
induced by generic supersymmetric fluxes (see e.g. \cite{GKP}) and
the lowest-order (i.e. $\xi = 0$) K\"ahler potential, whereas the
K\"ahler moduli are left undetermined in this procedure (hence are
``no scale" models).  Including both leading perturbative and
non-perturbative corrections and integrating out the dilaton and the
complex structure moduli, one obtains a potential for the K\"ahler
moduli which in general has two types of minima. The first type is the
KKLT minima \cite{KKLT} which requires significant fine tuning of $W_0$
($\sim 10^{-4}$) for their existence. As pointed out in \cite{CQS}, the 
KKLT approach has a few shortcoming, among which are the
limited range of validity of the KKLT effective action (due to
$\alpha'$ corrections) and the fact that either the dilaton or some of
the complex structure moduli typically become tachyonic at the minimum
for the K\"ahler modulus. (We note, however, that 
\cite{Aspinwall:2005ad} argued that a consistent KKLT-type model
with all moduli properly stabilized can be found.)  The second type is the
``large-volume" AdS minima studied in \cite{BB,BBCQ,CQS}. These
minima exist in a broad class of models and at arbitrary values of
parameters. An important characteristic feature of these models is
that the stabilized volume of the internal manifold is exponentially large, 
$\mc{V}_{min} \sim \exp{(a\tau_{min} )}$, and can be
$\mc{O}(10^5-10^{20})$ in string units. (Here $\tau_{min}$ is the value of
$\tau$ at its minimum.) The relation between the
Planck scale and string scale is 
\be M_P^2 = \frac{4 \pi \mc{V}_{min}}{g_s^2} m_s^2 \, , 
\ee
where $\mc{V}_{min}$ is the volume in string units at the minimum of
the potential. Thus these models can have $m_s$ in the range between
the GUT and TeV scale.  In these models one can compute the
spectrum of low-energy particles and soft supersymmetry breaking terms
after stabilizing all moduli, which makes them especially attractive
phenomenologically (see \cite{CQS,CAQS}).

Conlon and Quevedo \cite{CQ} studied inflation in these models and
showed that there is at least one natural inflationary direction in
the K\"ahler moduli space. The non-perturbative corrections in the
superpotential eq.(\ref{Super}) depend exponentially on the K\"ahler
moduli $T_i$, and realize by eq.(\ref{ScalP}) exponentially flat
inflationary potentials, the first time this has arisen from string
theory. As mentioned in \S~\ref{sec:intro}, higher (sub-leading)
$\alpha'$ and string loop corrections could, in principle, introduce a
small polynomial dependence on the $T_i$ which would beat
exponential flatness at large values of the $T_i$. Although the
exact form of these corrections is not known, we assume in this paper
that they are not important for the values of the $T_i$ during the
last stage of inflation (see \S~\ref{sec:corr}). 

After stabilizing the dilaton and the complex structure moduli we can
identify the string coupling as $g_s = e^{\phi}$, so the K\"ahler
potential (\ref{Kahler}) takes the simple form
\be
\label{KahlerSimple}
\frac{\mc{K}}{M_P^2} = - 2 \ln \left(\mc{V} + \frac{\xi}{2} \right) +
\ln g_s + \mc{K}_{cs} \,, 
\ee 
where $\mc{K}_{cs}$ is a constant. Using
this formula together with equations (\ref{ScalP}), (\ref{Super}), and
(\ref{vol}), one can compute the scalar potential. In our subsequent
analysis, we shall absorb the constant factor $e^{\mc{K}_{cs}}$ into the 
parameters $W_0$ and $A_i$. 

The volume of the internal CY manifold $M$ can be expressed in terms
of the 2-cycle moduli $t^i, \, i=1,...,n=h^{1,1}$:
\be
\label{Vt}
\mc{V} = \frac{1}{6} \kappa_{ijk}t^i t^j t^k \,,
\ee
where $\kappa_{ijk}$ is the triple intersection form of $M$. 
The 4-cycle moduli $\tau_i$ are related to the $t^i$ by 
\be
\label{tau-t}
\tau_i = \frac{\partial}{\partial t^i} \mc{V} = \frac{1}{2}
\kappa_{ijk} t^j t^k \,, \ee which gives $\mc{V}$ an implicit
dependence on the $\tau_i$, and thus $\mc{K}$ through
eq.(\ref{KahlerSimple}). It is known \cite{CdlO} that for a CY
manifold the matrix $\frac{\partial^2\mc{V}}{\partial t^i \partial
t^j}$ has signature $(1, h^{1,1} -1)$, with one positive eigenvalue and
$h^{1,1} -1$ negative eigenvalues. Since $\tau_i = \tau_i(t^j)$ is
just a change of variables, the matrix
$\frac{\partial^2\mc{V}}{\partial \tau_i \partial \tau_j}$ also has
signature $(1, h^{1,1} -1)$.  In the case where each of the 4-cycles
has a non-vanishing triple intersection only with itself, the matrix
$\frac{\partial^2\mc{V}}{\partial \tau_i \partial \tau_j}$ is diagonal
and its signature is manifest. The volume in this case takes a
particularly simple form in terms of the $\tau_i$: \be
\label{vol} 
\mc{V} = \alpha \left(\tau_1^{3/2} - \sum_{i=2}^{n} \lambda_i
\tau_i^{3/2}\right)\,.  
\ee
Here $\alpha$ and $\lambda_i$ are positive constants depending on the
particular model.\footnote{For example, the two-K\"ahler model with the
orientifold of $\mbb{P}^4_{[1,1,1,6,9]}$ studied
in \cite{DDF,BBCQ,CQS} has $\alpha = 1/9\sqrt{2}$, $\lambda_1=1$, and
$\lambda_2 = -1$.}  This formula suggests a ``Swiss-cheese" picture
of a CY, in which $\tau_1$ describes the 4-cycle of maximal size and
$\tau_2, \ldots, \tau_{n}$ the blow-up cycles. The modulus $\tau_1$
controls the overall scale of the CY and can take an arbitrarily large
value, whereas $\tau_2, \ldots, \tau_{n}$ describe the holes in the CY
and cannot be larger than the overall size of the manifold.  As argued
in \cite{BBCQ,CQS}, for generic values of the parameters
$W_0$, $A_i$, $a_i$ one finds that $\tau_1 \gg \tau_i$ and $\mc{V} \gg
1$ at the minimum of the effective potential. In other words, the
sizes of the holes are generically much smaller than the overall size
of the CY.

The role of the inflaton in the model of \cite{CQ} is the last modulus
among the $\tau_i$, $i=2,...,n$, to attain its minimum. As noted by
\cite{CQ}, the simplified form of the volume eq.(\ref{vol}) is not
really necessary to have inflation. For our analysis to be correct, it
would be enough to consider a model with at least one K\"ahler modulus
whose only non-zero triple intersection is with itself, {\it i.e.},
\be
\label{vol...} 
\mc{V}  = \alpha (... - \lambda_i \tau_i^{3/2}) \,, 
\ee   
and which has its own non-perturbative term in the superpotential
eq.(\ref{Super}).

\section{Perturbative Corrections}\label{sec:corr}

There are several types of perturbative corrections that
could modify the classical potential on the K\"ahler moduli space:
those related to higher string modes, or $\alpha'$-corrections, coming
from the higher derivative terms in both bulk and source (brane)
effective actions; and string loop, or $g_s$-corrections, coming from
closed and open string loop diagrams.

As we mentioned before, $\alpha'$-corrections are an important
ingredient of the ``large volume" compactification models of
\cite{BB,BBCQ,CQS}.  They are necessary for the existence of the large
volume minimum of the effective potential in the models with K\"ahler
moduli ``lifted" by instanton terms in the superpotential.  The
leading $\alpha'$-corrections to the potential arise from the higher
derivative terms in the ten dimensional IIB action at the order $\sim
\alpha'^3$, \be\label{high_deriv} S_{IIB} = - \frac{1}{2\kappa_{10}^2}
\,\int d^{10}x \sqrt{-g^{(10)}} e^{-2\phi} [R + 4(\partial\phi)^2 +
\alpha'^3 \frac{\zeta(3)}{3\cdot2^{11}} \, J_0 - \alpha'^3
\frac{(2\pi)^3 \zeta(3)}{4} \, Q + ...] \,, \ee where $J_0 \sim
(R^{MN}{}_{PQ})^4$ and $Q$ is a generalization of the six-dimensional
Euler integrand, \be\label{euler} \int_M d^6 x \sqrt{g}Q = \chi \,.
\ee Performing a compactification of (\ref{high_deriv}) on a CY
threefold, one finds $\alpha'$-corrections to the metric on the
K\"ahler moduli space, which can be described by the $\xi$-term in the
K\"ahler potential (\ref{Kahler}) (see \cite{a_pair,BBHL}).
We will see later that this correction introduces
a positive term $\sim \xi W_0^2/\mc{V}^3$ into the potential.  As
discussed in \cite{CQS}, further higher derivative bulk corrections at
$\mc{O}(\alpha'^4)$ and above are sub-leading to the $1/ \mc{V}^3$ term
and therefore suppressed. (Note that in the models we are dealing with,
there is effectively one more expansion parameter, $1/\mc{V}$, due to the large
value of the stabilized $\mc{V}$.)
Also, $\alpha'$-corrections from the D3/D7 brane actions depend on 4d
space-time curvatures and, therefore, do not contribute to the
potential.
String loop corrections to the K\"ahler potential come from the Klein
bottle, annulus and M\"obius strip diagrams computed in \cite{Kors}
for the models compactified on the orientifolds of tori.  The K\"ahler
potential including both leading $\alpha'$ and loop corrections can be
schematically written as \cite{Kors} 
\be\label{loop_corr} K = -2
\ln{\mc{V}} - {\xi\over \mc{V}} + {f_1 \over \mc{V}^{2/3}} + {f_2
\over \mc{V}^{4/3}} + ... \, .  
\ee 
(We have dropped terms depending
only on the brane and complex structure moduli.)  Here $f_1$ and $f_2$
are functions of the moduli whose forms are unknown for a generic CY
manifold. If they depend upon the inflaton $\tau_i$ polynomially, a
mass term will arise for $\tau_i$ with the possibility of an
$\eta$-problem. Further study is needed to decide.  Note that,
although the exact form of higher (sub-leading) corrections is unknown,
any correction which introduces dependence on $\tau_i$ only via
$\mc{V}$ will not generate any new mass terms for $\tau_i$. 

As well as $\tau$-corrections there are possible
$\theta$-corrections. Non-perturbative effects modify the
superpotential by breaking the shift symmetry, making it discrete. As
noted we do include these. Although leading perturbative terms leave
the K\"ahler potential $\theta$-independent, subleading corrections
can lead to $\theta$-dependent modifications, which we ignore here.

In this paper, we assume that the higher corrections, though possibly
destroying slow roll at large values of the inflaton, are not
important during the last stage of inflation.

\section{Effective Potential and Volume Stabilization}\label{sec:pot}

In this Section, we sketch the derivation of the effective field
theory potential starting from
equations (\ref{ScalP},\ref{Super},\ref{KahlerSimple}).  We choose $T_2$
to be the inflaton field and study its dynamics in the 4-dimensional
effective theory.  We first have to ensure that the volume modulus
$\mc{V}$ and other K\"ahler moduli are trapped in their minima and
remain constant or almost constant during inflation.  For this we have
to focus on the effective potential of all relevant fields.

Given the K\"ahler potential and the superpotential, it is
straightforward but tedious to compute the scalar potential as a
function of the fields $T_i$.  To make all computations we modified the
{\sc SuperCosmology} Mathematica package \cite{supercosmo} which
originally was designed for real scalar fields to manipulate complex
fields.

The K\"ahler potential (\ref{KahlerSimple}) gives rise to the 
K\"ahler metric
$K_{i\bar{j}}=\frac{\partial^2 K}{\partial T^i \partial \bar{T}^{\bar{j}}}$, with
\begin{eqnarray}
  K_{1\bar{1}} = \frac{3\alpha (4\mc{V}-\xi + 6\alpha (\sum_{k=2}^n
  \lambda_k \tau_k^{3/2}))}{4 (2\mc{V}+\xi)^2 (\frac{\mc{V}}{\alpha} +
  \sum_{k=2}^n \lambda_k\tau_k^{3/2})^{1/3}}\ , && K_{i\bar{j}} =
  \frac{9\alpha^2\lambda_i\lambda_j\sqrt{\tau_i\tau_j}}{2(2\mc{V}+\xi)^2}\
  ,\nonumber\\ K_{1\bar{j}} =
  -\frac{9\lambda_j\sqrt{\tau_j}(\alpha^5(\mc{V}+\alpha\sum_{k=2}^n
  \lambda_k\tau_k^{3/2}))^{1/3}}{2(2\mc{V}+\xi)^2}\ , && K_{i\bar{i}}
  =
  \frac{3\alpha\lambda_i(2\mc{V}+\xi+6\alpha\sum_{k=2}^n\lambda_k\tau_k^{3/2})}
  {4(2\mc{V}+\xi)^2\sqrt{\tau_i}}\ .\qquad
\end{eqnarray}
This can be inverted to give
\begin{eqnarray}
  K^{1\bar{1}} =
  \frac{4(2\mc{V}+\xi)(\mc{V}+\alpha\sum_{k=2}^n\lambda_k\tau_k^{3/2})^{1/3}(2\mc{V}
  +\xi+6\alpha\sum_{k=2}^n\lambda_k\tau_k^{3/2})}{3\alpha^{4/3}(4\mc{V}-\xi)}\
  , K^{i\bar{j}} = \frac{8(2\mc{V}+\xi) \tau_i \tau_j}{4\mc{V}-\xi}\ ,
  \nonumber\\ K^{1\bar{j}} =
  \frac{8(2\mc{V}+\xi)\tau_j(\frac{\mc{V}}{\alpha}+\sum_{k=2}^n\lambda_k\tau_k^{3/2})^{2/3}}{4\mc{V}-\xi}\
  , K^{i\bar{i}} =
  \frac{4(2\mc{V}+\xi)\sqrt{\tau_i}(4\mc{V}-\xi+6\alpha\sum_{k=2}^n\lambda_k\tau_k^{3/2})}{3\alpha(4\mc{V}
  -\xi)\lambda_2}\ .
\end{eqnarray}
This is the full expression for an arbitrary number of K\"ahler moduli
$T_i$. The entries of the metric contain terms of different orders in
the inverse volume. If we were to keep only the lowest order terms 
$\gsim \mc{O}({1\over\mc{V}^3})$,
the shape of the trajectories we determine in the following sections and our
conclusions would remain practically unchanged. That is, we are working with
higher precision than necessary. Note that the kinetic terms for
$\tau$ and $\theta$ are identical, appearing as $K_{2\bar{2}}
(\partial \tau\partial\tau + \partial\theta\partial\theta)$ in the
Lagrangian.

The resulting potential is 
\begin{eqnarray}
\label{pot}
  V(T_1, ..., T_{n})&=&\frac{12 W_0^2 \xi}{(4\mc{V}-\xi)(2\mc{V}+\xi)^2}+\sum_{i=2}^{n}
  \frac{12e^{-2 a_i \tau_i}\xi A_i^2}{(4\mc{V}-\xi)(2\mc{V}+\xi)^2}
  +\frac{16 (a_i A_i)^2 \sqrt{\tau_i} e^{-2 a_i \tau_i}}{3\alpha\lambda_2 (2\mc{V}+\xi)} \\ \nonumber
  &&{}+\frac{32 e^{-2 a_i\tau_i} a_i A_i^2\tau_i (1+a_i\tau_i)}{(4\mc{V}-\xi)(2\mc{V}+\xi)}
  + \frac{8 W_0 A_i e^{-a_i\tau_i}\cos(a_i\theta_i)}{(4\mc{V}-\xi)(2\mc{V}+\xi)}\left(\frac{3\xi}{(2\mc{V}+\xi)}
  +4 a_i\tau_i\right)
 \\ \nonumber
 &&{}+\sum_{{i,j=2\atop i<j}}^{n}
\frac{A_i A_j\cos(a_i \theta_i - a_j \theta_j)}{(4\mc{V} - \xi) (2\mc{V} + \xi)^2 } \, e^{-(a_i\tau_i+a_j\tau_j)}
 \left[32(2\mc{V}+\xi) (a_i\tau_i +a_j\tau_j  \right. \\  \nonumber
 &&{}\left. + 2a_i a_j\tau_i \tau_j) + 24 \xi \right] + V_{\mathrm{uplift}} \ .
\end{eqnarray}
We have to add here the uplift term $V_{\mathrm{uplift}}$ to get a
Minkowski or tiny dS minimum.  Uplifting is not just a feature needed
in string theory models. For example, uplifting is done in QFT to tune
the constant part of the scalar field potential to zero. At least in
string theory there are tools for uplifting, whereas in QFT it is a
pure tuning (see, {\it e.g.}, \cite{KKLT,Burgess:2003ic} and references therein). 
We will adopt the form 
\be\label{uplift}
V_{\mathrm{uplift}}=\frac{\beta}{\mc{V}^2} \, , 
\ee 
with $\beta$ to be adjusted. 

We now discuss the stabilization of all moduli $T_i$ plus the volume
modulus.  For this we have to find the global minimum of the potential
eq.(\ref{pot}), which we do numerically. However, it is instructive to give
analytic estimations. Following \cite{BBCQ,CQS}, we study an
asymptotic form of eq.(\ref{pot}) in the region where both $\mc{V}
\sim \exp{(a_i\tau_i)}$, $i=2,...,n$, and $\mc{V} \gg 1$.  The
potential is then a series of inverse powers of $\mc{V}$.  Keeping the
terms up to the order $\mc{O}(\frac{1}{\mc{V}^3})$ we obtain \be
\label{Vapprox} 
V = \frac{1}{ \mc{V}} \sum_{i=2}^{n} \frac{8 (a_i A_i)^2 \sqrt{\tau_i}}{ \lambda_i \alpha} e^{-2 a_i \tau_i} 
+\frac{4}{\mc{V}^2}
 \sum_{i=2}^{n}  a_i A_i W_0 \tau_i e^{-a_i \tau_i}\cos(a_i\theta_i) 
+ \frac{3 \xi W_0^2}{4 \mc{V}^3}  + \frac{\beta}{\mc{V}^2}\ .
\ee 
The cross terms for different $\tau_i$ do not appear in this asymptotic form, as they
would be of order $1/ \mc{V}^5$.
Requiring $\frac{\partial V}{\partial \tau_i}=0$ and $\cos (a_i\theta_i)=-1$ 
at the minimum of the potential eq.(\ref{Vapprox}), we get
\be
\label{nuexp}
\mc{V}_{min} \sim \frac{W_0 \lambda_2}{a_2 A_2} \sqrt{\tau_{2,min}} e^{a_2\tau_{2,min}} \sim 
\frac{W_0 \lambda_3}{a_3 A_3} \sqrt{\tau_{3,min}} e^{a_3\tau_{3,min}}\sim ...\sim
 \frac{W_0 \lambda_i}{a_n A_n} \sqrt{\tau_{n,min}} e^{a_n\tau_{n,min}}  \ , 
\ee
where $\tau_{i,min}$ are the values of the moduli at the global minimum.
The expression (\ref{Vapprox}) has the structure 
\be
\label{Vap} 
V = \frac{C_1}{ \mc{V}}+\frac{C_2}{ \mc{V}^2}+\frac{C_3}{ \mc{V}^3} \
, \ee 
where the coefficients $C_1, C_2, C_3$ are functions of $\tau_i$ and
$\theta_i$.  $C_1$ and $C_3$ are positive but $C_2$ can be of either
sign.  However, the potential for the volume $ \mc{V}$ has a minimum
only if $C_2 < 0$, which is achieved for $\cos (a_i\theta_i)<0$;
otherwise $V( \mc{V})$ would have a runaway character.  Also if all
$\tau_i$ are very large so that $e^{-a_i \tau_i} \to 0$, then $C_2 \to
0$ and $ \mc{V}$ cannot be stabilized.  Therefore to keep $C_2$ non-zero
and negative we have to require that some of the $i>2$ K\"ahler moduli
$\tau_i$ and their axionic partners are trapped in the minimum. For
simplicity we assume all but $T_2$ are already trapped in the minimum.

\begin{figure}
  (a)\includegraphics[width=\mypicturewidth]{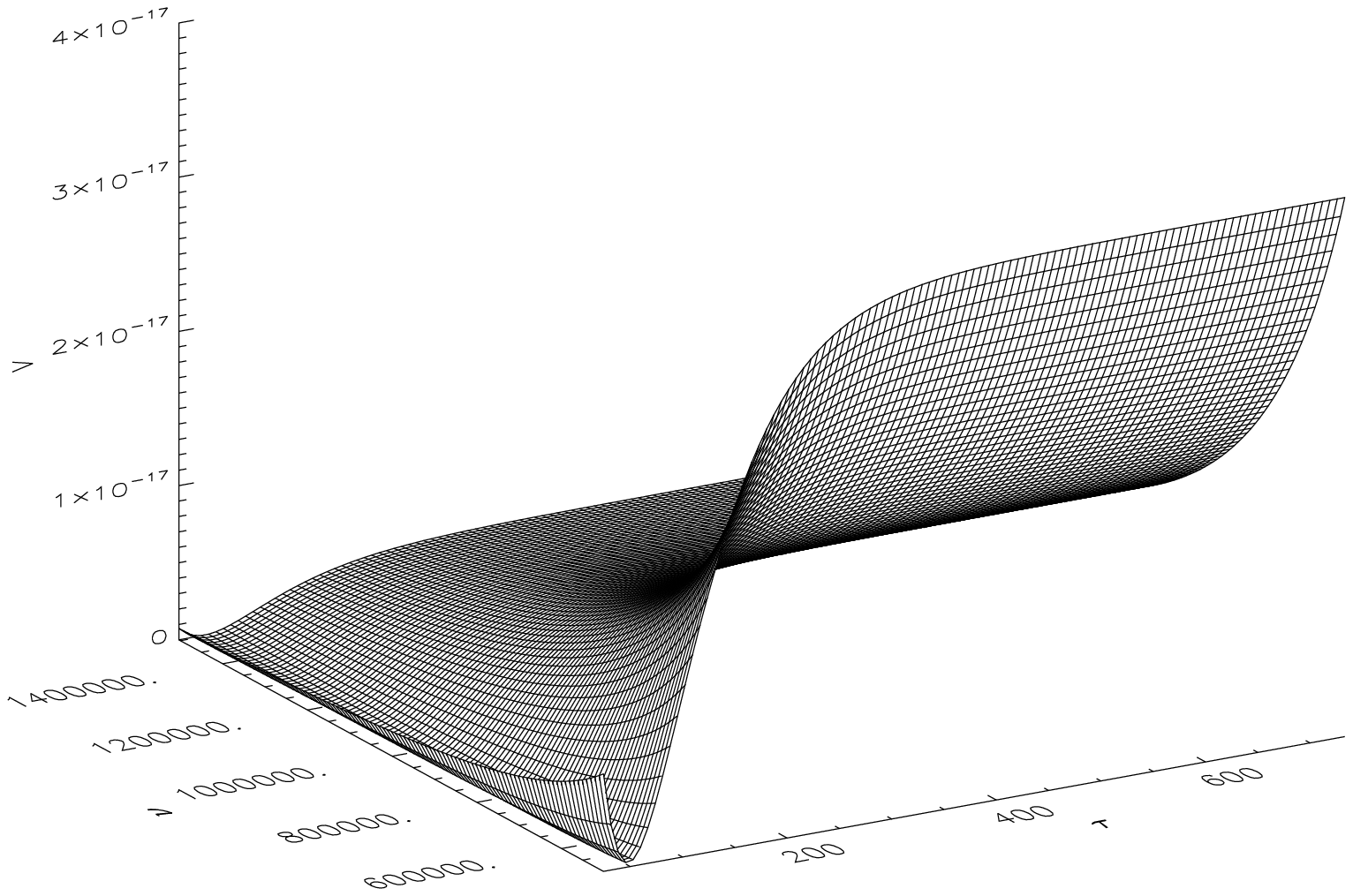}
  (b)\includegraphics[width=\mypicturewidth]{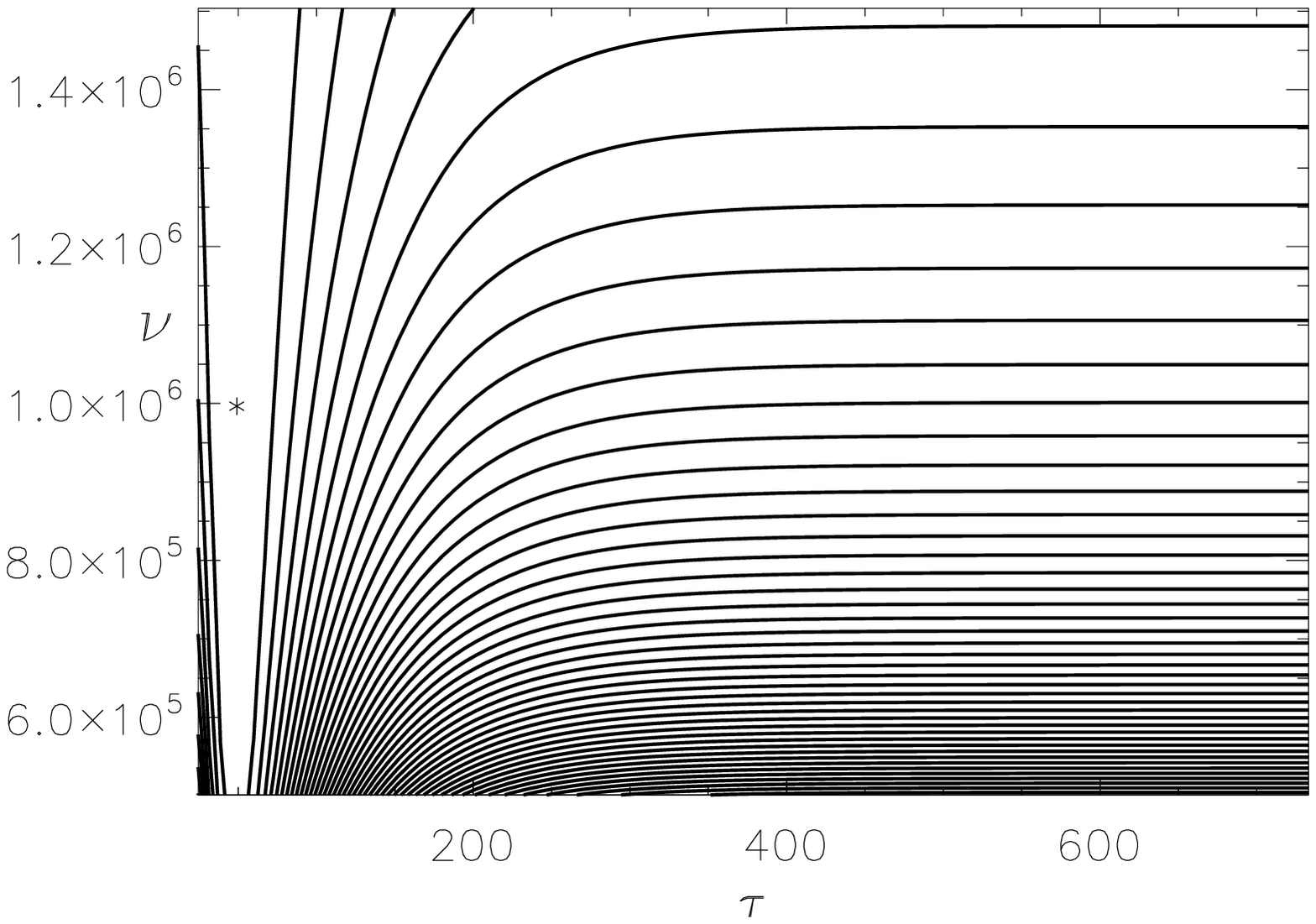}
  \caption{(a) The potential surface $V(\tau_1, \tau_2)$ in a two-K\"ahler 
  model, with the axionic components $\theta_1$ and
    $\theta_2$ fixed at their minima. (b) shows the related contour
    plot of the volume $\mc{V}$ against $\tau_2$. Although it may be
    possible to find a local (very shallow) minimum for the volume in this model
    (marked by a star), the
    generic situation is that both $\tau_1$ and $\tau_2$ will be
    dynamical, and indeed, the evolving $\tau_2$ could force $\tau_1$
    out of a local minimum, thereby destabilizing what may have once
    been stabilized. For this reason, we have focused on models with
    three or more K\"ahler moduli, with all but one mutually enforcing
    their respective stabilizations, and in particular that of the
    volume. This large volume multi-K\"ahler approach to stabilization
    differs from the KKLT stabilization mechanism. }
  \label{fig:destable}
\end{figure}

It is important to recognize that trapping all moduli but one in the
minimum cannot be achieved with only two K\"ahler moduli $\tau_1$ and
$\tau_2$, because $\tau_1$ effectively corresponds to the volume, and
$\tau_2$ is the inflaton which is to be placed out of the minimum.
The Fig.~\ref{fig:destable} shows the potential as a function of $\tau_1$ and
$\tau_2$ for the two-K\"ahler model. One can see from this plot that a
trajectory starting from an initial value for $\tau_2$ larger than a
critical value will have runaway behavior in the $\tau_1$ (volume)
direction.  Thus, as shown by \cite{CQ}, one has to consider a model with
three or more K\"ahler moduli. 

By contrast, the ``better racetrack'' inflationary model based on the
KKLT stabilization is achieved with just two K\"ahler moduli
\cite{Blanco-Pillado:2006he}. However, in our class of models with
three and more K\"ahler moduli we have more flexibility in parameter
space in achieving both stabilization and inflation. Another aspect
of the work in this paper is that a ``large volume" analog of the
``better racetrack'' model may arise.

We have learned that to be fully general we would allow all other moduli including the
volume $\mc{V}$ to be dynamical. This will lead to even richer
possibilities than those explored here, where we only let $T_2$
evolve, and assume that varying it does not alter the values of the
other moduli which we pin at the global minimum.  To demonstrate this
is viable, we need to show the contribution of $T_2$ to the position
of the minimum is negligible. Following \cite{CQ}, we set all
$\tau_i$, $i=2,...,n$, and their axions $\theta_i$ to their minima and
use equations (\ref{Vapprox}) and (\ref{nuexp}) to obtain the potential for
$\mc{V}$: 
\be
\label{Vnu}
V(\mc{V}) = -\frac{3 W_0^2}{2 \mc{V}^3} \left( \alpha\sum_{i=2}^{n} \left[\frac{\lambda_i}{a_i^{3/2}} \right]
 (\ln \mc{V})^{3/2} -\frac{\xi}{2} \right) + \frac{\beta}{\mc{V}^2} \,. 
\ee 
As one can see from eq.(\ref{Vapprox}), the contribution of $T_2$ to
the potential is maximal (by absolute value) when $\tau_2$ and
$\theta_2$ are at their minimum, and vanishes as $\tau_2 \to
+\infty$. This gives a simple criterion for whether the minimum for the
volume $\mc{V}$ remains stable during the evolution of $T_2$: 
the functional form of the potential for $\mc{V}$
(\ref{Vnu}) is insensitive to $T_2$ provided \cite{CQ}: 
\be \label{rho}
\sum_{i=3}^{n}\frac{\lambda_i}{a_i^{3/2}} \gg
\frac{\lambda_2}{a_2^{3/2}} \ .  \ee 
For a large enough
number of K\"ahler moduli this condition is automatically satisfied
for generic values of $a_i$ and $\lambda_i$. We conclude that 
with many K\"ahler moduli the volume does not change during
the evolution of the inflaton $T_2$ because the other $T_i$ stay
at their minimum and keep the volume stable.

\begin{figure}
  (a)\includegraphics[width=\mypicturewidth]{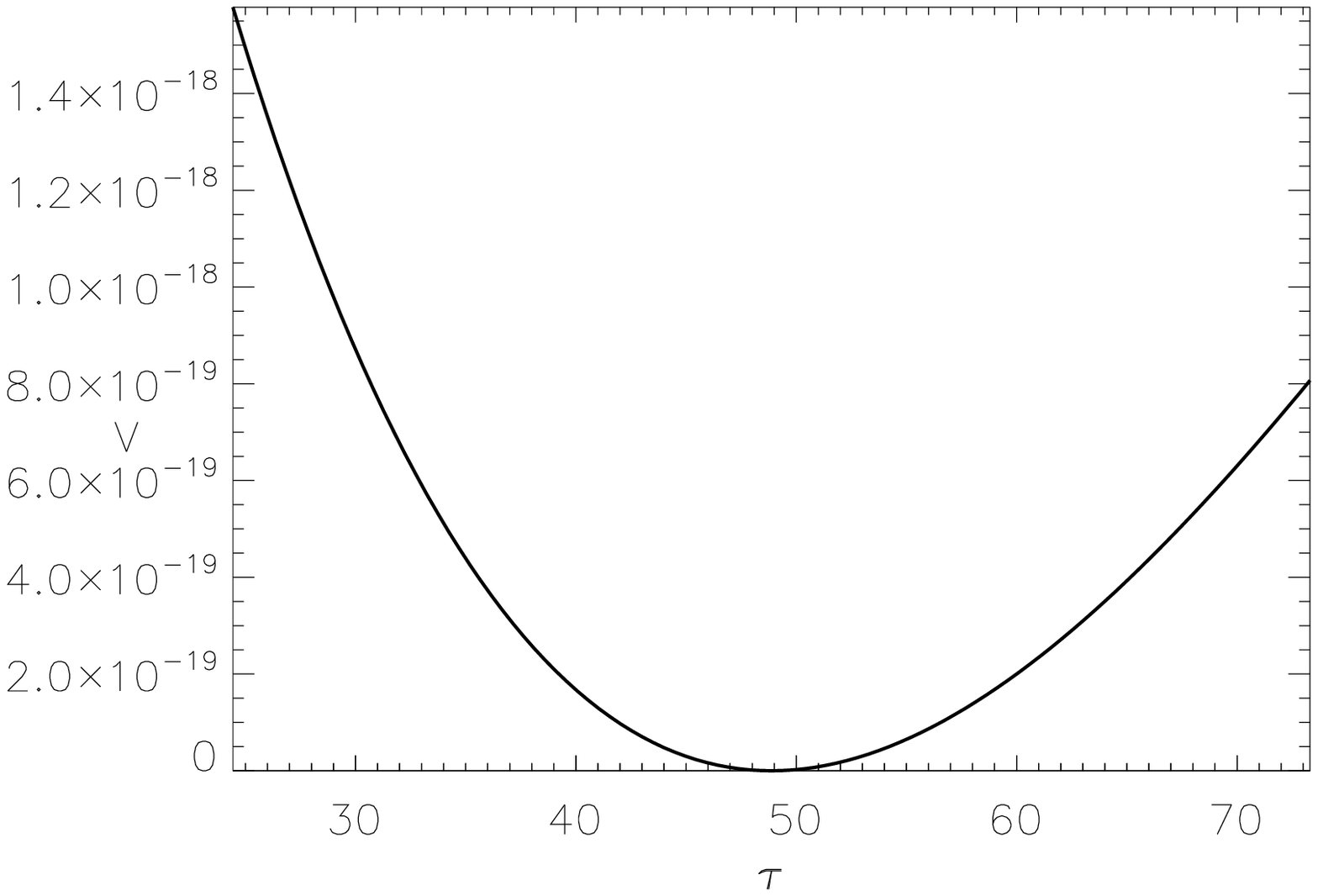}
  (b)\includegraphics[width=\mypicturewidth]{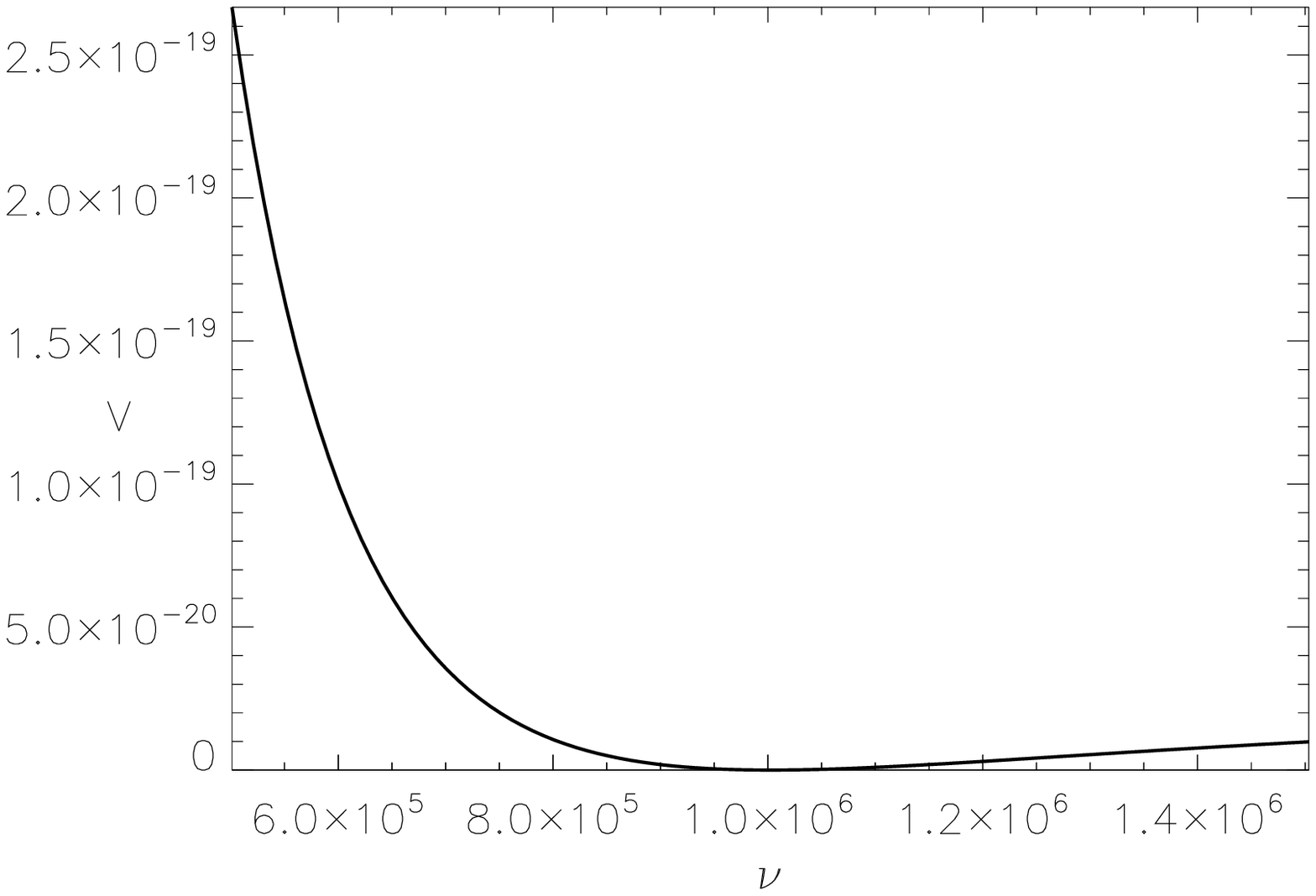}
  \caption{One dimensional sections of the uplifted potential for parameter set 1. 
    We perform a proper uplift procedure by explicitly introducing the additional 
    field $\tau_3$ which is responsible for stabilizing the volume during inflation.
    The parameters for $\tau_3$ are chosen in such a way that the stability condition
    is fulfilled and at the same time we recover the desired value of 
    $\mc{V}_{min}\approx10^6$. The minimum in $\tau_3$ direction is clearly visible at 
    $\tau_{3,min}\approx 49$.}
  \label{fig:uplift_sections}
\end{figure}

Consider a toy model with three K\"ahler moduli in which $T_2$ is the 
inflaton and $T_3$ stays at its own minimum to provide
an unvarying minimum for $\mc{V}$. We choose parameters as in set 1 in
Table 1 (which will be explained in detail below in Sec.~\ref{sec:inf}), and also let $a_3 = 2\pi/300$,
$A_3 = 1/200$, and $\lambda_3 =10$.  Eq.(\ref{rho}) is strongly
satisfied, ${\frac{\lambda_2}{a_2^{3/2}}}\Big\slash
{\frac{\lambda_3}{a_3^{3/2}}}=10^{-4}$, under this choice of parameters.  Therefore
we can drop the $\tau_2, \theta_2$-dependent terms in the potential
(\ref{Vapprox}) and use it as a function of the two fields $\mc{V}$
and $\tau_3$ to find their values at the minimum (after setting also
$\theta_3 = \pi/a_3$ to its minimum). The minimization procedure
should also allow one to adjust the uplift parameter $\beta$ in a way
that the potential vanishes at its global minimum. With our choice of
parameters we found the minimum numerically to be at $\tau_3=49$ and
$\mc{V}=10^6$ with $\beta = 8.5\times 10^{-6}$, as shown in 
Fig.~\ref{fig:uplift_sections}.

\section{Inflaton Potential}\label{sec:inf}

We now take all moduli $T_i$, $i > 3$, and the volume $\mc{V}$ (hence
$T_1$) to be fixed at their minima, but let $T_2$ vary, since it is
our inflaton. For simplicity in the subsequent sections we drop the
explicit subscript, setting $T_2= \tau+i \theta$. The scalar potential
$V(\tau, \theta)$ is obtained from eq.(\ref{pot}) with the other
K\"ahler moduli stabilized:
\begin{eqnarray}
\label{pot1}
  &{}& V(\tau,\theta)=\frac{12 W_0^2 \xi}{(4\mc{V}_m-\xi)(2\mc{V}_m+\xi)^2}
  +\frac{D_1+12e^{-2 a_2 \tau}\xi A_2^2}{(4\mc{V}_m-\xi)(2\mc{V}_m+\xi)^2}
  +\frac{D_2+\frac{16 (a_2 A_2)^2}{3\alpha\lambda_2} \sqrt{\tau} e^{-2 a_2 \tau}}{(2\mc{V}_m+\xi)} \\\nonumber
  &&+\frac{D_3+32 e^{-2 a_2 \tau} a_2 A_2^2\tau (1+a_2\tau)}{(4\mc{V}_m-\xi)(2\mc{V}_m+\xi)}
  + \frac{D_4+8 W_0 A_2 e^{-a_2\tau}\cos(a_2\theta)}{(4\mc{V}_m-\xi)(2\mc{V}_m+\xi)}\left(\frac{3\xi}{(2\mc{V}_m+\xi)}
  +4 a_2\tau\right) +\frac{\beta}{\mc{V}_m^2} \ . 
\end{eqnarray}
Here the terms $D_1, ..., D_4$ contain contributions from the
stabilized K\"ahler moduli other than the inflaton.  We dropped cross
terms between $\tau$ and other $\tau_i$, $i>3$, since these are
suppressed by inverse powers of $\mc{V}_m$.  We can trust
eq.(\ref{pot1}) only up to the order $1/\mc{V}^3$; at higher orders in
$1/\mc{V}$, higher perturbative corrections to the K\"ahler potential
eq.(\ref{loop_corr}) start contributing. Explicitly expanding to order
$1/\mc{V}^3$ yields the simpler expression
\be
\label{appr_pot}
V(\tau, \theta)= \frac{8 (a_2 A_2)^2 \sqrt{\tau} e^{-2a_2 \tau}}{3\alpha \lambda_2 \mc{V}_m}
-\frac{4 W_0 a_2 A_2 \tau e^{-a_2 \tau} \cos{(a_2 \theta)}}{\mc{V}_m^2}
+\Delta V \ ,
\ee
where 
\be
\Delta V = \frac{3 W_0^2 \xi}{4 \mc{V}_m^3}+ \frac{D_2}{\mc{V}_m} + \frac{\beta-D_4}{\mc{V}_m^2} 
\ee
is a constant term, since $\mc{V}$ and $\tau_i$, $i > 3$ are all stabilized at the minimum, 
and $D_2$, $D_4$  depend only on these $\tau_i$.

\begin{table}
  \begin{center}
  \begin{tabular}{|c|c|c|c|c|c|c|c|c|c|c|}
    \hline
    Parameter& $W_0$ & $a_2$ & $A_2$ & $\lambda_2$ & $\alpha$ & $\xi$ & $g_s$ & $\mathcal{V}$& $\Delta \varphi \slash M_p$\\
    \hline
    Parameter set 1  & $300$ & $2\pi\slash 3$ & $0.1$ & $1$ & $1\slash 9\sqrt{2}$ & $0.5$ & $1\slash 10$ & $10^6$ & $2\times10^{-3}$\\ 
    Parameter set 2  & $6\times 10^4$ & $2\pi\slash 30$ & $0.1$ & $1$ & $1\slash 9\sqrt{2}$ & $0.5$ & $1\slash 10$ & $10^8$ & $ 1\times 10^{-3}$ \\  
    Parameter set 3  & $4\times 10^5$ & $\pi\slash100$ & $1$ & $1$ & $1\slash 9\sqrt{2}$ & $0.5$ & $1\slash 10$ & $10^9$ & $1.4\times 10^{-3}$\\ 
    Parameter set 4  & $200$ & $\pi$ & $0.1$ & $1$ & $1\slash 9\sqrt{2}$ & $0.5$ & $1\slash 10$ & $10^6$ & $1.5\times10^{-3}$\\
    Parameter set 5  & $100$ & $2\pi\slash 3$ & $0.1$ & $1$ & $1\slash 9\sqrt{2}$ & $0.5$ & $1\slash 10$ & $10^6$ & $1.9\times 10^{-3}$ \\ 
    Parameter set 6  & $75$  & $2\pi\slash 6$ & $1$   & $1$ & $1\slash 9\sqrt{2}$ & $0.5$ & $1\slash 10$ & $10^8$ & $4\times 10^{-4}$\\
    \hline
  \end{tabular}
  \end{center}
  \label{tab:valuesforconstants}
  \caption{Sample parameter sets for the $T_2$ inflation model. Sets 1
    to 4 define models in which the primordial power spectra of
    perturbations are approximately compatible with observations,
    whereas sets 5 and 6 do not. There are approximate scaling
    relations which map one set of parameters without changing the
    power spectrum (see explanation in the text regarding $W_0$). The
    value of $W_0$ in sets 2 and 3 is so high that corrections to the
    potential seem likely to appear. We note however that these
    large values can be reduced using the scaling transformation, but
    the undesirable cost is that $\tau$ can drop below the string
    scale or $a_2$ can become too large. The last column shows
    approximate values for the variation of the canonically-normalized
    inflaton eq.(\ref{eq:canon}) over the observable e-fold range,
    appropriately small {\it cf.} the Planck scale. }
\end{table}

\begin{figure}
  (a)\includegraphics[width=\mypicturewidth]{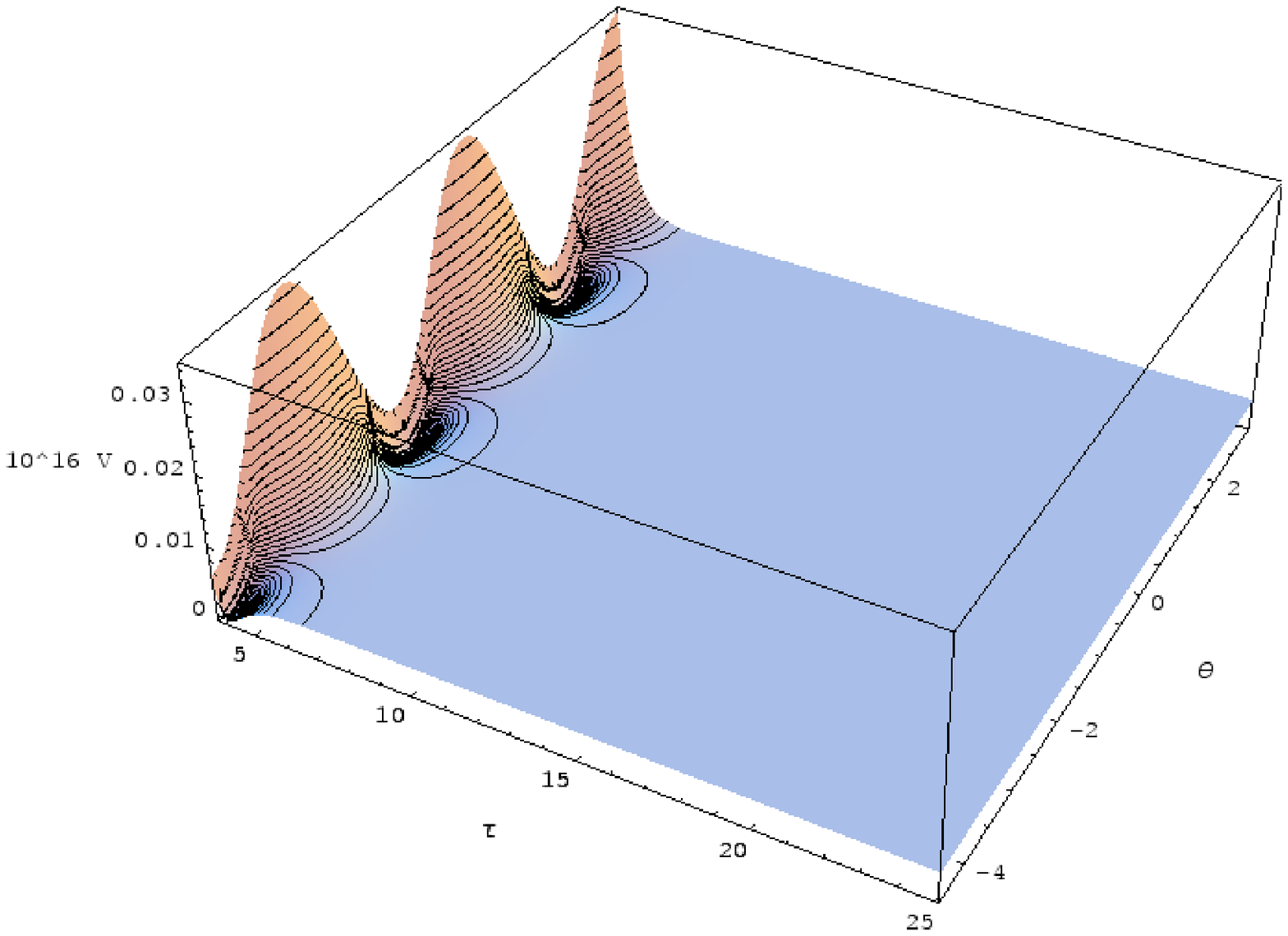}
  (b)\includegraphics[width=\mypicturewidth]{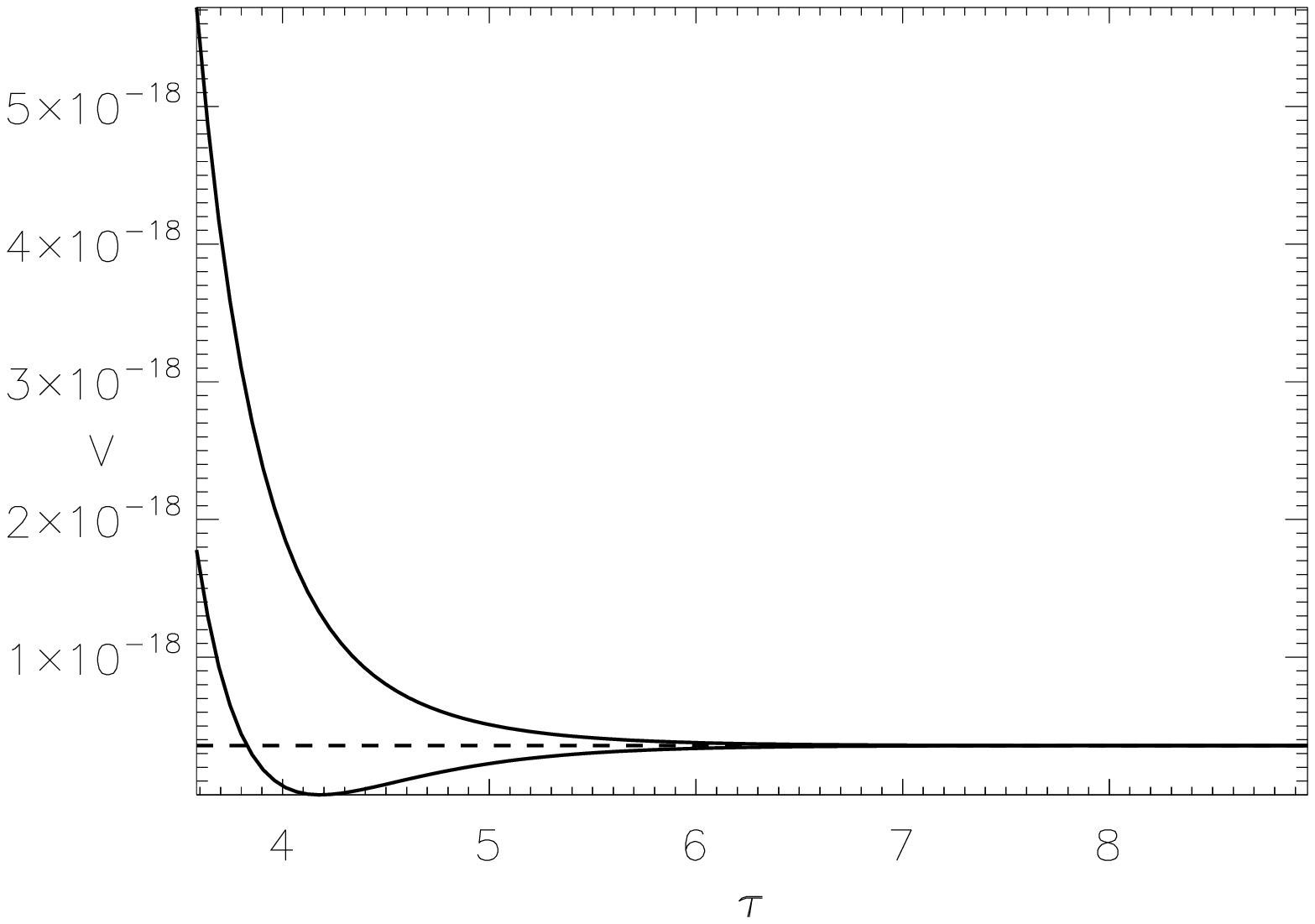}
  \caption{(a) The $T_2$-potential surface $V(\tau ,\theta)$ for the
    parameter set 1 of Table~1. 
    Equipotential contour lines
    are superposed. Both $\tau$ and $\theta$ are multiplied by the
    characteristic scale $a_2$. The surfaces for the other parameter
    sets in the Table are very similar when $a_2$-scaled, even
    sets 5 and 6 with high $W_0$. In all cases, we manually uplifted
    the potential to have zero cosmological constant at the minimum.
    The periodicity in the axionic direction and the constancy at
    large $a_2 \tau$ are manifest. If instead of $\tau$, we used the
    canonically-normalized field eq.(\ref{eq:canon}) which is amplified
    by $\sqrt{\mc{V}}$, the undulating nature in $\theta$ at large
    $\varphi$ becomes more evident. However, the
    canonically-normalized inflaton is trajectory dependent and not a
    global function.  (b) $V(\tau)$ for $\theta$ at $\cos(a_2 \theta)
    = \pm 1$ shows how the flow for the positive value from large
    $\tau$ would be inward, but the flow would be outward for the
    negative value (and be unstable to $\theta$ perturbations). The
    dashed line is $V_\infty$, the $\tau\rightarrow\infty$ asymptote.}
  \label{fig:potential3d_good}
\end{figure}

The potential eq.(\ref{appr_pot}) has seven parameters $W_0, a_2,
A_2, \lambda_2,\alpha, \xi$, and $g_s$ whose meaning was explained in
\S~\ref{sec:model}.  We have investigated the shape of the potential
$V(\tau, \theta)$ for a range of these parameters.  $W_0$, $a_2$,
$A_2$ control the low energy phenomenology of this model (see
\cite{CAQS}) and are the ones we concentrate on here for our inflation
application.  We shall not deal with particle phenomenology aspects in
this paper. Some choices of parameters $W_0$, $a_2$, $A_2$ seem to be
more natural (see \cite{CAQS}). To illustrate the range of potentials,
we have chosen the six sets of parameters given in Table~1. There is
some debate on what are likely values of $W_0$ in string theory. We
chose a range from intermediate to large. Since there are scaling
relations among parameters, we can relate the specific ones we have
chosen to others. An estimate of the magnitude of $W_0$ comes from a
relation of the flux 3-forms $F_3$ and $H_3$ which appear in the
definition of $W_0$ (eq.\ref{Super}) to the Euler characteristic
$\tilde{\chi}$ of the F-theory 4-fold, which is $\tilde{\chi}\sim \int
F_3 \wedge H_3$ from the tadpole cancellation condition.  This
suggests an approximate upper bound $W_0 \sim \sqrt{\tilde{\chi}}$
\cite{BB}. For typical values of $\tilde{\chi} \sim 10^3$, we
would have $W_0 \sim 10-100$.  There are examples of manifolds with
$\tilde{\chi}$ as large as $10^6$, which would result in $W_0 \sim
10^3$. Further, the bound itself can be evaded by $F_3 \wedge
H_3=0$ terms. However, we do not wish to push $W_0$ too high so that we can
avoid the effects of higher perturbative corrections.  We can use the
scaling property for the parameters to move the value of $W_0$ into a
comfortable range. This should be taken into account while examining
the table.

The  parameter sets in the table  can be
divided into two classes: Trajectories in sets $1\dots4$ produce a
spectrum of scalar perturbations that is comparable to the
experimentally observed one (good parameter sets), whereas
trajectories in parameter set $5$ and $6$ produce spectra whose
normalization is in disagreement with observations (bad parameter
sets). Sets 3 and 4 were chosen to large values of $W_0$ to illustrate
how things change with this parameter, but we are wary that with such
large fluxes, other effects may come into play for determining the
potential over those considered here. A typical potential surface
$V(\tau, \theta)$ is shown in Fig.~\ref{fig:potential3d_good} with the
isocontours of $V(\tau, \theta)$ superimposed.

The hypersurface $V(\tau, \theta)$ has a rich structure. It is
periodic in $\theta$ with period $2\pi/a_2$, as seen in
Fig~\ref{fig:potential3d_good}. Along $\theta=\frac{\pi (2l+1)}{a_2}$,
where $l$ is integer, the profile of the potential in the $\tau$
direction is that considered in \cite{CQ}. It has a minimum at some
$\tau=\tau_{min}$ and gradually saturates towards a constant value at
large $\tau$, $V(\tau) \to V_\infty (1-C a_2\tau e^{-a_2 \tau})$,
where $C$ is a constant.\footnote{This type of potential is similar to
that derived from the Starobinsky model of inflation
\cite{Starobinsky:1980te} with a $\frac{M_p^2}{2}\left(R-
\frac{1}{6M^2}R^2\right)$ Lagrangian via a conformal transformation,
$V(\phi)=6\pi M_p^2 M^2 [1-exp(-\frac{\phi}{\sqrt{12\pi} M_p})]^2$.}
Along $\theta=\frac{2\pi l}{a_2}$, $V(\tau, \theta)$ falls gradually
from a maximum at small $\tau$ towards the same constant value,
$V(\tau) \to V_\infty (1+Ca_2\tau e^{-a_2 \tau})$. Trajectories
beginning at the maximum run away towards large $\tau$.
Fig.~\ref{fig:Hubble_epsilon}(b) shows these two one-dimensional
sections of the potential. For all other values of $\theta$ the
potential interpolates between these two profiles. Thus, at large
$\tau$, the $V$ surface is almost flat but slightly rippled. At small
$\tau$, the potential in the axion direction is highly peaked. Around
the maximum of the potential it is locally reminiscent of the
``natural inflation'' potential involving a pseudo Goldstone boson
\cite{natural} (except that $\theta$ and $\tau$ must be simultaneously
considered), as well as the racetrack inflation potential
\cite{Blanco-Pillado:2004ns, Blanco-Pillado:2006he}.

There are two approximate scaling symmetries 
in this model (in the asymptotics of (\ref{appr_pot}) for $\mc{V}\gg\xi$), 
similar to those in \cite{Blanco-Pillado:2004ns}, 
\begin{eqnarray}
a_i\rightarrow u^{-1}a_i\ ,\  (A_i, W_0, \xi)\rightarrow u^{3/2} (A_i,
W_0, \xi)\ ,\  (\tau_i, \theta_i)\rightarrow u (\tau_i, \theta_i)\ ,\  (\lambda_i, {\cal P}_s, \epsilon)\rightarrow (\lambda_i, {\cal P}_s, \epsilon)\ ; \nonumber\\
a_i\rightarrow a_i\ ,\  (A_i, W_0)\rightarrow v^{-3/2}(A_i, W_0)\ ,\
(\tau_i, \theta_i)\rightarrow(\tau_i, \theta_i)\ ,\  {\cal P}_s\rightarrow v^3{\cal P}_s\ ,\  (\lambda_i, \xi, \epsilon)\rightarrow(\lambda_i, \xi, \epsilon)\ .
\end{eqnarray}
which can be used to generate families of models, trading for example
large values of $W_0$ for small values of $\tau$. For instance,
applying the $u$ scaling to parameter set 1, we can push the value of
$W_0$ down to $W_0\approx2$, but at the same time pushing $\tau$ to
lie in the range $\tau=0.1\dots1.0$ during inflation, a range which is
quite problematic since at such small $\tau$ higher order string
corrections would become important.

More generally, for the supergravity approximation to be valid the
parameters have to be adjusted to have $\tau_{min}$ at least a few:
$\tau$ is the four-cycle volume (in string units) and the supergravity
approximation fails when it is of the string scale. However, even if
the SUGRA description in terms of the scalar potential is not valid at
the minimum, it still can be valid at large $\tau$, exactly where we
wish to realize inflation.  The consequence of small $\tau_{min}$ is
that the end point of inflation, i.e. preheating, would have to be
described by string theory degrees of freedom. We will return to this
point in the discussion.

\subsection{The Canonically-normalized Inflaton}\label{sec:inflaton}

If we define a canonically-normalized field $\varphi$ by
$d\varphi^2/2 = K_{22}d\tau^2$, then
\be
\label{eq:canon}
\varphi=\sqrt{\frac{4}{3}\frac{\alpha\lambda_2}{\mc{V}+\frac{\xi}{2}}}\left(\frac{\tau}{\ell_s^4}\right)^{3/4} M_p.
\ee
It is therefore volume-suppressed. For inflation restricted to the
$\tau_2$ direction, we identify $\varphi$ with the inflaton. 
The field change $\Delta
\varphi$ over the many inflationary e-foldings $N\equiv -\ln
a/a_{end}$ is given in the last column in Table~1 
for a typical radial ($\tau$) trajectory. It is much less than
$M_p$. (Here the scale factor at the end of inflation is
$a_{end}$ so $N$ goes in the opposite direction to time.)  
The variations of the inflaton and the Hubble parameter {\it
wrt} $N$, 
\be
\label{eq:inflaton}
{d\varphi/M_p \over dN} = \sqrt {2\epsilon}\, , 
\ee 
\be
\label{eq:deceleration}
{d\ln H/M_p \over dN} =\epsilon\ , \epsilon \equiv 1+q \, ,
\ee
suggest we must have a deceleration parameter $q$ nearly the de Sitter
$-1$ and the Hubble parameter $H$ nearly constant over the bulk of the
trajectories.  This is shown explicitly in \S~\ref{sec:plots} and
Figures \ref{fig:Hubble_epsilon} and \ref{fig:Hubble_epsilon_bad}. The
parameter $\epsilon(N)$ is the first ``slow-roll parameter'', although
it only needs to be below unity for inflation. With $\epsilon$ so
small, we are in a {\it very} slow-roll situation until near the end of
inflation when it rapidly rises from approximately zero to unity and
beyond.

Equation~(\ref{eq:inflaton}) connects the change in the inflaton field
to the tensor to scalar ratio $r={\cal P}_t/{\cal P}_s$, since to a good
approximation $r\approx 16\epsilon$. Since we find $\epsilon \ll 1$,
we get very small $r$.  The following relation
\cite{Efstathiou:2005tq,Lyth:1996im} gives a lower limit on the field
variation in order to make tensor modes detectable \be
\frac{\varphi}{M_{\mathrm{pl}}} \approx 0.46 \sqrt{\frac{r}{0.7}}\ .
\ee We are not close to this bound. If $\varphi/M_p$ is restricted to
be $< 1$ in stringy inflation models, getting observable gravity
wave signals is not easy. (A possible way out is to have many
fields driving inflation in the spirit of assisted inflation
\cite{Liddle:1998jc}.) 

When the trajectory is not in the $\tau$ direction, the field
identified with the canonically-normalized inflaton becomes
trajectory-dependent as we describe in the next section and there is
no global transformation. That is why all of our potential contour plots have
focused on the K\"ahler modulus and its axion rather than on the inflaton.

\section{Inflationary Trajectories}\label{sec:plots}

\subsection{The Inflaton Equation of Motion} \label{sec:EOM}

We consider a flat FRW universe with scalar factor $a(t)$ and real
fields $(\tau, \theta)$. To find trajectories, we derive their
equations of motion in the Hamiltonian form starting from the four
dimensional Lagrangian (see \cite{supercosmo} and references therein)
\begin{eqnarray}
  {\cal L}&=&\sqrt{-g} \left(R+  G_{ij} \dot{\phi}^i \dot{\phi}^j - V\right),
\end{eqnarray}
with canonical momentum 
$ P_i = \frac{\partial {\cal L}}{\partial \dot{\phi}^i} = 2 a^3 G_{ij} \dot{\phi}^i$, 
where we used $\sqrt{-g} = a^3$ and $\phi_i$, $i=1,2$ stands for 
$(\tau, \theta)$. (The usual field momentum is $P_i/a^3$.) 
Here the non-canonical kinetic term is 
\be
G_{ij} = K_{2\bar{2}}\delta_{ij},\  K_{i{\bar{j}}} = \frac{\partial^2 K}{\partial T^i \partial \bar{T}^{\bar{j}}}.
\ee
The Hamiltonian is
\be
  \mc{H} = P_i \dot{\phi}^i - {\cal L }= \frac{1}{a^3}  G^{ij} P_i P_j
  + V \, ,
\ee
where $G^{ij}=G_{ij}^{-1}$. The equations of motion follow from 
$\dot{\phi}^i = \frac{\partial \mc{H}}{\partial P_i},
\dot{P_i}=-\frac{\partial \mc{H}}{\partial \phi^i}$, which reduce to 
\be
\label{eom}
\dot{\phi}^i = \frac{1}{2 a^3} G^{ij} P_j\ ,\ \dot{P}_i = -\frac{1}{4a^3} \frac{\partial G^{kl}}{\partial \phi^i} P_k P_l -
 a^3 \frac{\partial V}{\partial \phi^i}\ ,\ \dot{a} = a H\ ,\ \dot{H} = -\frac{1}{4a^6} G^{ij} P_i P_j \ .
\ee
We also use the constraint equation 
\begin{eqnarray}
  M_P^2 H^2 &=& \frac{1}{3} \left(\frac{1}{4 a^6} G^{ij} P_i P_j + V\right) = M_P^2 H^2 \epsilon/3 + V/3
\end{eqnarray}
to monitor the accuracy of the numerical integration routine.  Note
that the deceleration, eq.(\ref{eq:deceleration}), is related to the
non-canonical kinetic term in the implicit manner indicated.  

The inflaton is defined to be the field combination along the
classical (unperturbed) trajectory. Isocurvature (isocon) degrees of
freedom are those perpendicular to the classical trajectory. The
kinetic metric in $M_P$ units is $d\psi^2 = G_{ij}d\phi^i
d\phi^j/M_P^2$. If $\theta$ is fixed, then $\psi$ is related to
$\varphi$ introduced in \S~\ref{sec:inflaton} by $\psi = \varphi
/(\sqrt{2}M_P)$. More generally, the inflaton between the initial condition
$\phi^i(N_0) $ and the value $\phi^i(N_1) $ is the distance along the
path, $\int d\psi$. For our case with diagonal
$G_{ij}$, it is 
\be \psi = \int \sqrt{K_{2\bar{2}}} d\tau (1+ d\theta^2
/d\tau^2 )^{1/2} \, .  
\ee

\subsection{Stochastic Fluctuations and CMB 
and LSS Constraints} \label{sec:stochastic}

The classical trajectory is perturbed by zero point fluctuations in
all fields present, but only degrees of freedom with small mass will
be relevant, hence in $\tau$ and $\theta$, which in turn influence the
scalar metric fluctuations encoded in $\ln a =-N$, and in the
gravitational wave degrees of freedom.  Structure formation depends
upon the fluctuations in the scalar 3-curvature, which are related to
those in $\ln a$ measured on uniform Hubble surfaces \cite{sb91,bh95},
$\delta {}^{(3)}R = 4(k/a)^2 \delta \ln a\vert_H$. The usual result
for single-field inflation is motivated by stochastic inflation
considerations: the zero point fluctuations in the inflaton at
``horizon crossing'' when the three-dimensional wavenumber $k \approx
Ha$ are $\delta \psi = [H/(2\pi
M_P)]/\sqrt{2}$. Equation~(\ref{eq:inflaton}) gives the mapping,
$\delta \ln a\vert_H = \delta \psi /\sqrt{\epsilon}$ along the
inflaton direction. Thus the scalar power spectrum is \be
\label{eq:ps} {\cal P}_s \equiv k^3 /(2\pi^2) <\vert \delta \ln
a\vert_H (k) \vert^2 > = [H/(2\pi M_P)]^2/(2\epsilon)\ e^{2u_s} \, ,
\ee where $u_s$ encodes small corrections to this simple stochastic
inflation Hawking temperature formula.

The graviton zero point oscillations are, like those in a massless
scalar field, proportional to the Hawking temperature at $k=Ha$, 
\be \label{eq:pt}
{\cal P}_{t} (k) =16\, [H/(2\pi M_P)]^2/2\ e^{2u_t}\, .
\ee
where $u_t$ also encodes small
corrections. The ratio $r(k)$ is therefore $\approx 16 \epsilon$. 

The scalar spectral index is given by $n_s -1 = d\ln {\cal P}_s /d\ln
k$. At lowest order, and for small $\epsilon$, it is $n_s -1 = -2
\epsilon - \epsilon^\prime /\epsilon$, where $\epsilon^\prime \equiv
d\epsilon /d\ln a$. To have $\epsilon$ nearly zero as our trajectories
do, and yet have $n_s$ differing at the 2-sigma level from unity as the
current cosmic microwave background (CMB) and large scale clustering
data indicate, imposes a constraint on $\epsilon^\prime /\epsilon
$ which might seem to require a fine-tuning of the potential.

The current best estimate of $n_s$ in flat universe models
characterized by six parameters, in which gravity waves are ignored,
is $0.96\pm 0.017$ with CMB only, and $0.958\pm 0.015$ with CMB and
large scale structure (LSS) clustering data \cite{Acbar06}. The errors
are Bayesian 1-sigma ones. For future reference, we note that the CMB
data prefer a running of the scalar index at about the 2-sigma level,
$dn_s/d\ln k = -0.047 \pm 0.021$ at a pivot point $k_p=0.05 \, {\rm
Mpc}^{-1}$. However, the 6 parameter case with no running is a very
good fit, except in the low $\ell$ regime.  The preferred amplitude of
the scalar perturbations is $ {\cal P}_s (k_p) = [21^{+1.3}_{-1.0}]
\times 10^{-10}$. It is interesting to note that this number has been
stable for a long time: the estimate from the COBE DMR experiment
without the addition of any LSS or smaller scale CMB experiments was
$[21 \pm 3] \times 10^{-10}$ when extrapolated with this $n_s =0.96$
slope, and only slightly higher with no tilt \cite{bh95}.

The current constraint on the gravity wave contribution is ${\cal
P}_t/{\cal P}_s < 0.6$ at the 95\% confidence limit with CMB data
(with the powers evaluated at the pivot point $0.002 \, {\rm
Mpc}^{-1}$). When LSS data is added to the CMB, this drops to an upper
limit of 0.28 but requires a single slope connection of the low $k$
regime in which the tensors can contribute to the CMB signal and high
$k$ where the amplitude of LSS fluctuations is set. Relaxing this
allows for higher values \cite{BCKP}.

With the CMB-determined ${\cal P}_s$ estimate, we have $[H/(2\pi M_P)]
\approx 6.5 \times 10^{-5} \sqrt{\epsilon}$. If the acceleration $\epsilon$
were uniform over the observable range and gave rise to this $n_s$, we
would have $\epsilon =0.02$, $[H/(2\pi M_P)]\approx 10^{-5}$ and $r \sim
0.3$. But our trajectories have $\epsilon$ nearly zero and $H$ almost
flat, so to get the observed $n_s$ the observable range would have to
be well into the braking period towards preheating: {\it i.e.}, we
would need $\epsilon^\prime /\epsilon \approx 0.04$ over the
CMB+LSS window, which seems like it would require a very
finely-tuned potential. Rather remarkably, the first cases we tried
gave values near this at the relevant number of e-foldings before the
end of inflation. This point was also made by Conlon and Quevedo
\cite{CQ} for pure $\tau$ trajectories. We also find there is room for
modest running of the scalar index over the observable window. (Note
that the tensor slope $n_t \equiv d\ln {\cal P}_s /d\ln k$ is $\approx
-2 \epsilon$, hence very small.) A consequence of the small $\epsilon$ is 
$[H/(2\pi M_P)]\sim 10^{-10}$ and $V^{1/4} \sim 10^{14} \, {\rm Gev}$. 

We caution that the single field model we used to estimate ${\cal
P}_s$ includes fluctuations only along the trajectory. There will also
be fluctuations perpendicular to it, and fluctuations in the parallel
and perpendicular directions can influence each other. The latter are
isocurvature degrees of freedom. Although these will leave the ${\cal
P}_{t}$ formula unaffected, we expect modifications in the ${\cal
P}_{s}$ formula, a point we return to after making an inventory of the
sorts of trajectories that will arise.

\begin{figure}
  (a)\includegraphics[width=\mypicturewidth, height=5.5cm]{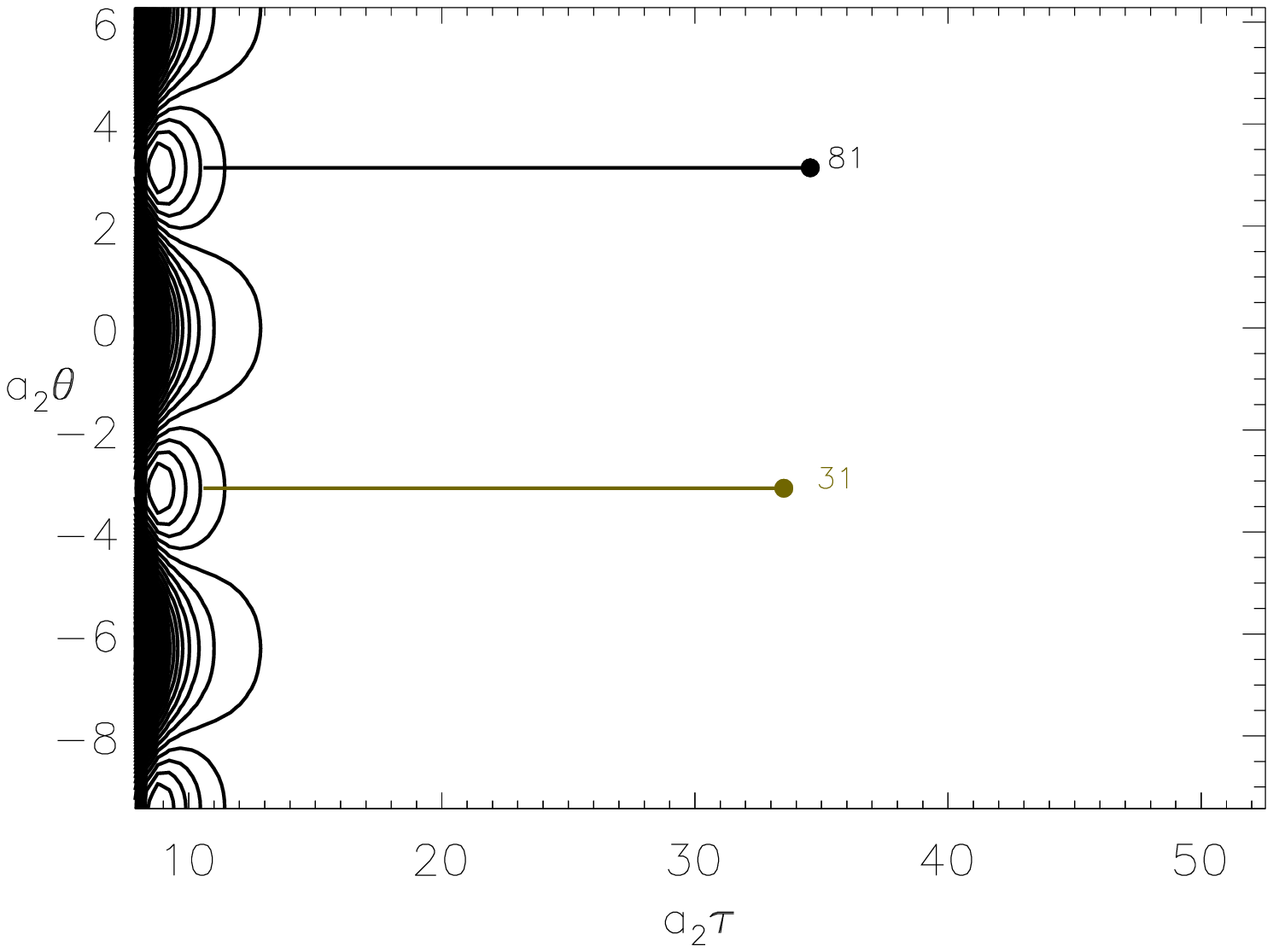}
  (b)\includegraphics[width=\mypicturewidth, height=5.5cm]{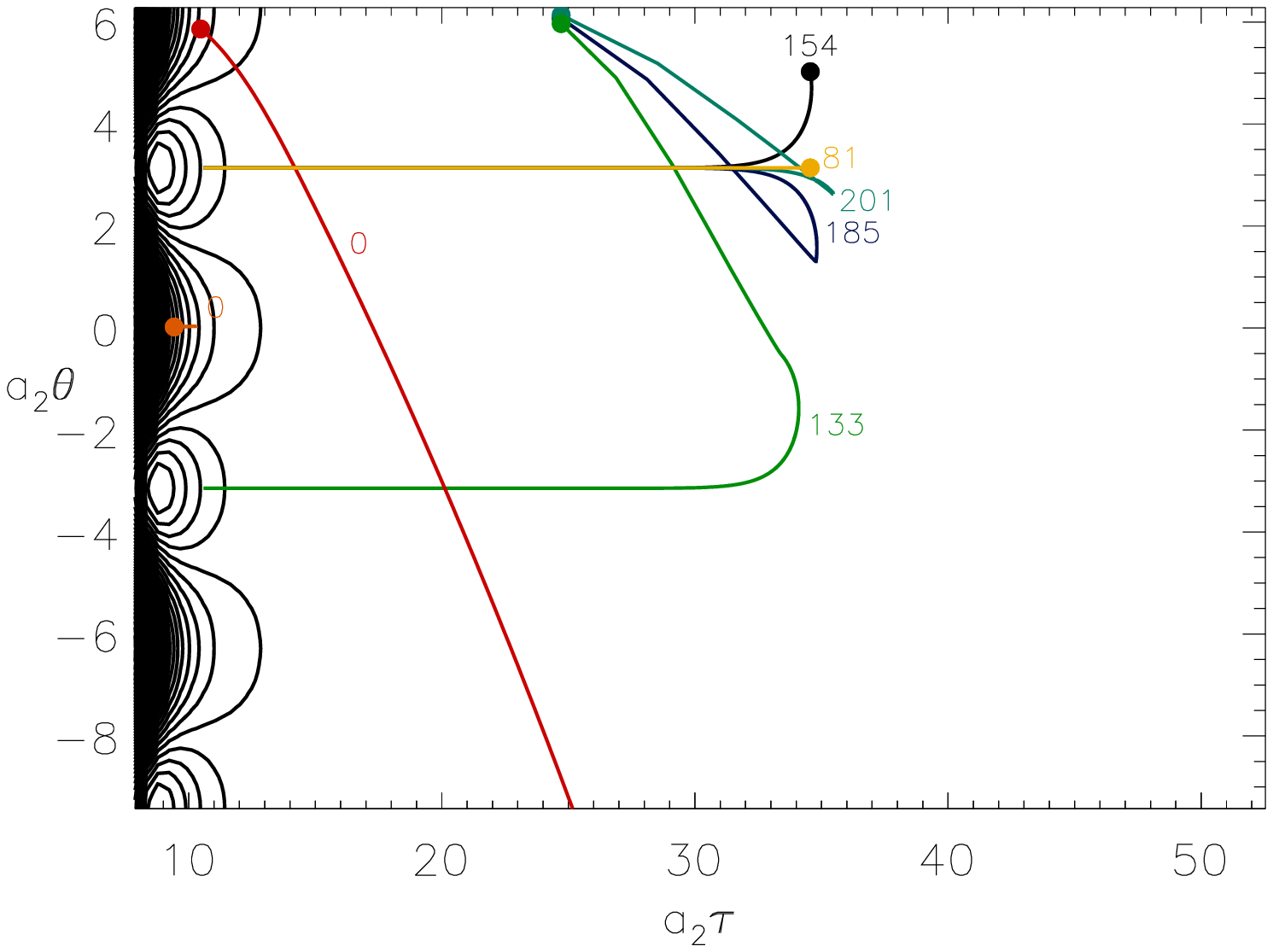}\\
  (c)\includegraphics[width=\mypicturewidth, height=5.5cm]{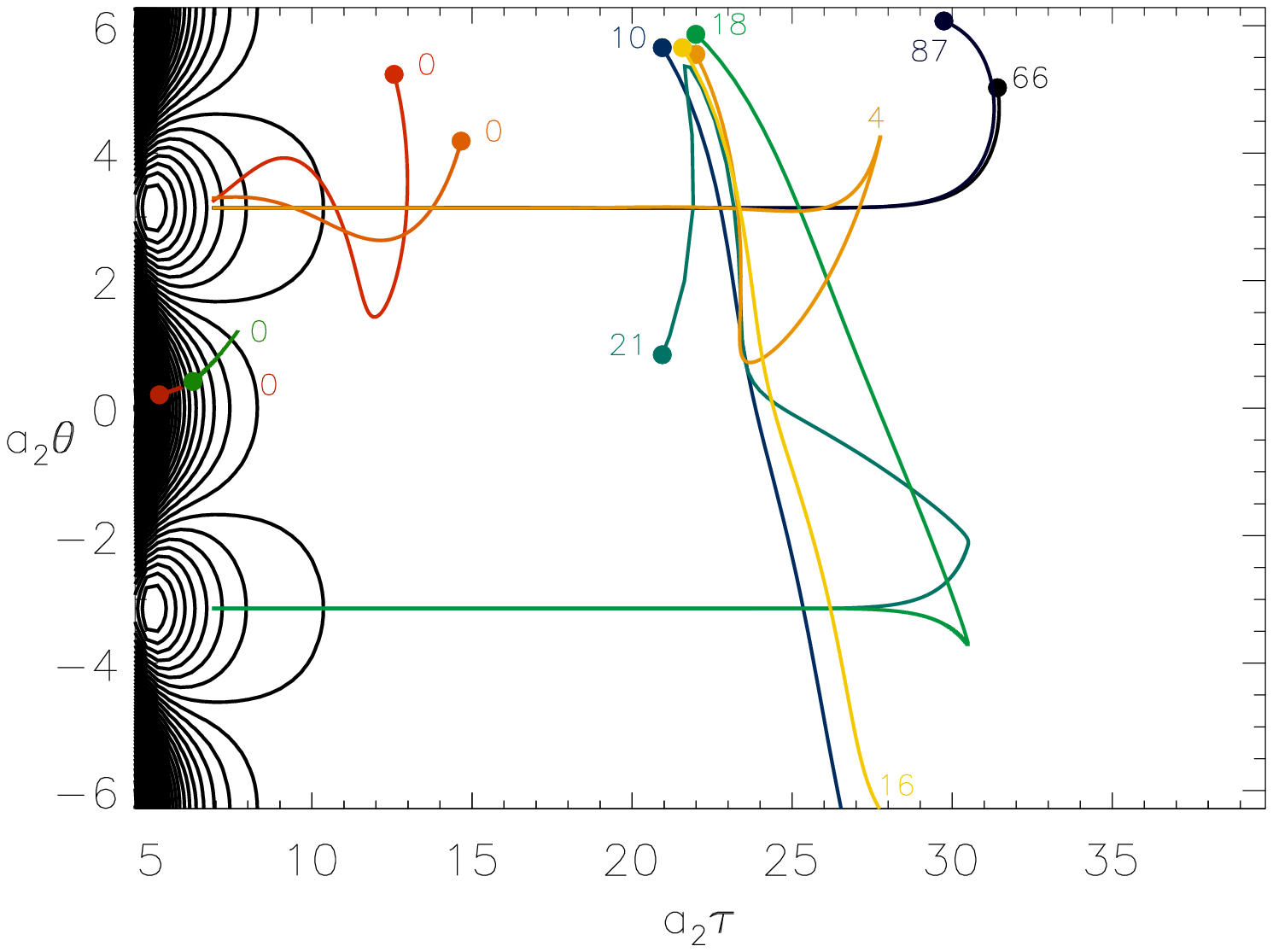}
  (d)\includegraphics[width=\mypicturewidth, height=5.5cm]{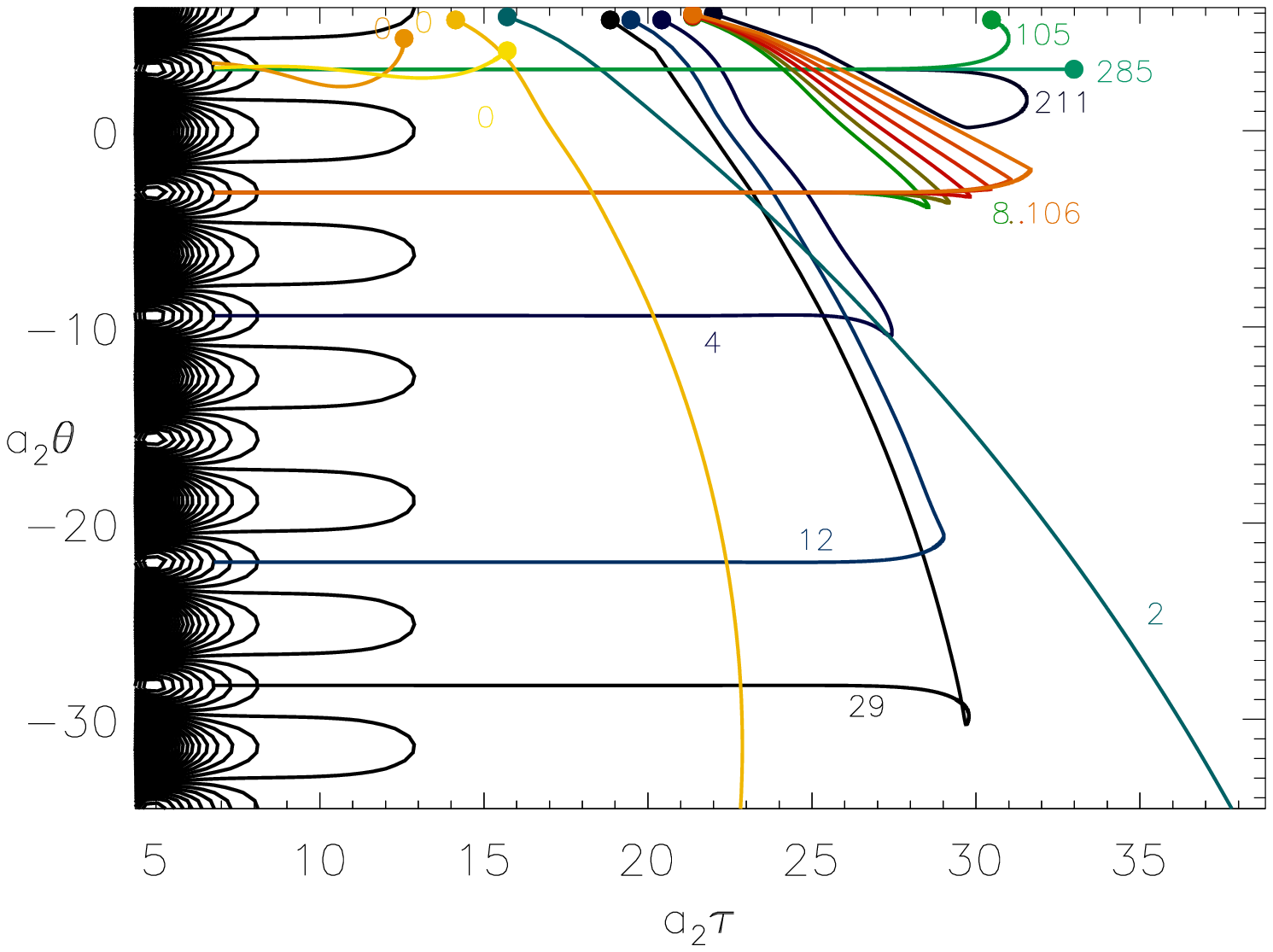}\\
  (e)\includegraphics[width=\mypicturewidth, height=5.5cm]{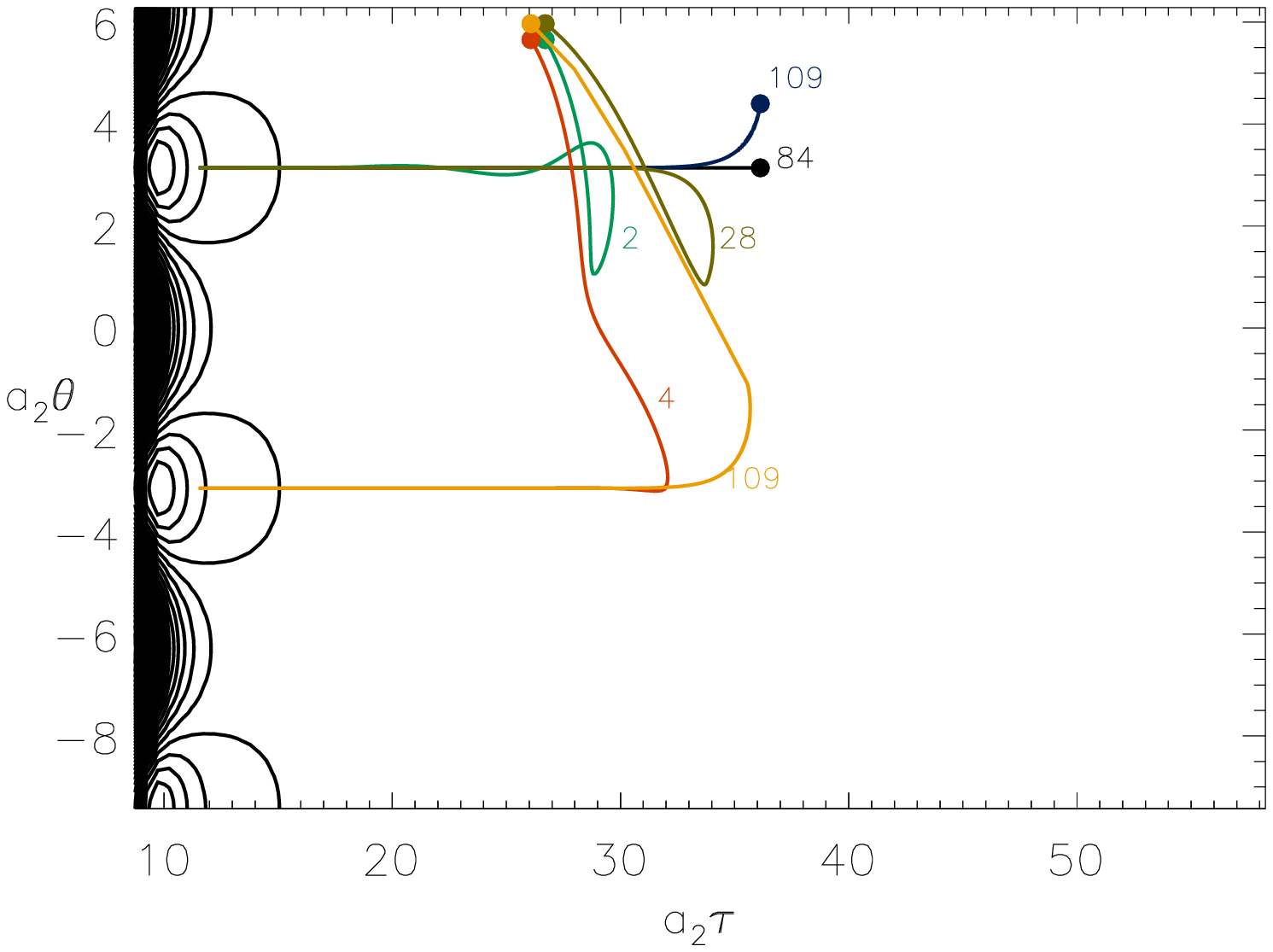}
  \caption{Contour-plots of the potential including trajectories for
    several choices of initial values (denoted by filled circles) in
    field space $(\tau, \theta)$. The trajectories are evolved
    numerically until inflation ends at $\epsilon=1$. The number of
    e-folds is indicated next to the corresponding trajectory. During
    the last stages of inflation, the field always rolls along one of
    the valleys towards the minima which are located in the centre of
    the white circles. The maxima are located in the dark spots.
    Inflation in the axion direction can significantly
    enhance the amount of inflation over that obtained in pure $\tau$
    inflation.  Trajectories starting at large $\tau$ roll to the
    nearest valley, then to the minimum. But starting at intermediate
    $\tau$ with the axion sufficiently far from its minimum, we find
    the field can cross several $\theta$-ridges before settling into a
    valley. Another manifestation is the run-away character for $\tau$
    if the axion is placed close to its maximum.  (a) shows the simple
    pure-$\tau$ inflation if $\theta$ is set to its minimum, as in
    Conlon and Quevedo, for parameter set 1. (b) shows the complex
    evolution for sample general starting conditions, for parameter set 1, 
    (c), (d) and (e) show the same for sets 2, 3 and 4, respectively.}
  \label{fig:contourplot}
\end{figure}

\begin{figure}
  (a)\includegraphics[width=\mypicturewidth, height=5.5cm]{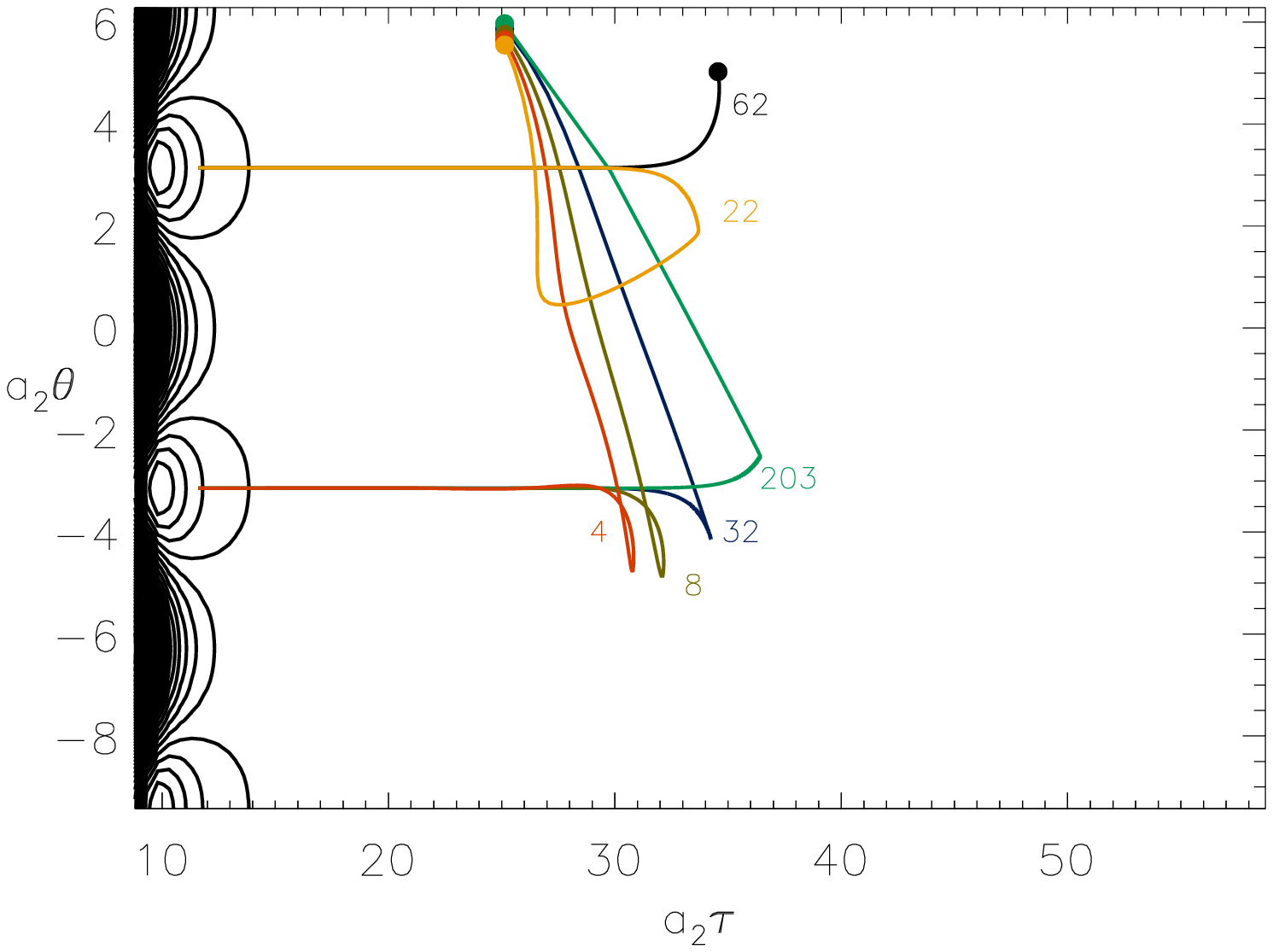}
  (b)\includegraphics[width=\mypicturewidth, height=5.5cm]{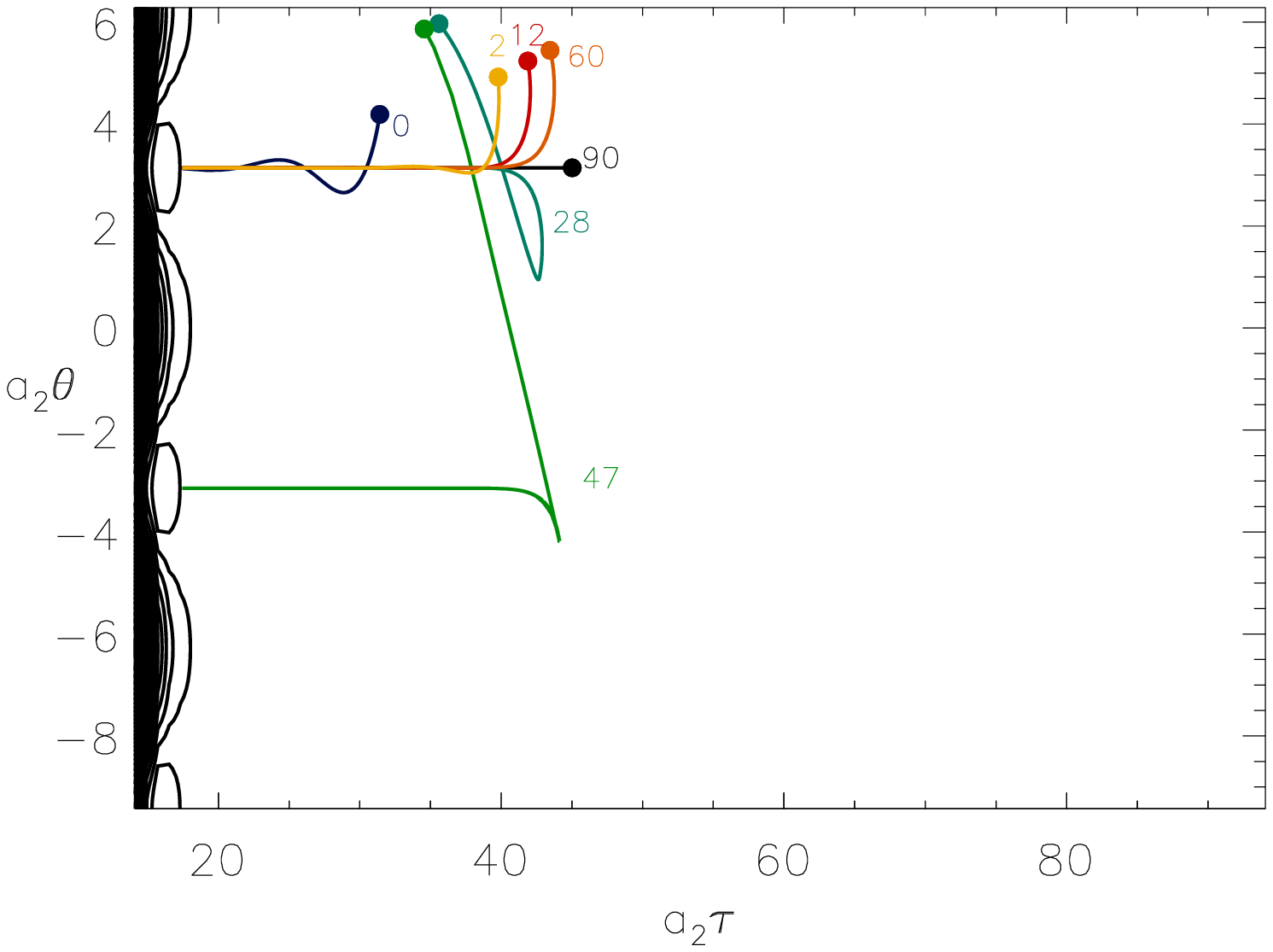}
  \caption{Sample trajectories for parameter sets 5 (a) and 6 (b),
    which have scalar power spectra amplitudes incompatible with the
    data. These still look similar to those in
    Fig.~\ref{fig:contourplot}. }
  \label{fig:contourplot_bad}
\end{figure}

\begin{figure}
  \includegraphics[width=\textwidth]{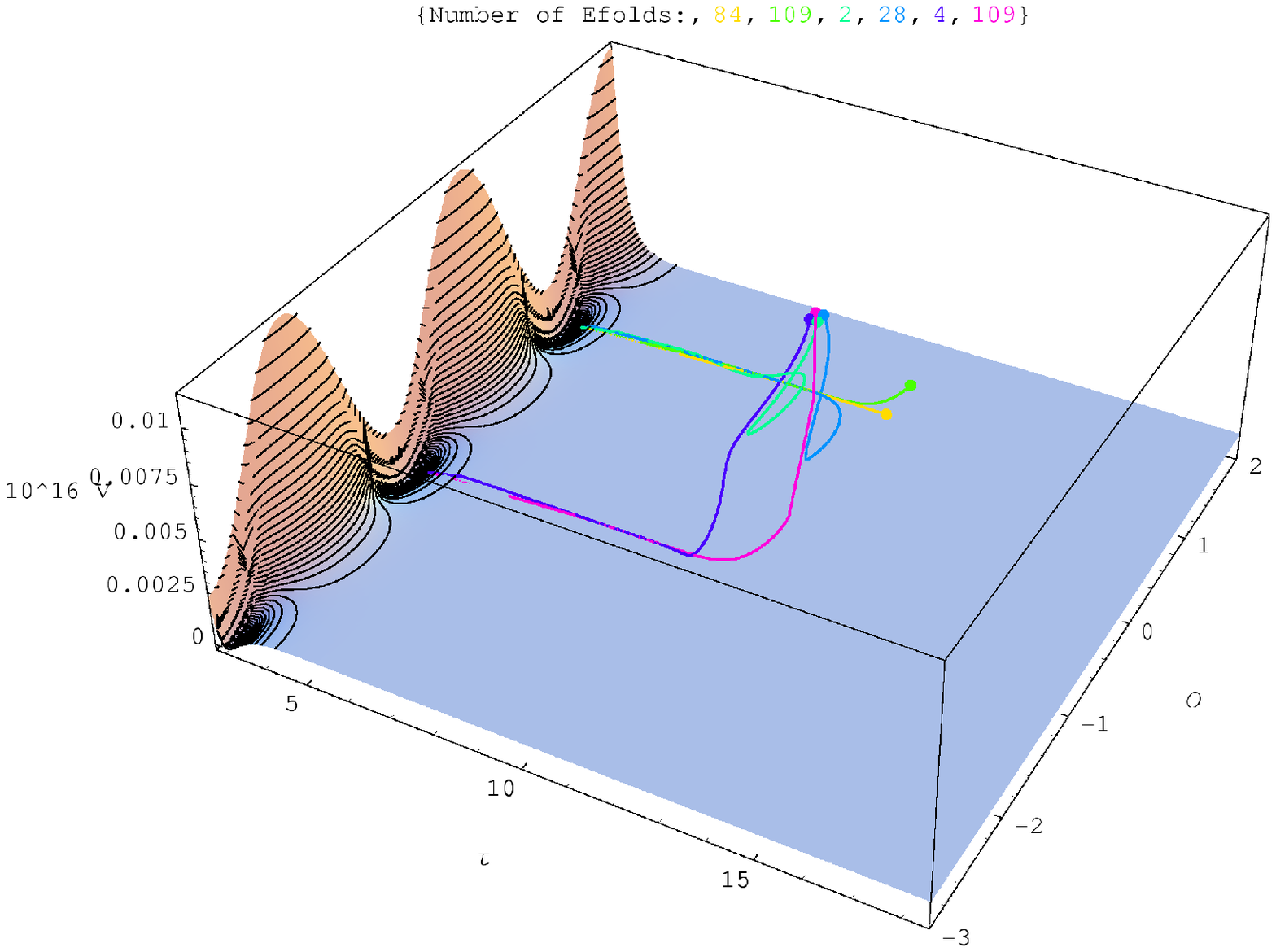}
  \caption{Potential with inflationary trajectories for parameter set 4. Shown
    is a plot of the potential in the $(\tau, \theta)$-plane, overlaid with 
    equipotential contours. Note that for $\tau > 5$ there are no contours 
    because the potential is exponentially flat but it still has the periodic 
    structure in the $\theta$-direction. The starting points of the trajectories
    are indicated by filled circles.}
\end{figure}

\subsection{Trajectories with General K\"ahler modulus and Axion Initial Conditions}

\subsubsection{The $\tau$-valley Attractor}

We first restrict ourselves to trajectories in $\tau$ to connect with
the Conlon and Quevedo \cite{CQ} treatment. The stable flow is in the
$\theta=\frac{2\pi l}{a_2}$ trough. As can be seen in
Fig.~\ref{fig:contourplot}(a), enough e-foldings for successful
inflation are possible provided one starts at large enough $\tau$. The
dashed line in Fig.~\ref{fig:Hubble_epsilon}(a) shows $\epsilon$ is
very small for this case, of order $\epsilon\approx 10^{-10}$. For the
parameters we have considered, no effective inflation is possible if
we start inward of $\tau_{min}$ rather than outward. 

Another class of $\tau$ trajectories are those along the ``ridge''
where $\theta=\frac{2\pi l}{a_2}$ gives a positive contribution to the
potential. These are unstable to small displacements in the axion
direction. 

The $\tau$-trough trajectories serve as late-time attractors for
initial conditions that begin with $\theta$ out of the trough. The
very flat profile of the potential at large $\tau$ allows for a regime
of self-reproducing inflation (\S~\ref{sec:stoch}). Trajectories which
originate in the self-reproducing regime invariably flow to the
$\tau$-valley attractor and the observed e-folds would be just those
of the Conlon and Quevedo sort.

\subsubsection{The Variety of ``Roulette'' $\tau$-$\theta$ Trajectories}

\begin{figure}
  (a1)\includegraphics[width=\mypicturewidth, height=3.5cm]{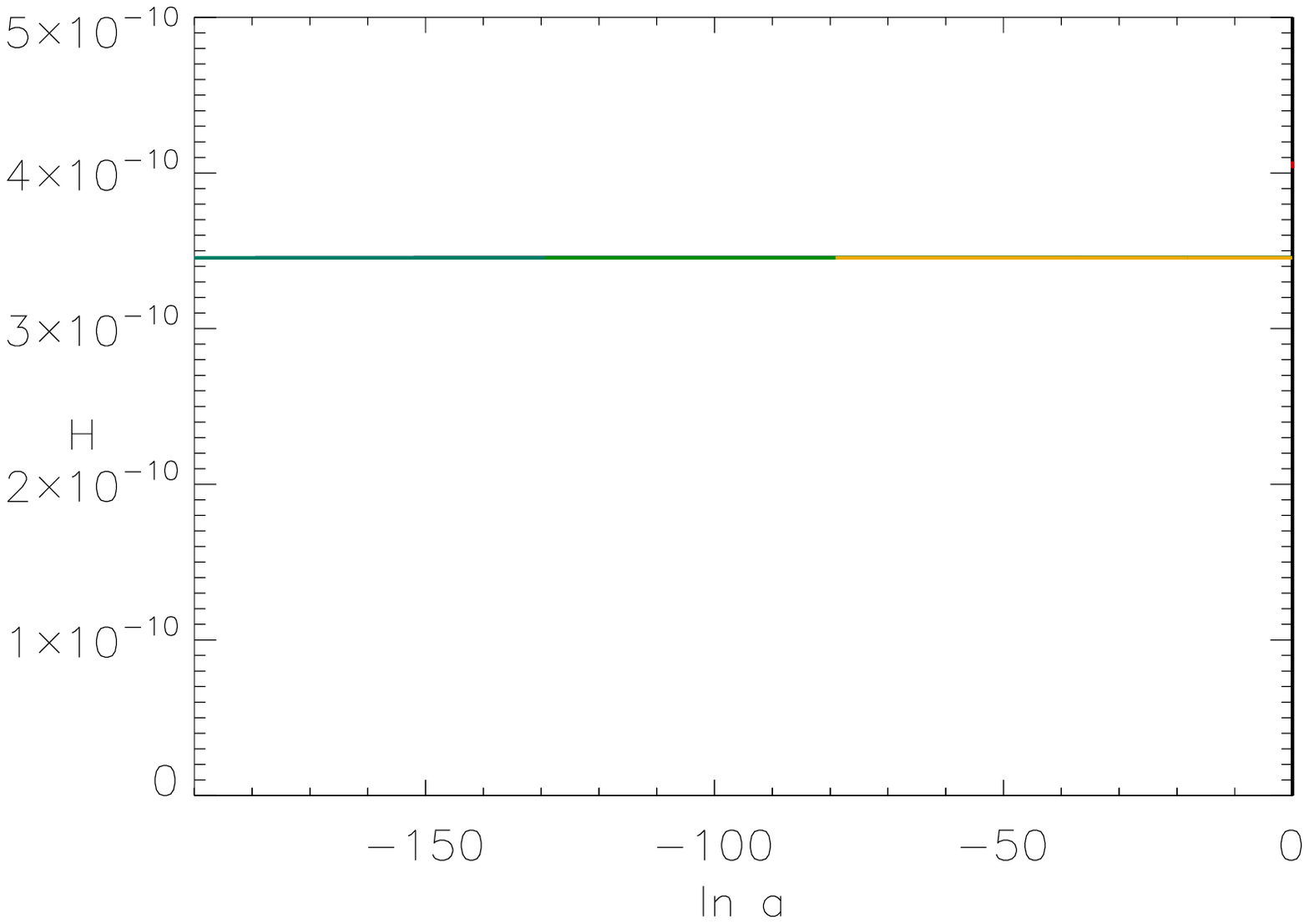}
  (a2)\includegraphics[width=\mypicturewidth, height=3.5cm]{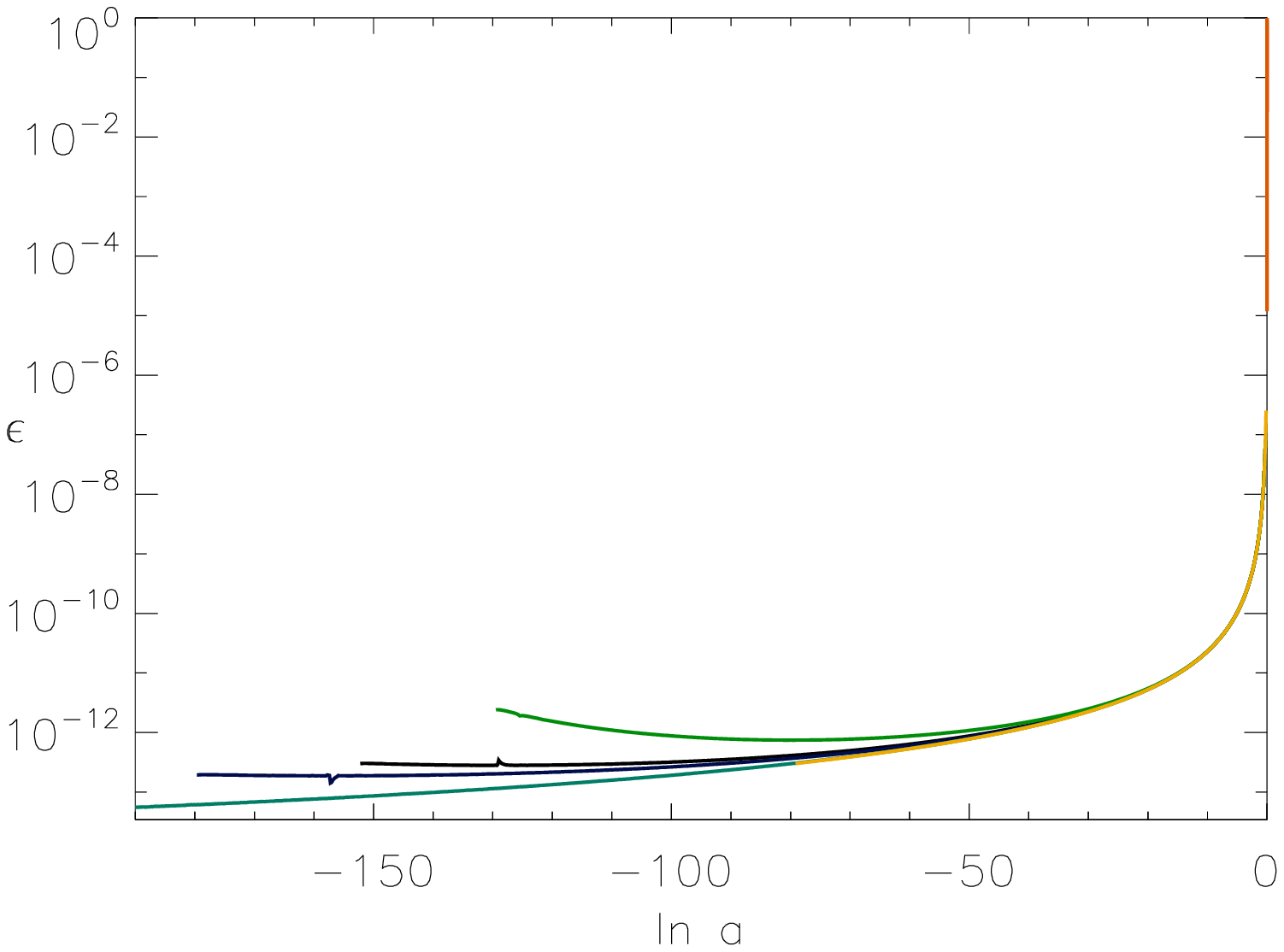}\\
  (b1)\includegraphics[width=\mypicturewidth, height=3.5cm]{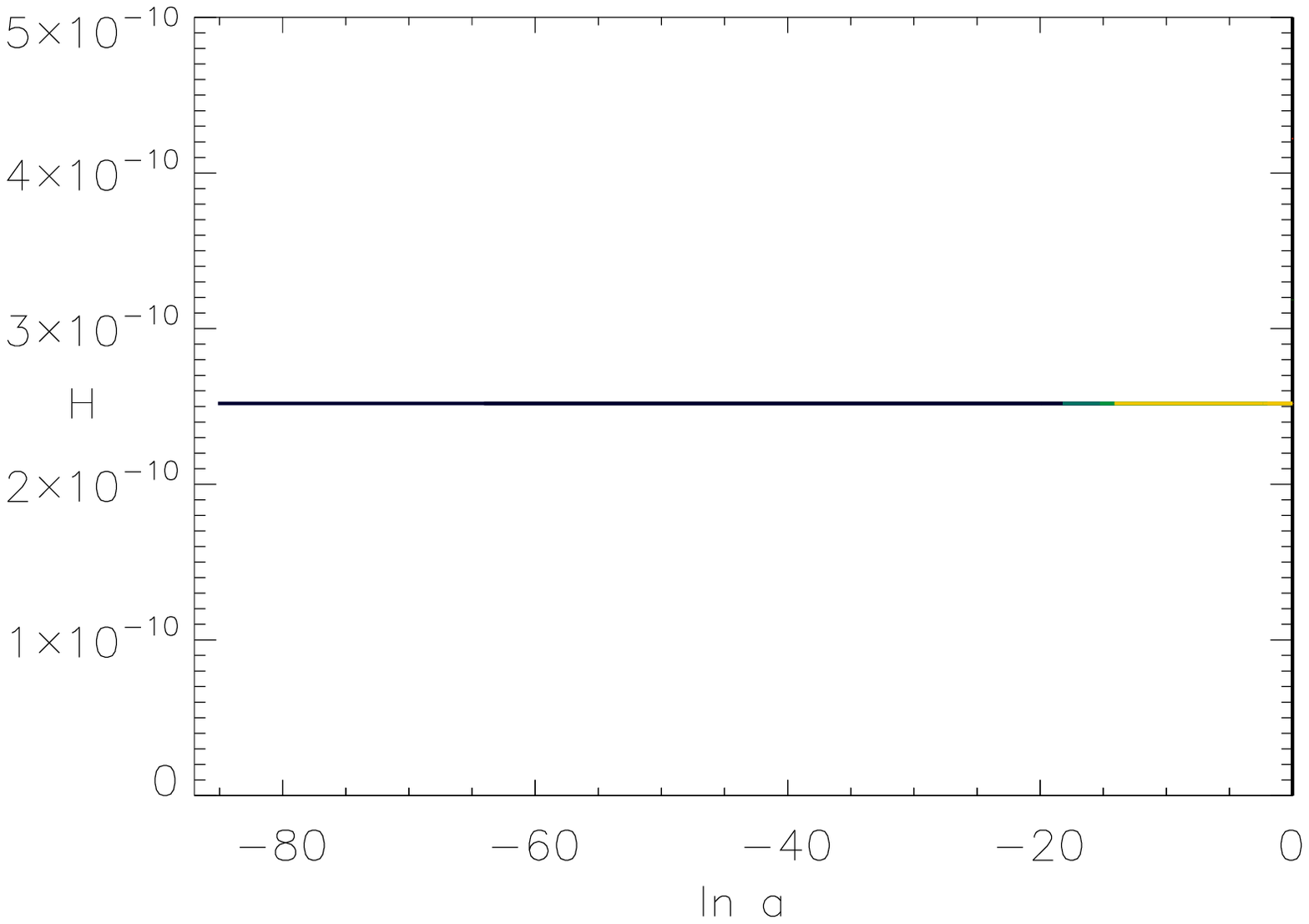}
  (b2)\includegraphics[width=\mypicturewidth, height=3.5cm]{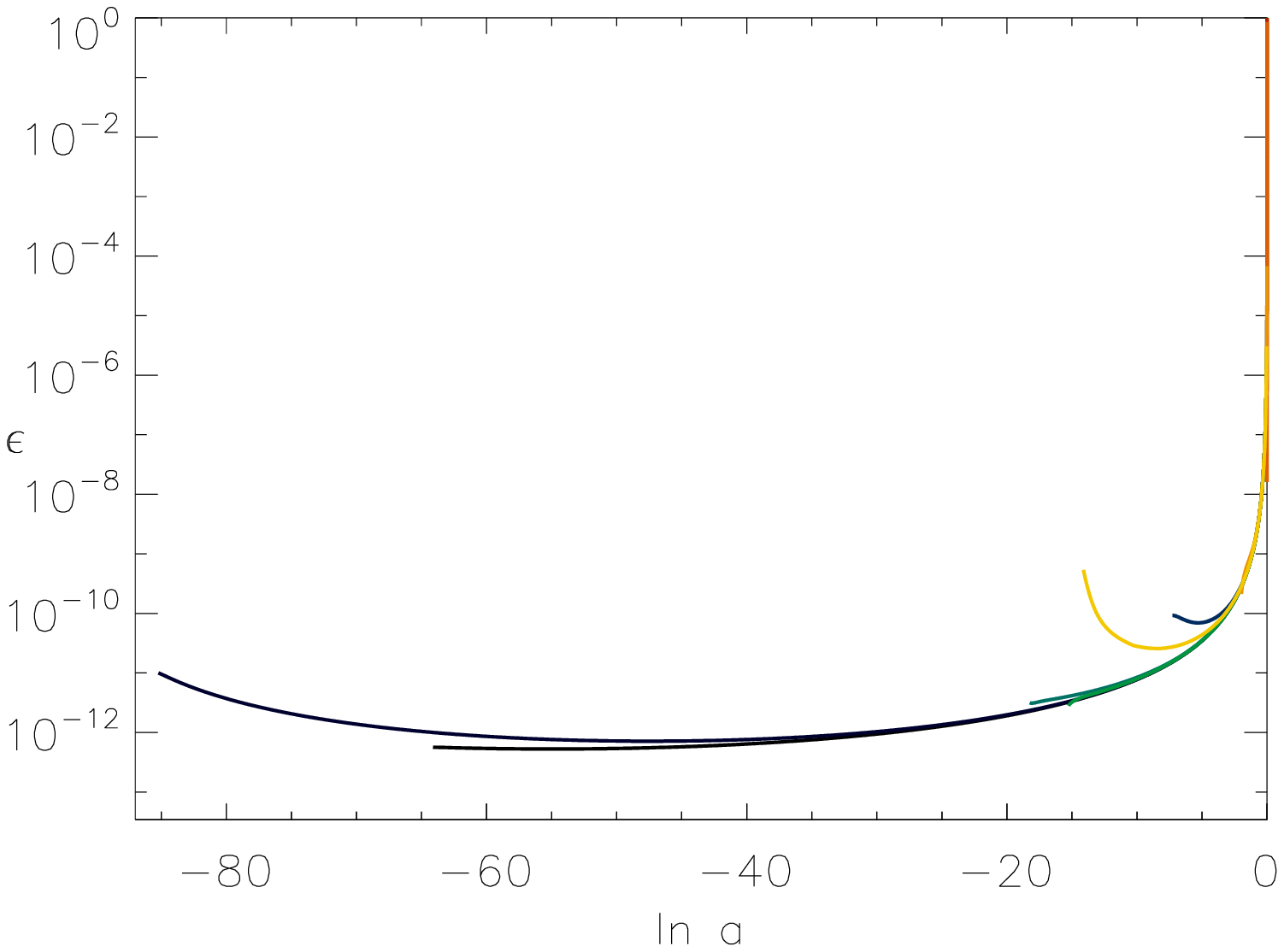}\\
  (c1)\includegraphics[width=\mypicturewidth, height=3.5cm]{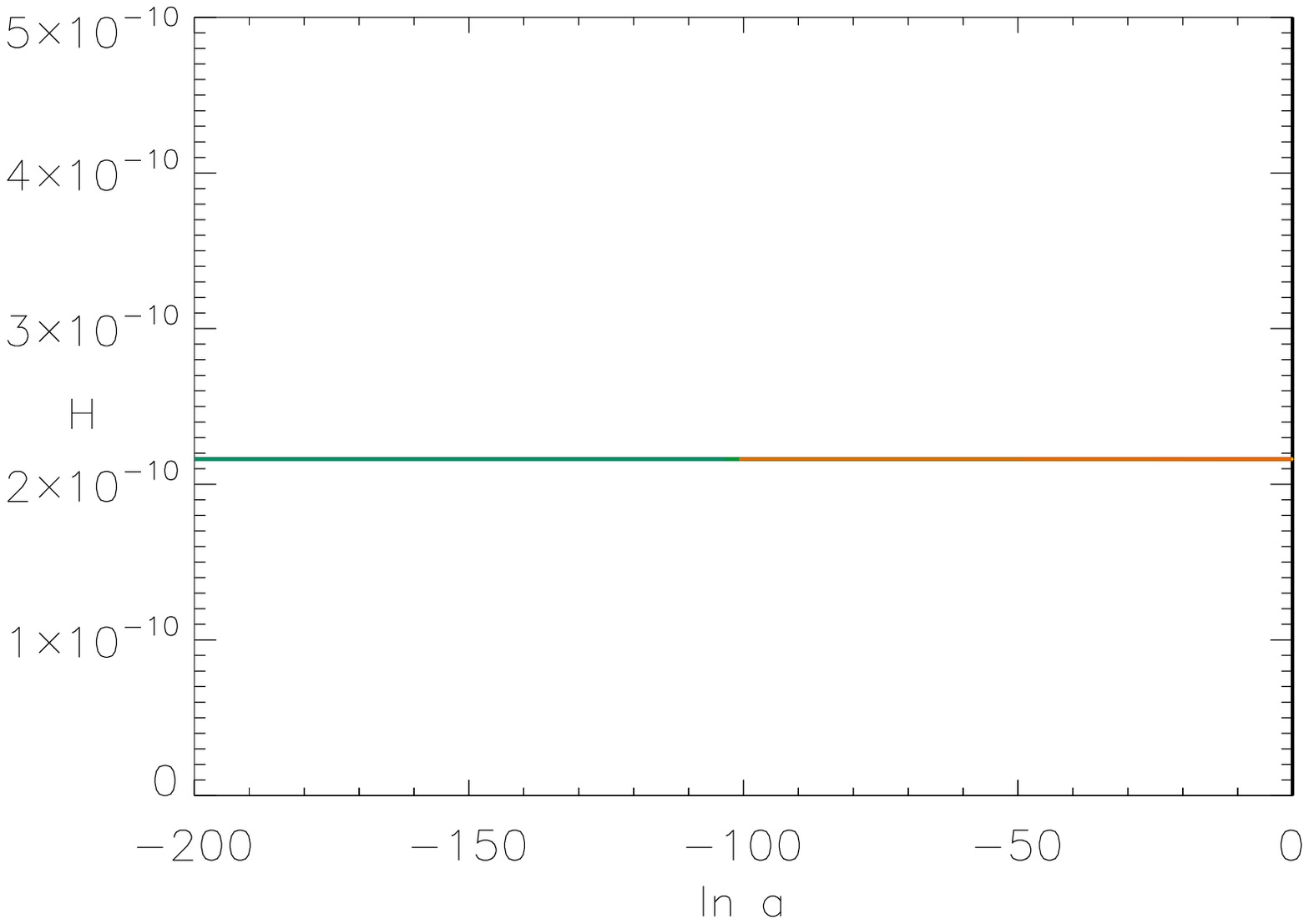}
  (c2)\includegraphics[width=\mypicturewidth, height=3.5cm]{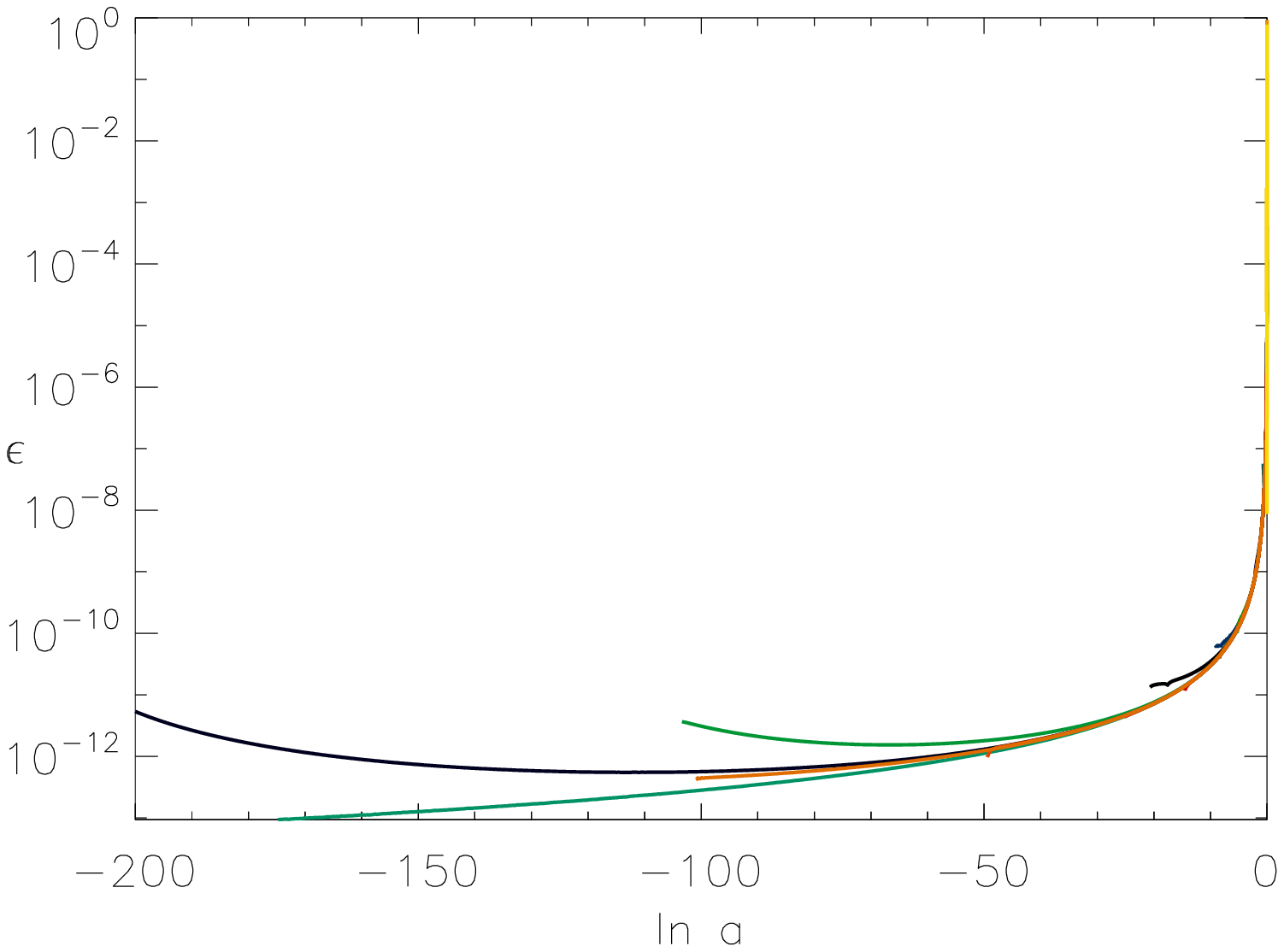}\\
  (d1)\includegraphics[width=\mypicturewidth, height=3.5cm]{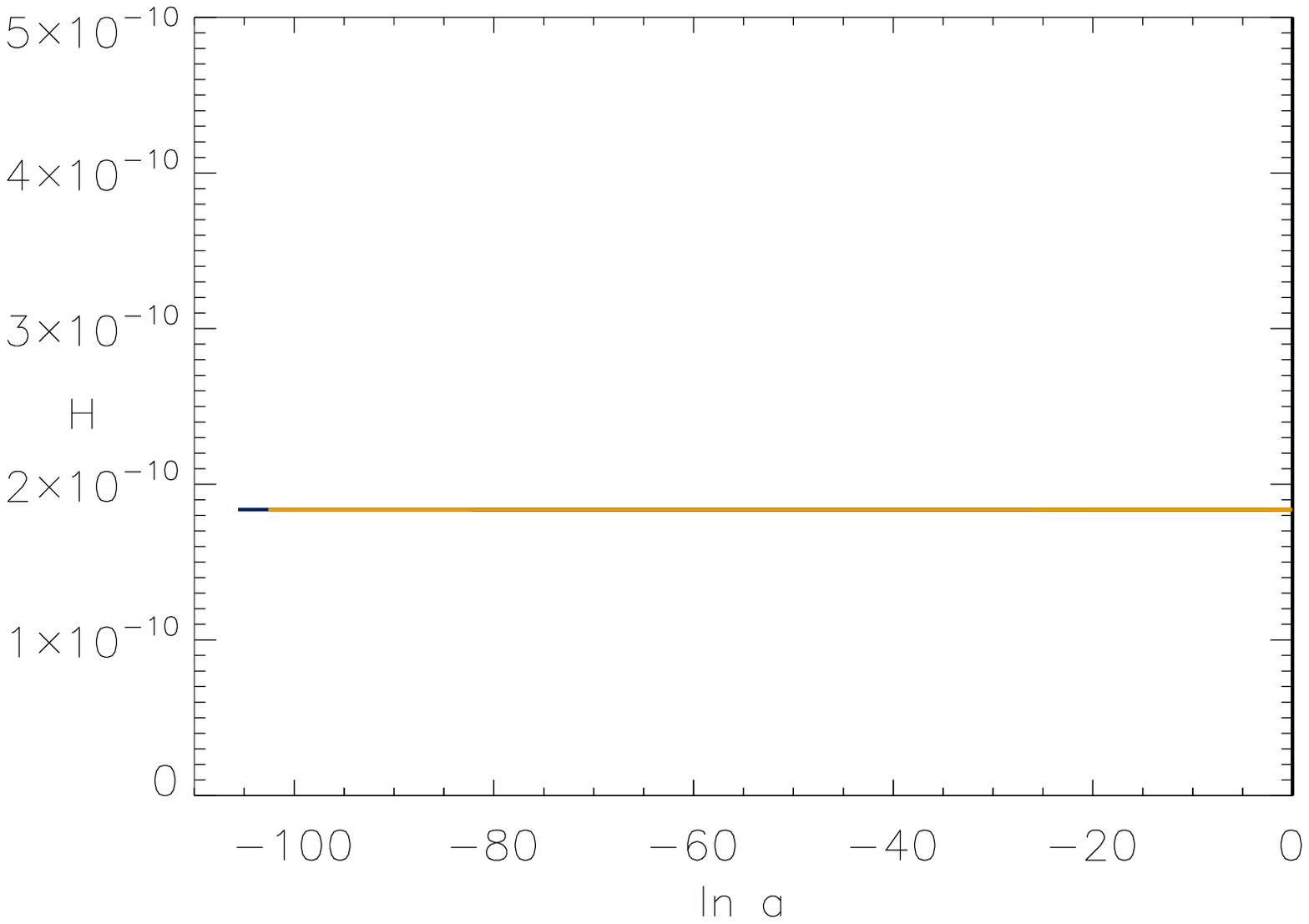}
  (d2)\includegraphics[width=\mypicturewidth, height=3.5cm]{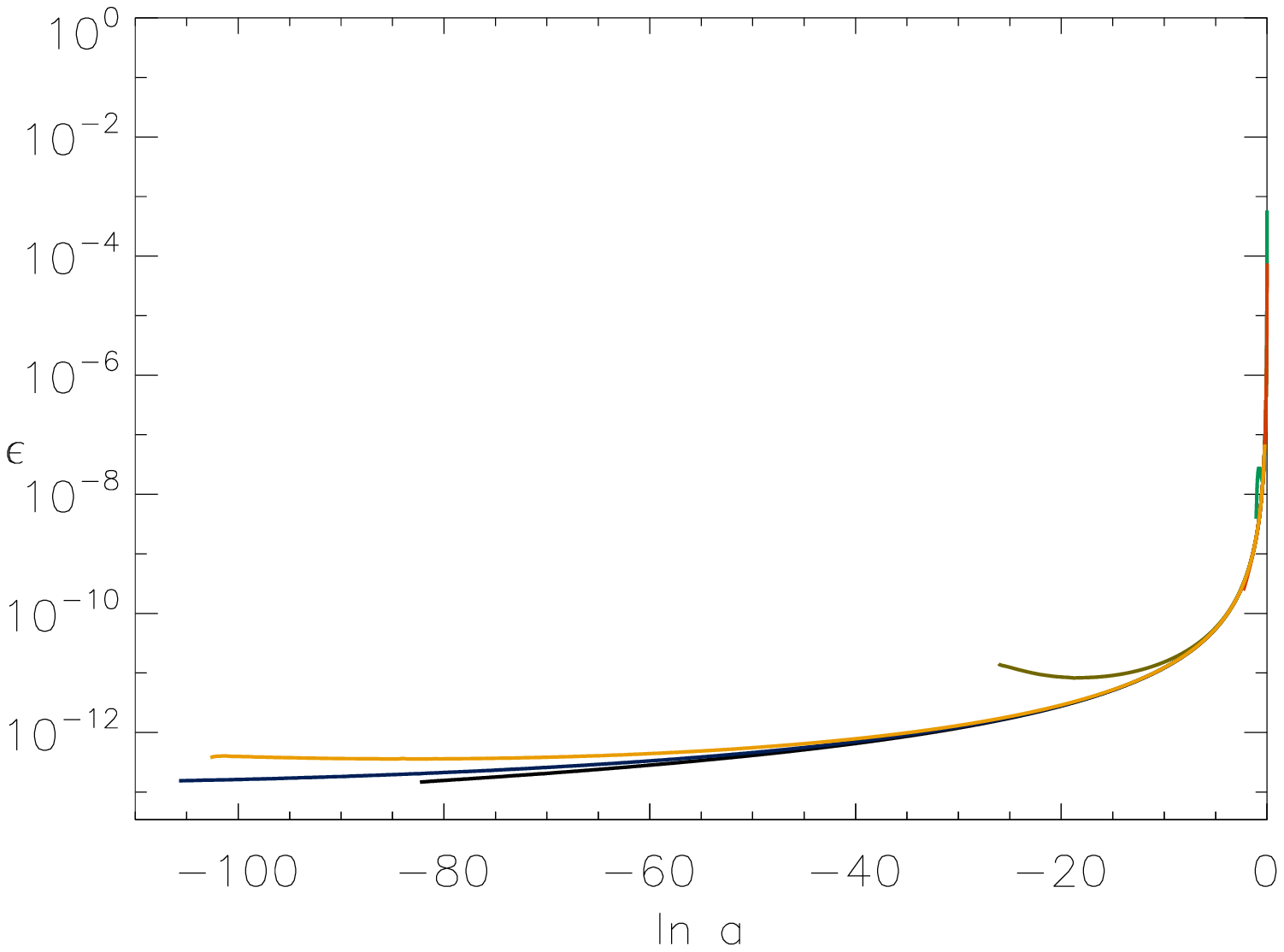}\\
  \caption{Hubble parameter $H(\ln a)$ (left column) and deceleration
    parameter $\epsilon(\ln a)$ (right column) as a function of the number
    of e-folds. $H(\ln a)$ is practically constant during
    inflation. $\epsilon$ during inflation is tiny and generally turns
    up towards the reheating phase rather rapidly.  (a) Parameter set
    1: the coloured trajectories are for general $(\tau,\theta)$
    inflation, the dashed trajectory is inflation strictly along
    $\tau$ (b,c,d) are the same for sets 2, 3, 4, respectively. Note
    that units along the x-axes are $\ln a=-N\approx \ln k$ here and
    in the plots for the power spectra. There is a very rapid phase in
    which the trajectory settles down to an attractor for the $\tau$
    and $\theta$ field momenta which we do not show.}
  \label{fig:Hubble_epsilon}
\end{figure}

\begin{figure}
  (a1)\includegraphics[width=\mypicturewidth]{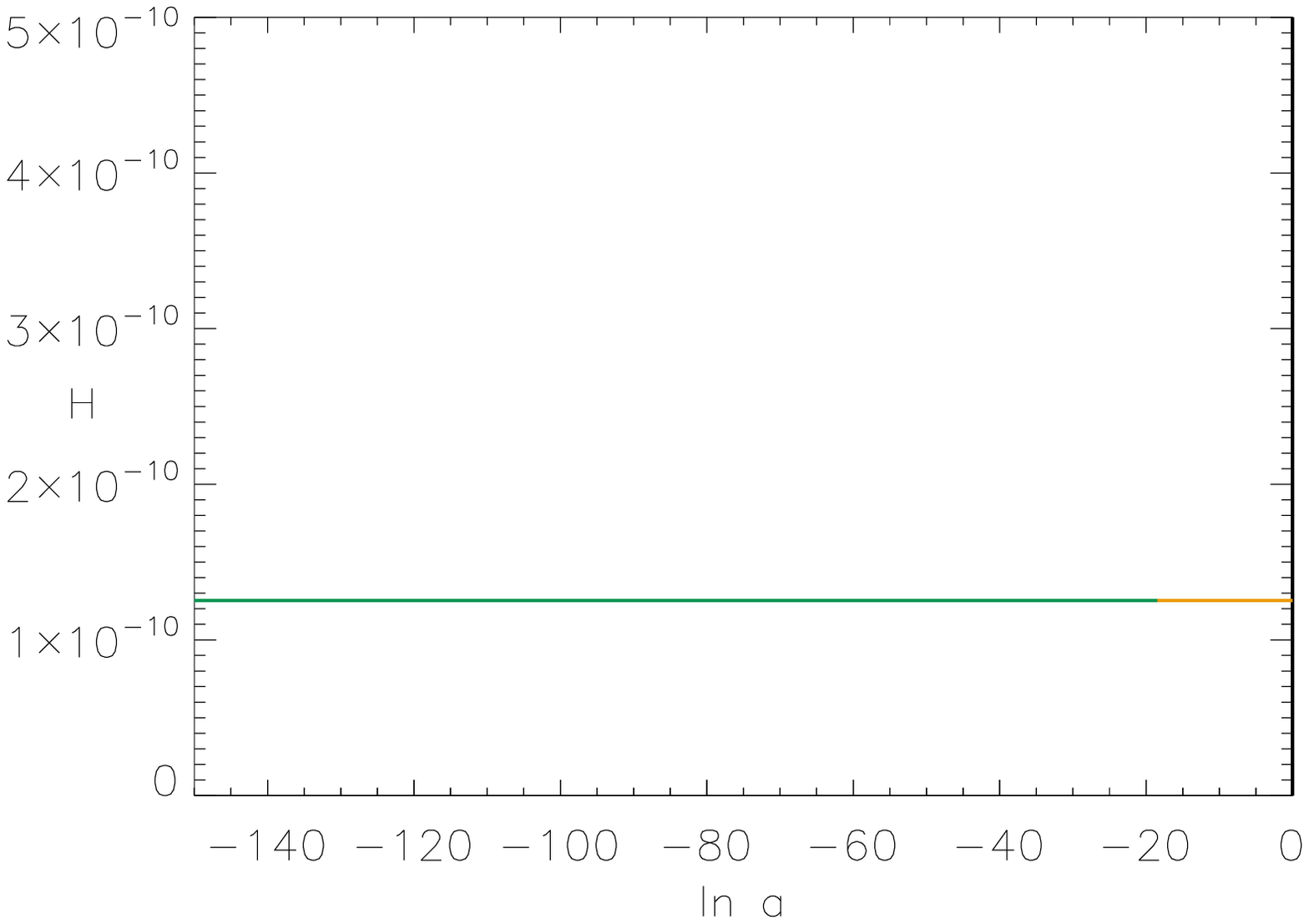}
  (a2)\includegraphics[width=\mypicturewidth]{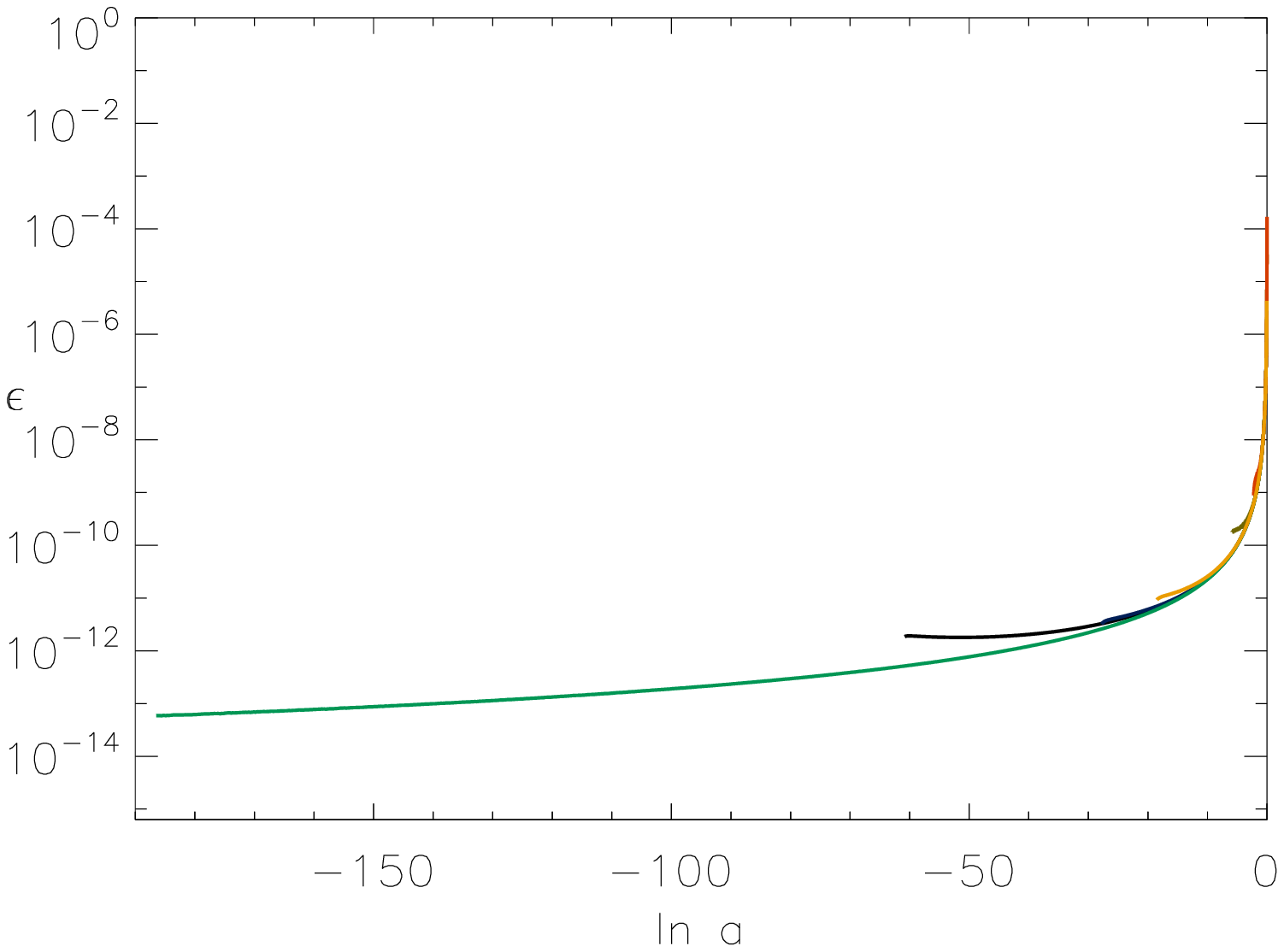}\\
  (b1)\includegraphics[width=\mypicturewidth]{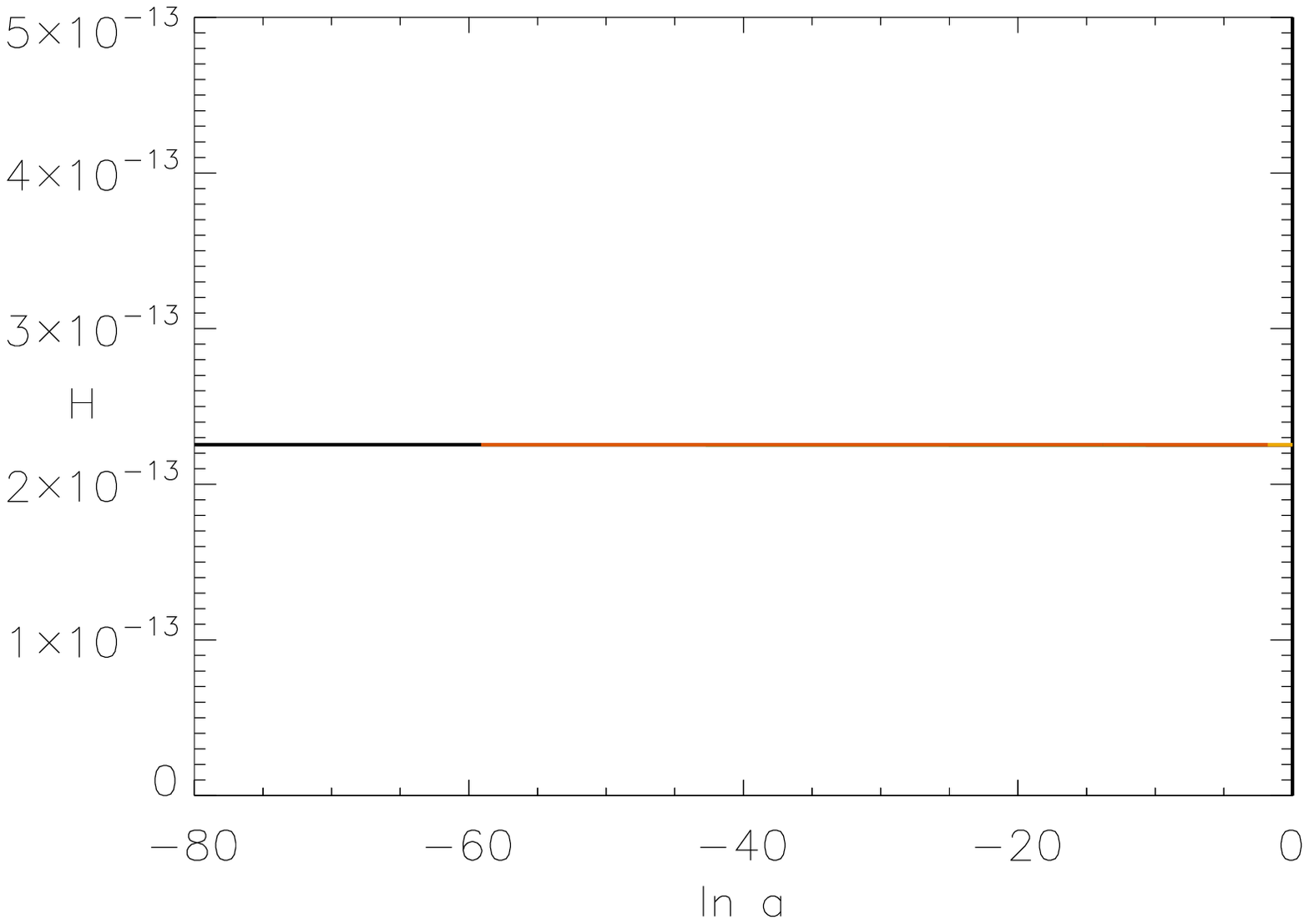}
  (b2)\includegraphics[width=\mypicturewidth]{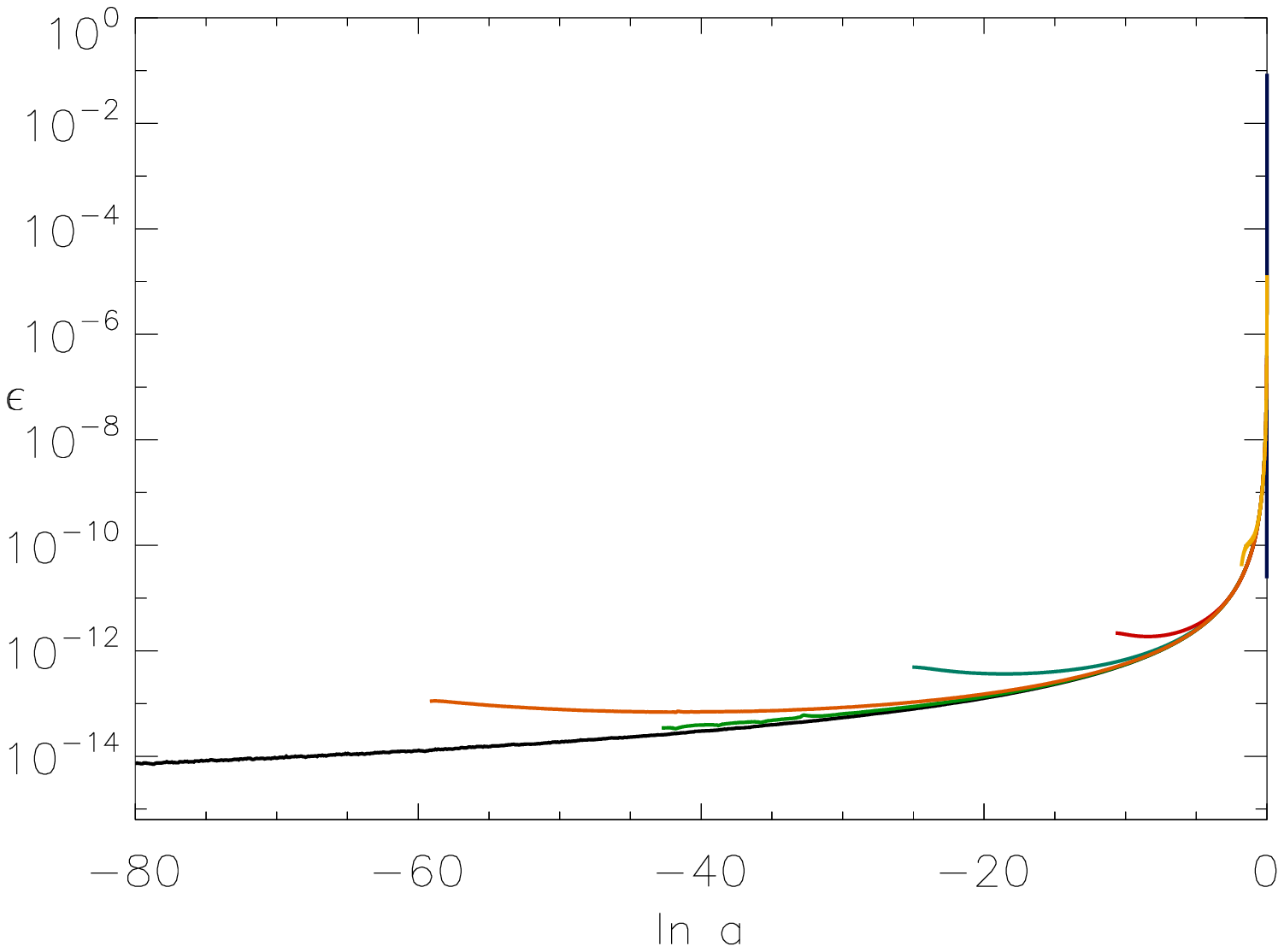}
  \caption{Same as Fig.~\ref{fig:Hubble_epsilon}, but for parameter sets 5 (a) and 6 (b).}
  \label{fig:Hubble_epsilon_bad}
\end{figure}

When we allow the initial values of $\tau$ and $\theta$ to be
populated with an equal a priori probability prior, given a set of
parameters defining the $V(\tau, \theta )$ surface, we encounter a
wide range of inflationary trajectories. Examples of the variety of
behaviours for the parameter sets given in
Table~1 are shown for $\tau(\ln a), \theta
(\ln a)$ in Figures \ref{fig:contourplot} and
\ref{fig:contourplot_bad}, and for $H(\ln a),\epsilon (\ln a)$ in
Figures \ref{fig:Hubble_epsilon} and \ref{fig:Hubble_epsilon_bad}.
Some of the trajectories are predominantly $\theta$ ones before
settling into a $\tau$-valley, similar to a roulette ball rotation
before locking into a slot with a specific number, hence the name
roulette inflation. 

We began with the momenta of the fields set to zero, but the momenta are very
quickly attracted to their slow-roll lock-in values, on of order an
e-fold. We do not show this settling down phase in the trajectories we
have plotted. We find a large fraction of trajectories are indeed
inflating, and have the required $> 40-50$ e-folds of inflation
(\S~\ref{sec:kNrelation}) to give homogeneity and isotropy over our
observable Hubble patch. Large enhancements of the number of e-folds
over $\tau$-only inflation can occur because of significant flows in
the $\theta$-direction, while $\tau$ evolves slowly. As the figures
show, initial values starting far out in $\tau$ generally
roll towards the nearest $\tau$-valley and then proceed along the
$\tau$-attractor. Trajectories starting at intermediate $\tau$ 
have enough energy to pass through the $\tau$-valley in the axionic
direction, but often turn around before reaching the neighbouring
ridge, and roll back into the valley. They can move to larger
$\tau$ while in the axion-dominated flow. If the initial values are
chosen such that the initial potential energy is just a little higher,
the inflaton can climb over the next ridge and settle in the adjacent
valley, or the next one, or the next. Thus there exists bifurcation
points which divide the phase space into solutions ending up in
different valleys. Another feature in Fig.~\ref{fig:contourplot} is
the existence of areas in which tiny changes in initial positions can
lead to dramatic changes in the number of e-folds produced.

There are also cases in which the $\tau$-attractor is not reached
before the end of inflation. When starting from moderate values in
$\tau$, relatively far away from the valley, the field oscillates in
the $\theta$ direction, albeit producing only a very small number of
e-folds $N<1$ (see Fig.~\ref{fig:contourplot} for examples).

\begin{figure}
  (a1)\includegraphics[width=\mypicturewidth, height=3.5cm]{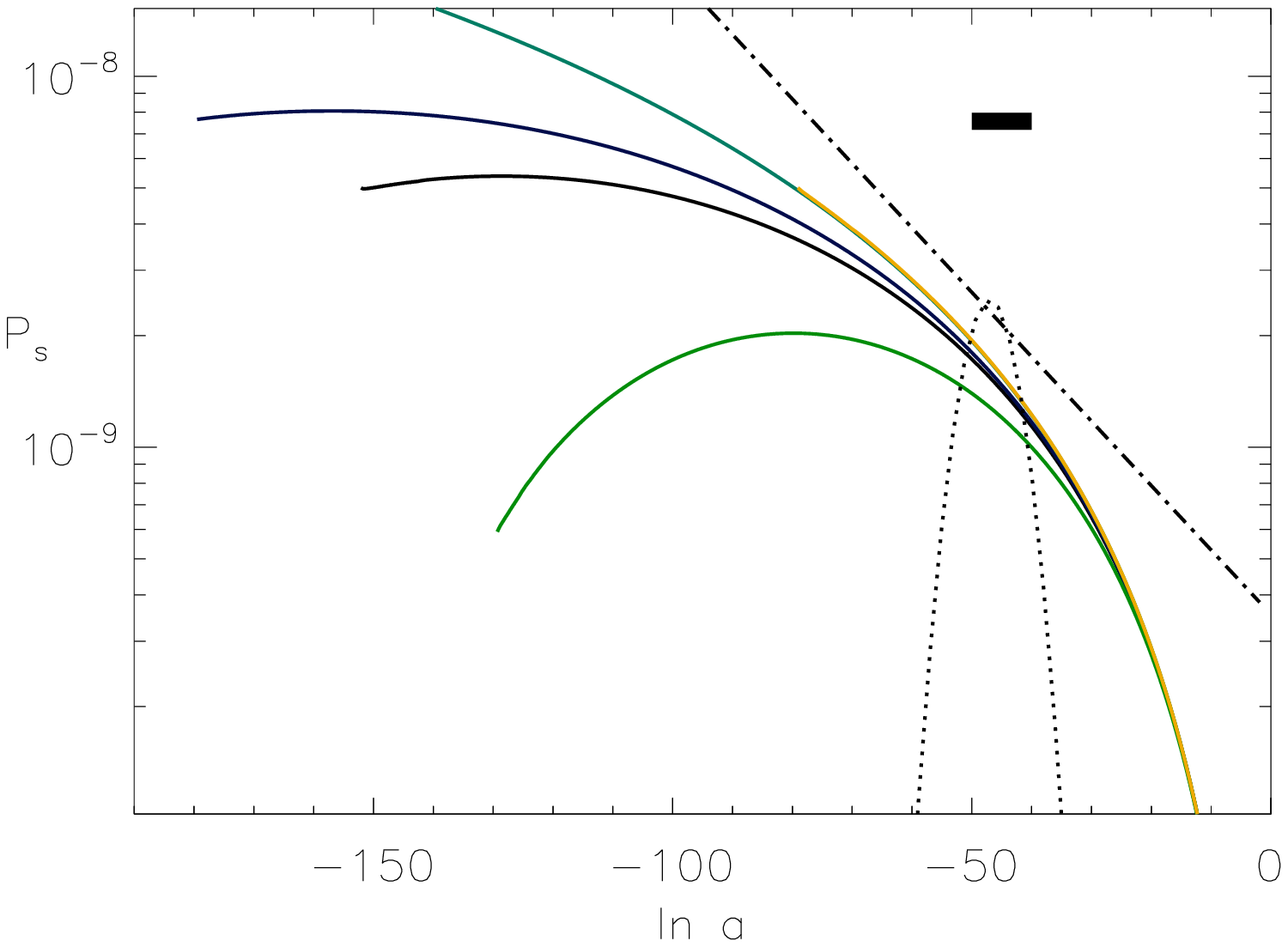}
  (a2)\includegraphics[width=\mypicturewidth, height=3.5cm]{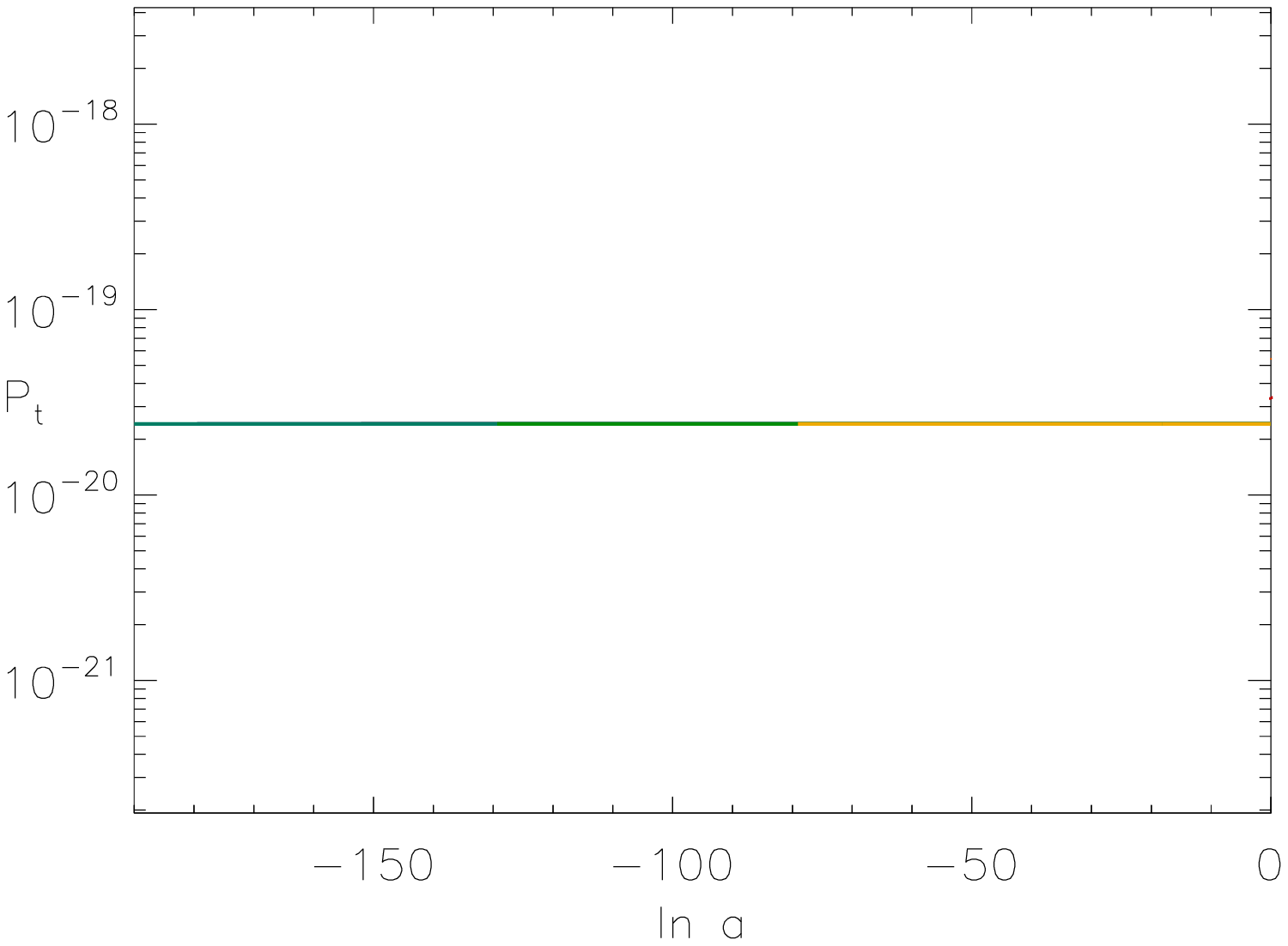}\\
  (b1)\includegraphics[width=\mypicturewidth, height=3.5cm]{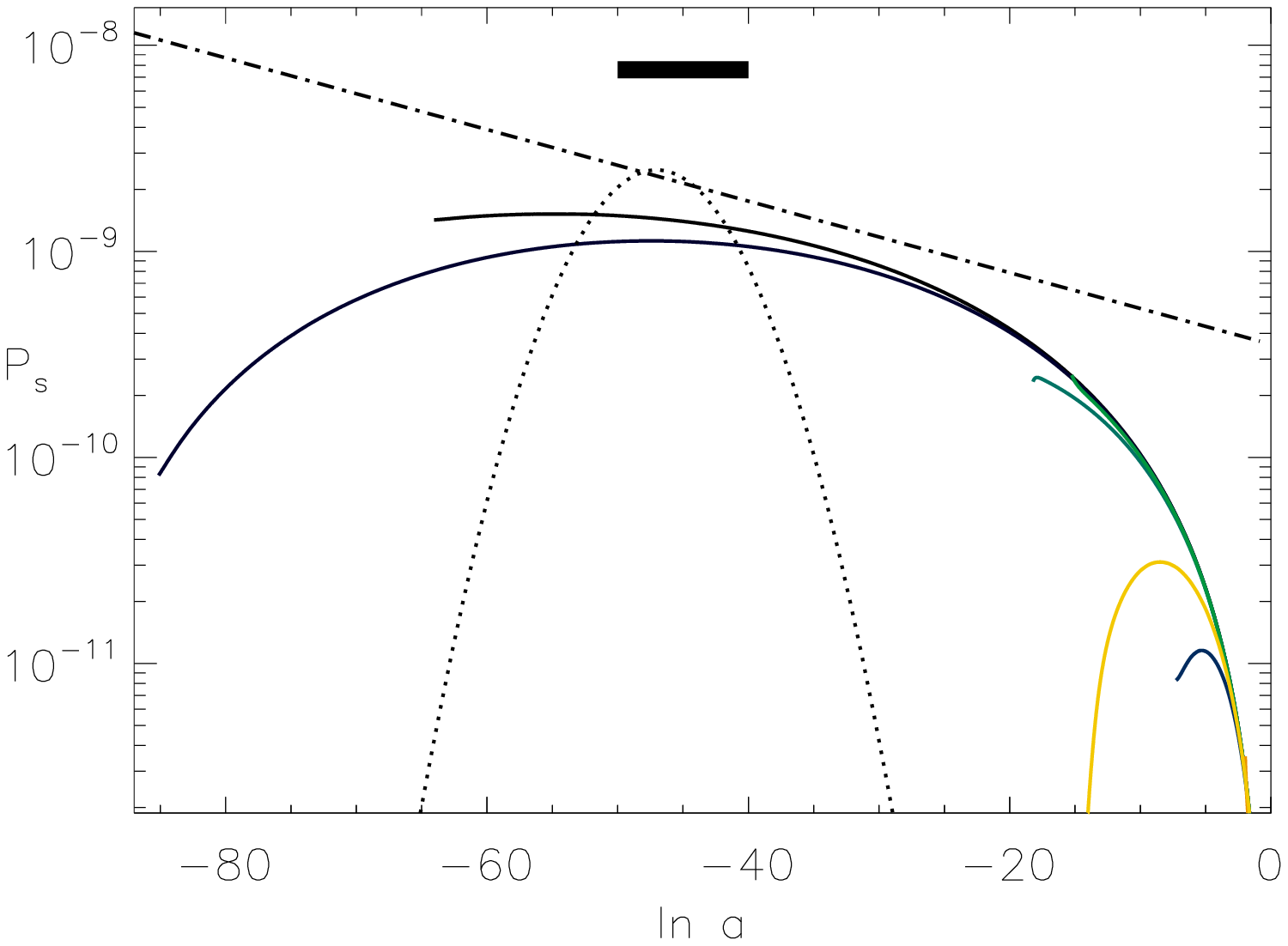}
  (b2)\includegraphics[width=\mypicturewidth, height=3.5cm]{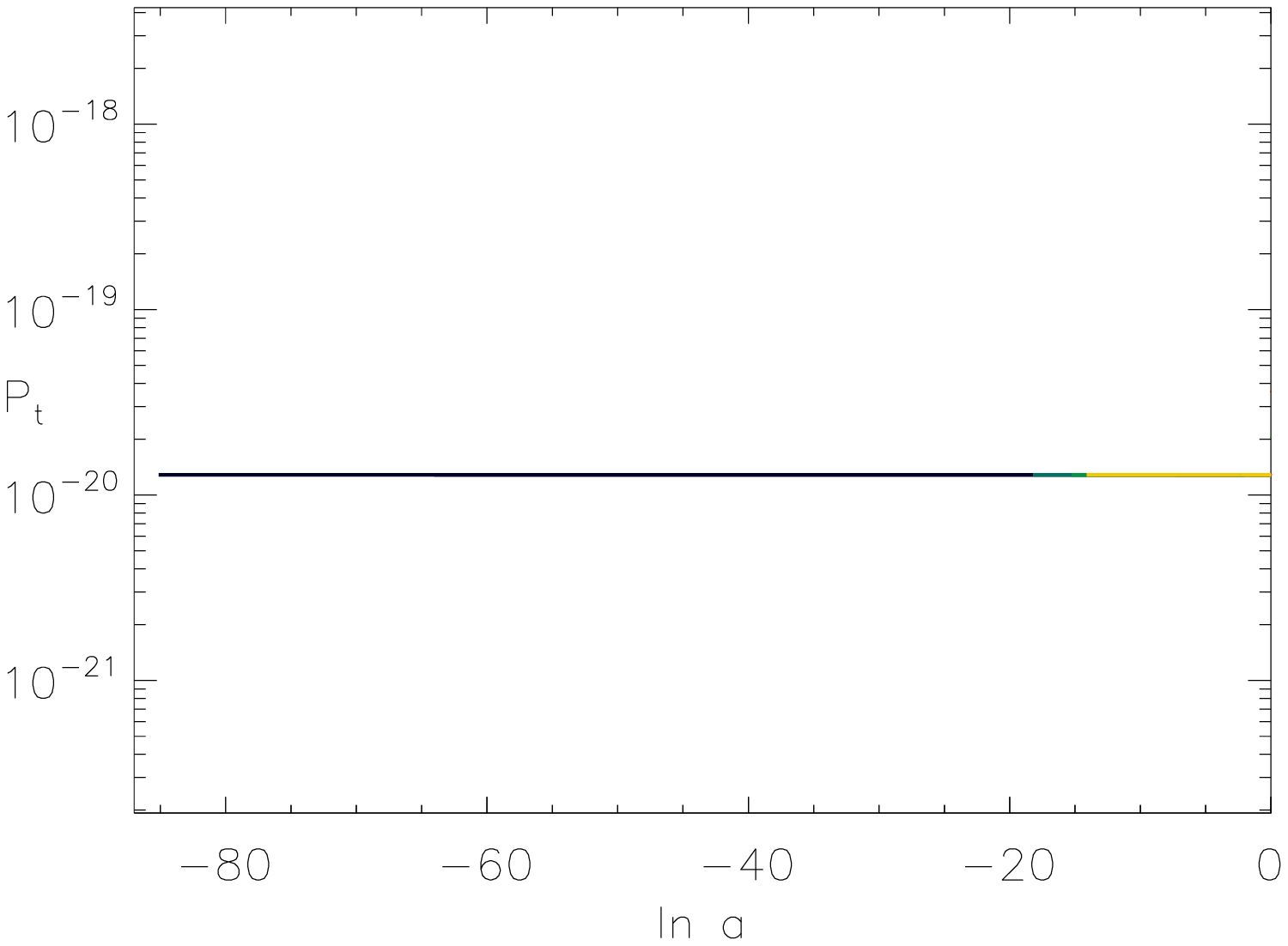}\\
  (c1)\includegraphics[width=\mypicturewidth, height=3.5cm]{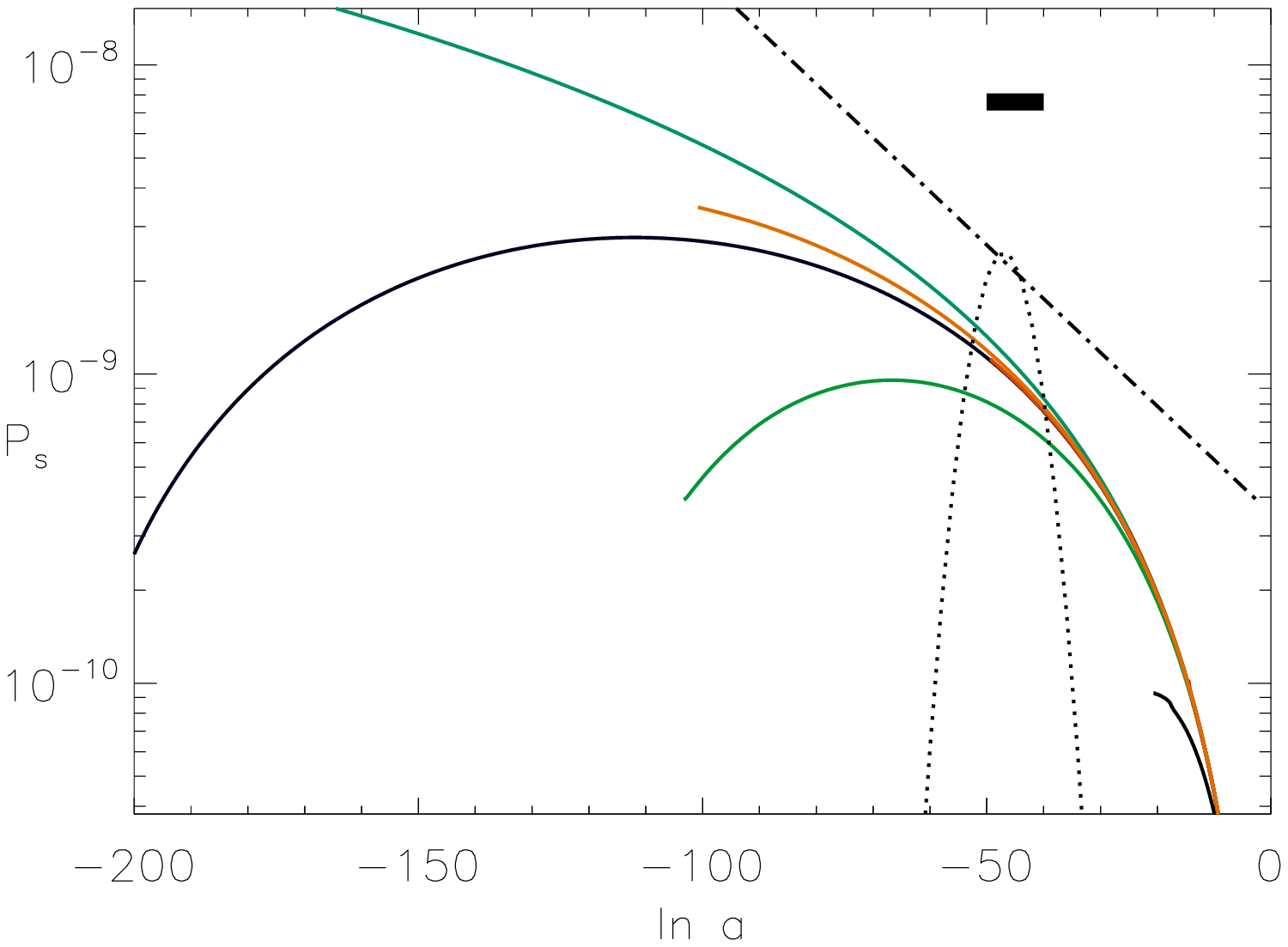}
  (c2)\includegraphics[width=\mypicturewidth, height=3.5cm]{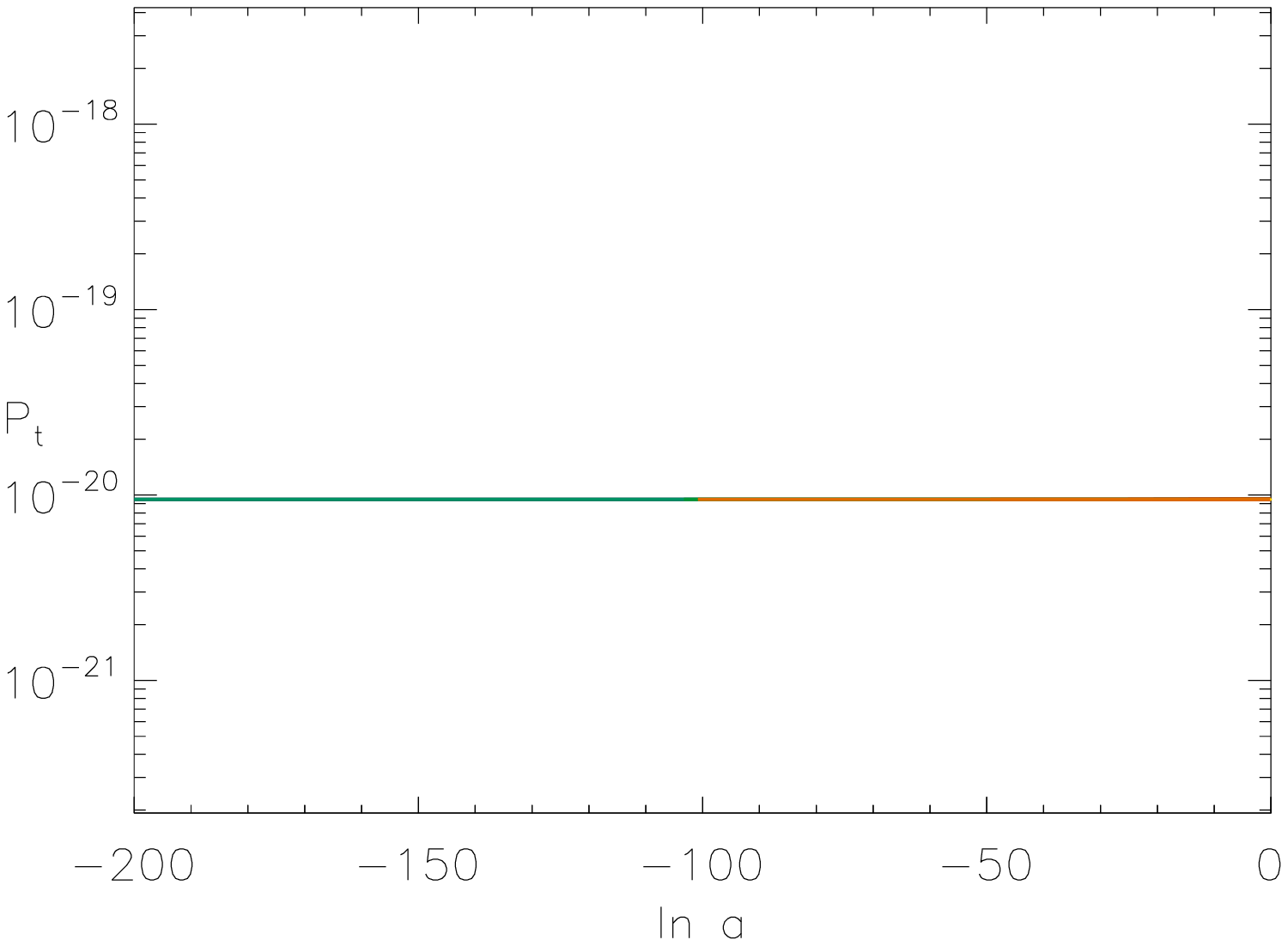}\\
  (d1)\includegraphics[width=\mypicturewidth, height=3.5cm]{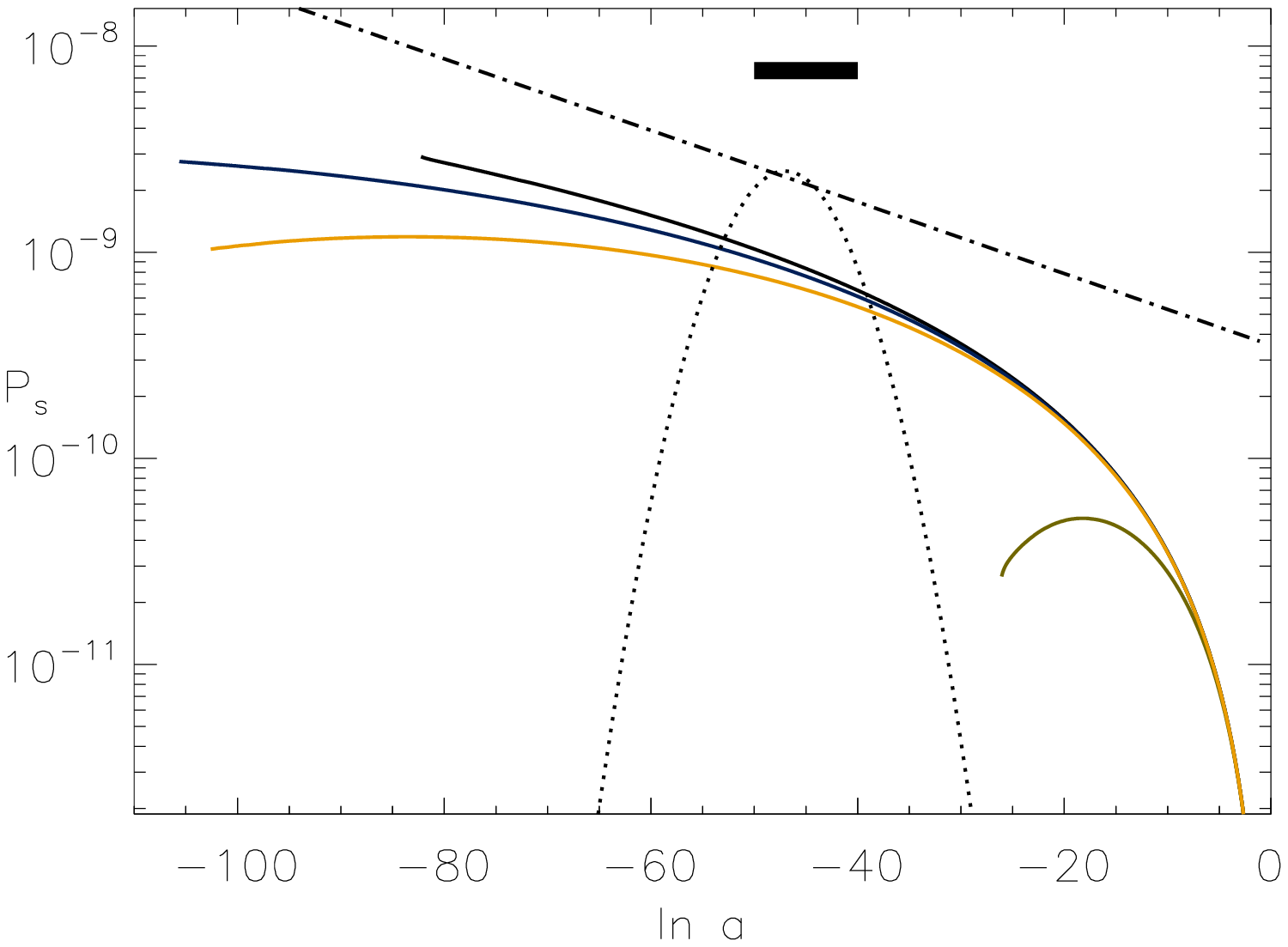}
  (d2)\includegraphics[width=\mypicturewidth, height=3.5cm]{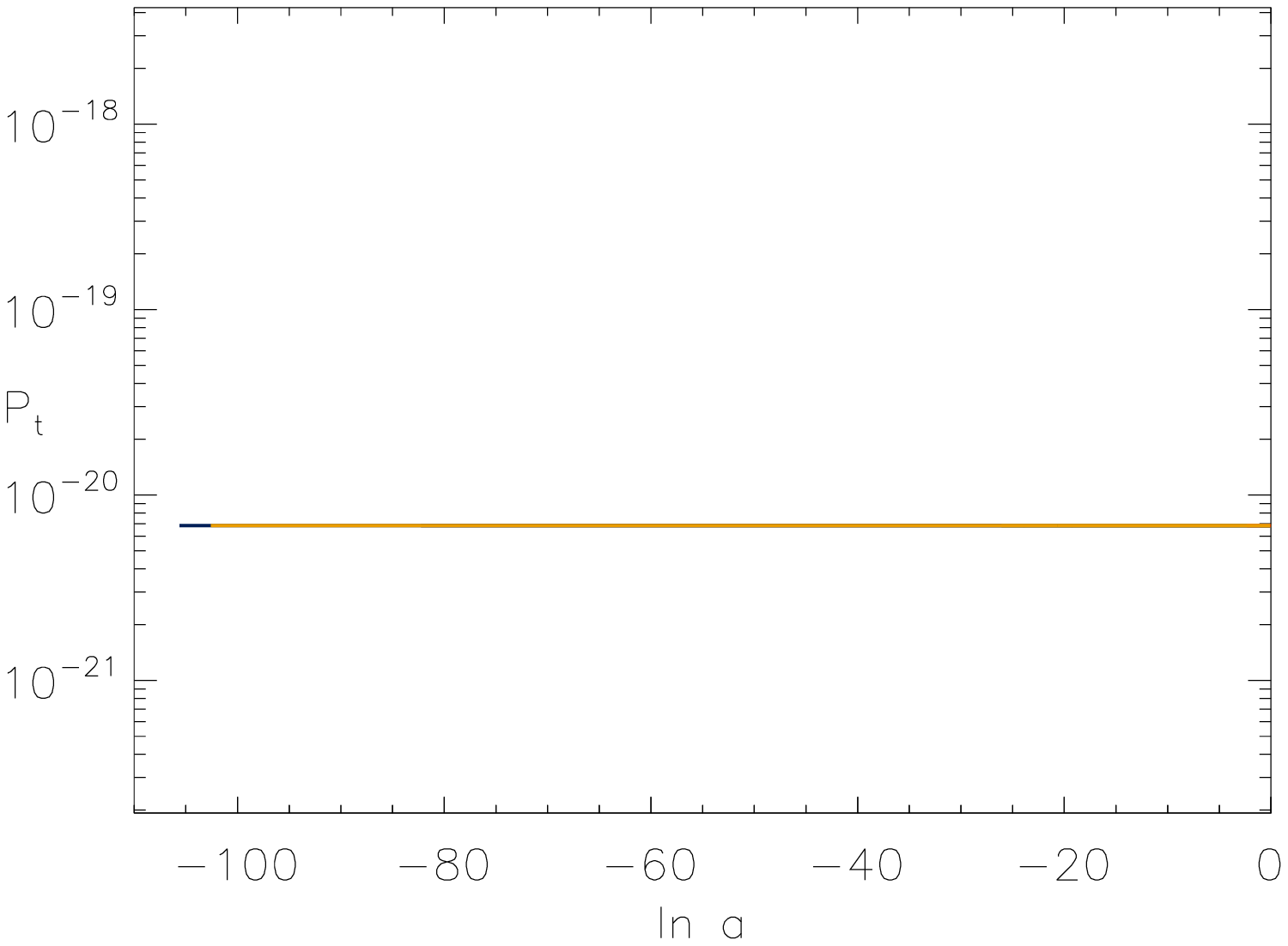}\\
  (e)\includegraphics[width=\mypicturewidth, height=3.5cm]{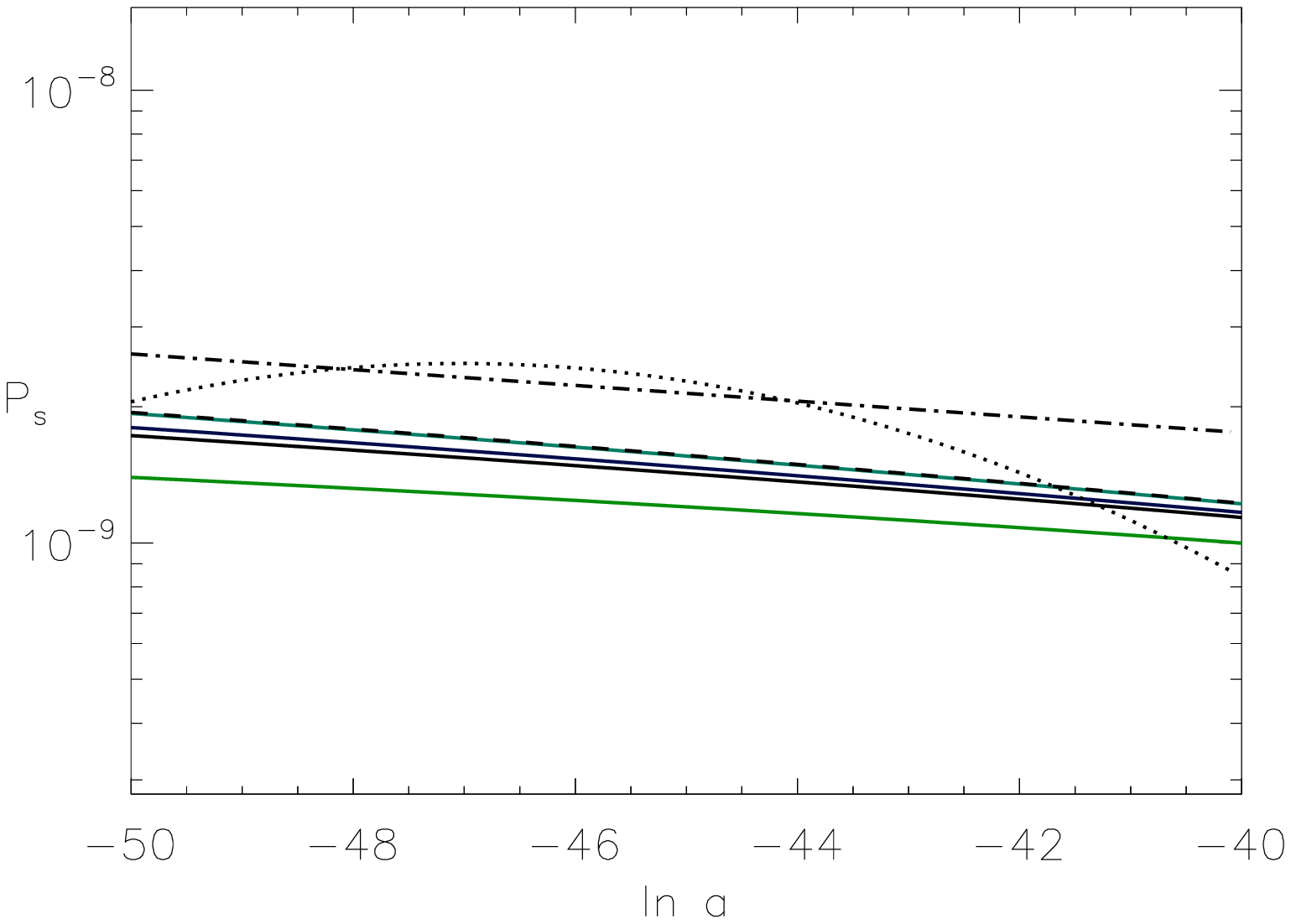}
  (f)\includegraphics[width=\mypicturewidth, height=3.5cm]{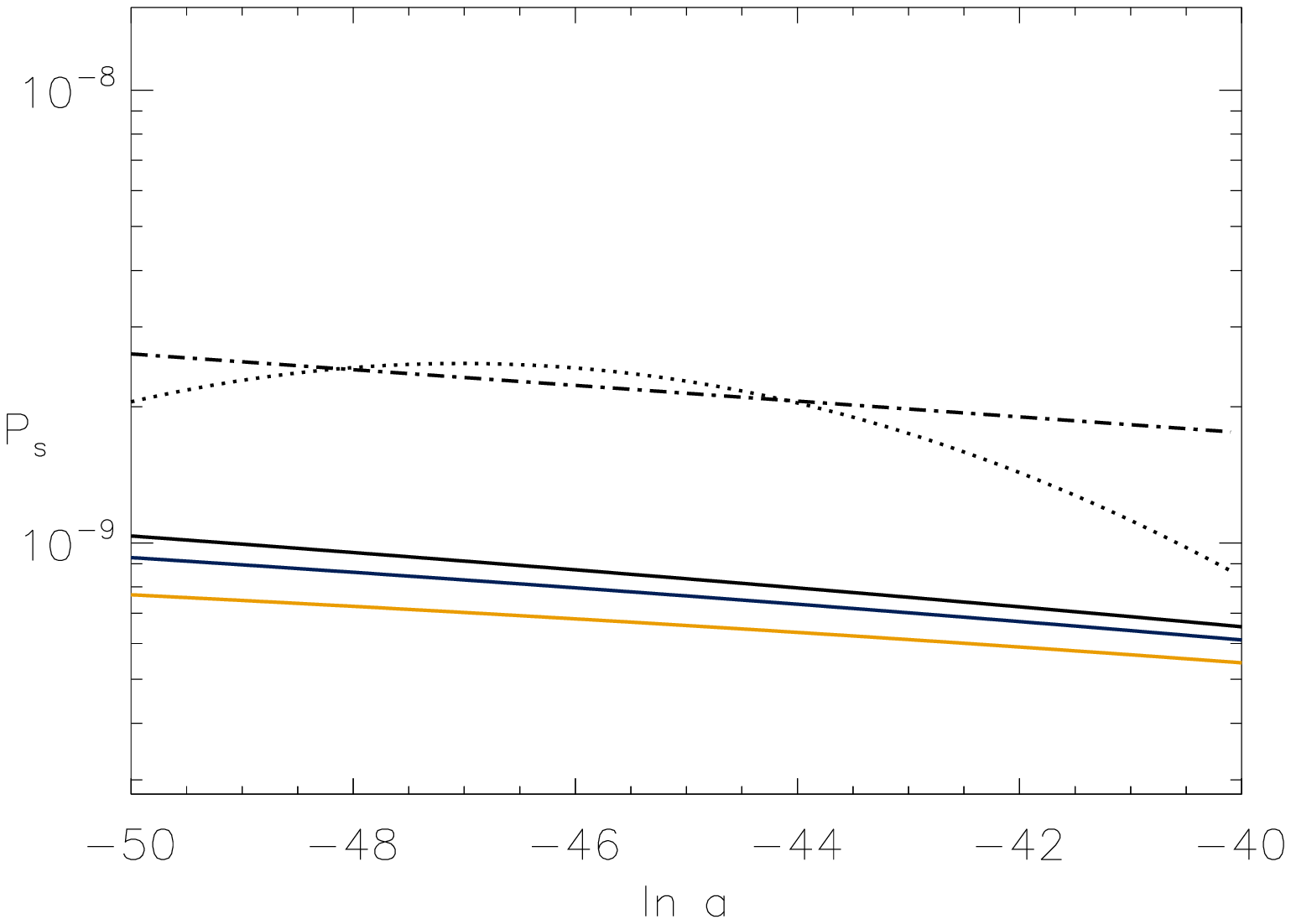}
  \caption{Power spectra for scalar ${\cal P}_s$ and tensor ${\cal
    P}_t$ fluctuations, derived assuming that only stochastic kicks
    along the trajectory are relevant for their determination,
    neglecting the influence of ``isocurvature'' fluctuations
    transverse to the inflaton trajectory. Both have amplitudes given
    by the instantaneous Hawking temperature. For comparison we
    present two template spectra: The dashed-dotted line shows a
    simple spectrum with no running of the spectral index $n_s=0.96$.
    The dotted line is a simple spectrum with running of the spectral
    index $n_s=0.91, dn_s/d\ln k =-0.044$, using values obtained by
    the ACBAR collaboration \cite{Acbar06}. Both these spectra have
    the normalization set to ${\cal P}_s=2.1\times 10^{-9}$ at
    $N=45$. (a,b,c,d) are for parameter sets 1, 2, 3 and 4. The dashed
    line corresponds to inflation with $\theta=$const in the
    valley. Even though there is significant running for all models
    over large scales, the spectra are mostly featureless in the
    observable interval which panels (e) and (f) zoom into for sets 1
    and 4.}
  \label{fig:powerspectra}
\end{figure}

\begin{figure}
  (a1)\includegraphics[width=\mypicturewidth, height=3.5cm]{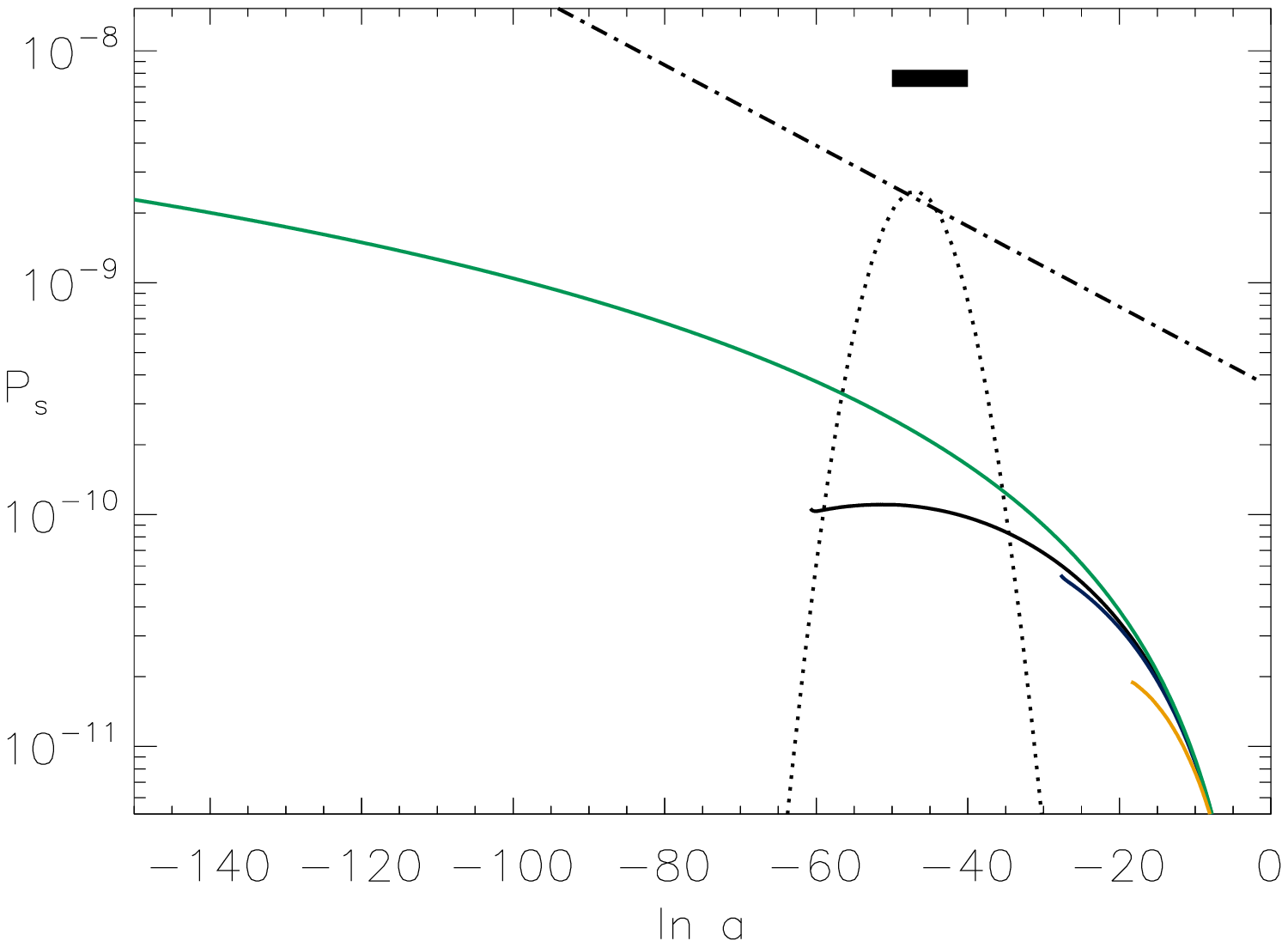}
  (a2)\includegraphics[width=\mypicturewidth, height=3.5cm]{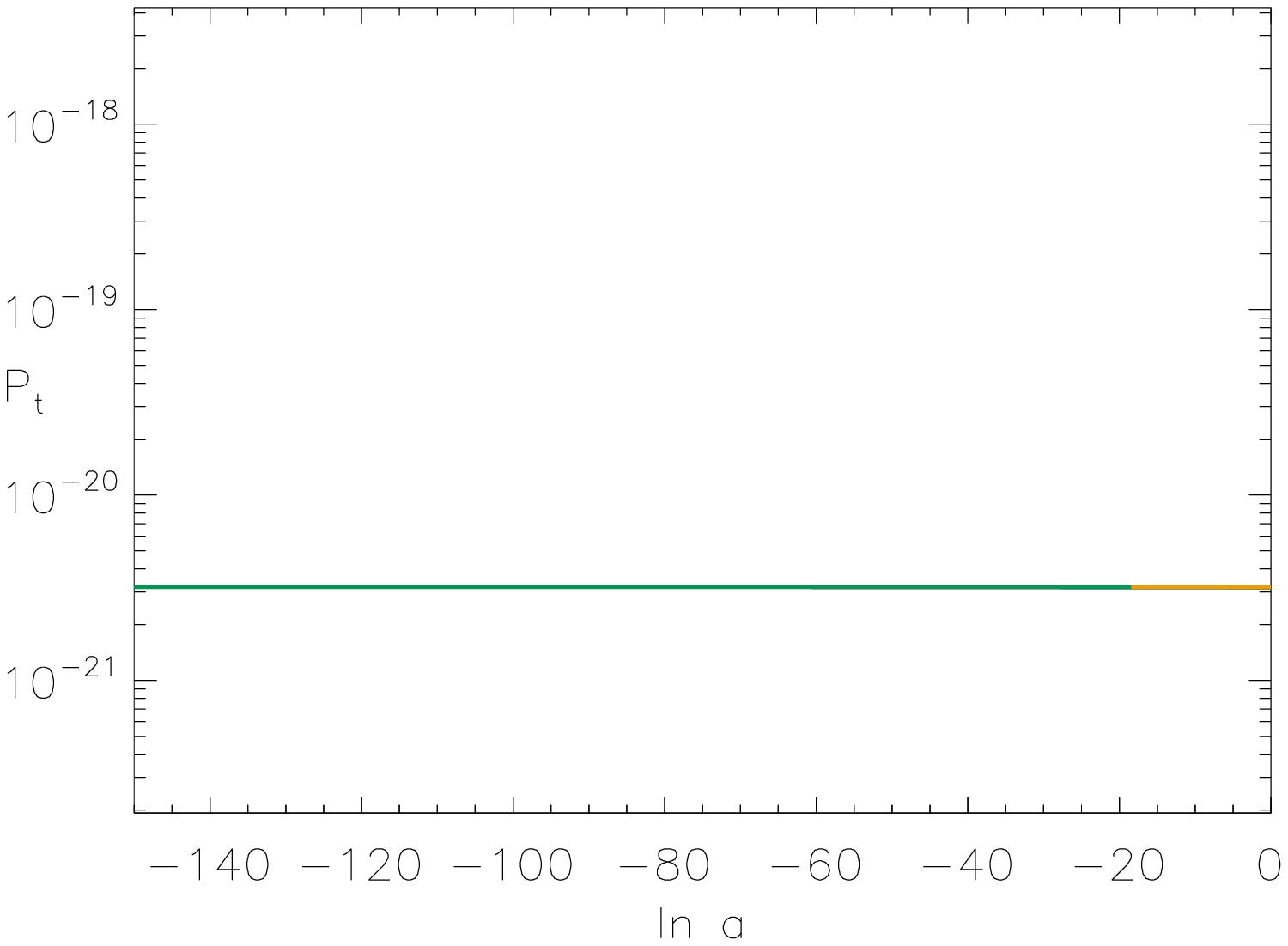}\\
  (b1)\includegraphics[width=\mypicturewidth, height=3.5cm]{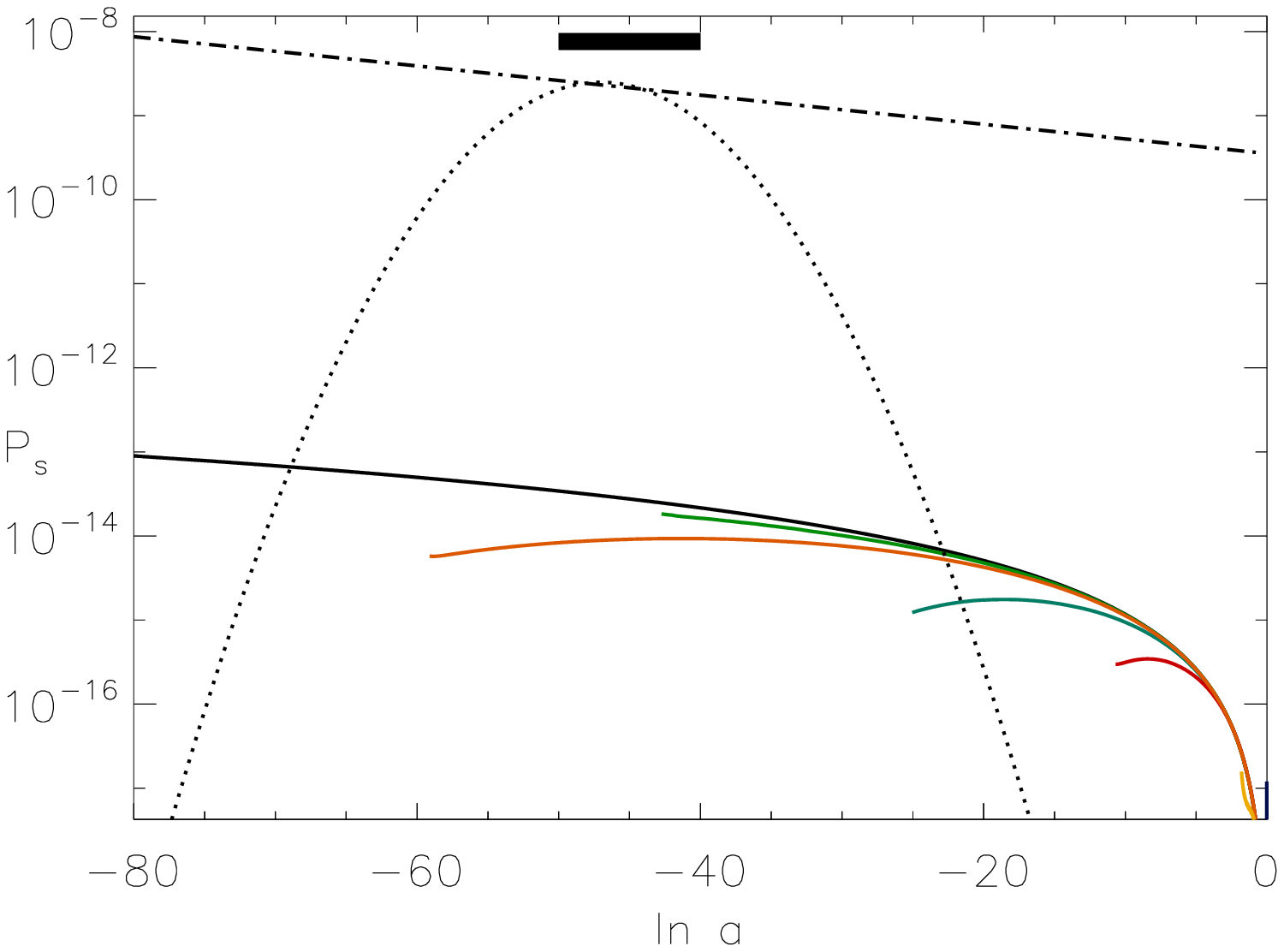}
  (b2)\includegraphics[width=\mypicturewidth, height=3.5cm]{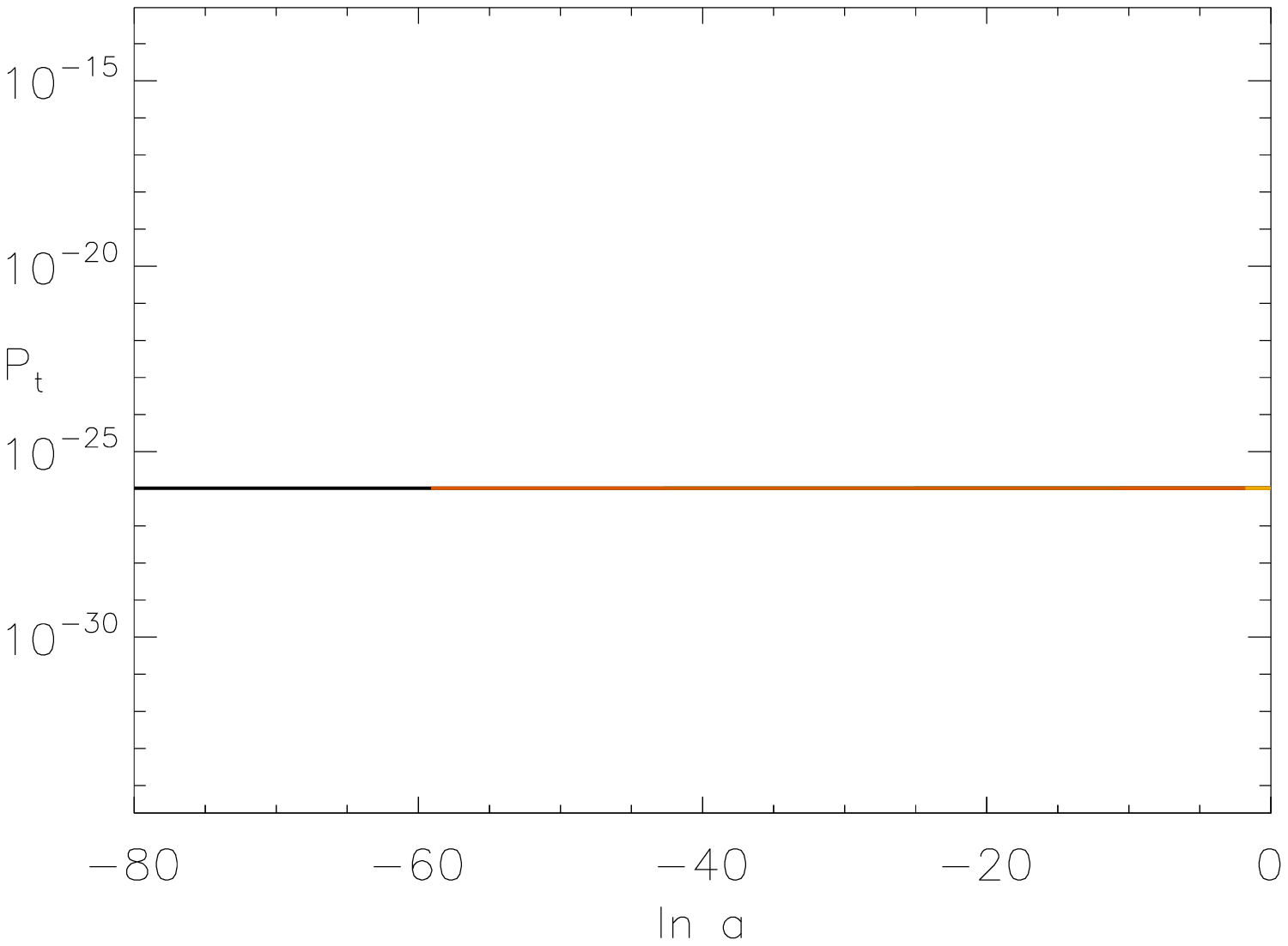}\\
  \caption{Power spectra for potentials with parameter sets 5 and 6
    that do not give a scalar amplitude near to the observed value.}
  \label{fig:powerspectra_bad}
\end{figure}

\subsubsection{The Scalar and Tensor Power Spectra and Isocurvature Effects}

We estimate the power spectra $ {\cal P}_{s} (k)$ using the single
inflaton approximation, eq.(\ref{eq:ps}), and ${\cal P}_{t} (k)$ using
eq.(\ref{eq:pt}), which only require $H$ and $\epsilon$ for each
trajectory. These are shown in Figures \ref{fig:powerspectra} and
Figures \ref{fig:powerspectra_bad}. As expected from the very small
$\epsilon$'s we have encountered, to match $ {\cal P}_{s}$ and hence
$H^2/\epsilon$ to the data would require small $H$, low energy inflation, and
hence very small gravity wave power, ${\cal P}_{t} \propto
[H/M_P]^2$. 

For comparison, Figures \ref{fig:powerspectra} and Figures
\ref{fig:powerspectra_bad} also show the best-fit scalar power spectra
that were obtained by the ACBAR collaboration \cite{Acbar06} using CMB
and LSS clustering data. The fixed power law case has $n_s=0.96$. The
spectrum with running has best-fits $n_s=0.911, d n_s /d\ln
k =-0.044$. Obviously the running model can at best be valid only over
a limited number of e-foldings. We indicate the observable range of
e-foldings by the heavy black line in the power spectrum figures. As
we show below, the mapping of wavenumber to number of e-foldings is
imprecise but because the energy of inflation is low, the relevant
range for CMB+LSS observables is in the range $N=-\ln a\in [40,50]$. For 
definiteness we take the pivot point $k_p$ for running to be at $N=45$.

To compute the scalar power spectra we assumed that the generation of
fluctuations is driven by small perturbations along the trajectories,
neglecting the effect of isocurvature fluctuations coming from
perturbations perpendicular to this direction. We will now discuss why
this might not necessarily catch the whole picture. In
Fig~\ref{fig:contourplot}d, there are a set of trajectories with
initial values around $\tau\approx21$ very close to each other where
all fields end up in the second valley from the top. Two things are
striking about those realizations of inflation: (1) the trajectories
flange out when rolling down the axion towards larger $\tau$'s during
the first part of their evolution; (2) the number of e-folds of
inflation varies tremendously from as little as $N\approx8$ all the
way to $N\approx 106$ when going towards larger initial $\theta$'s
while keeping $\tau$ fixed. The scalar spectra we compute for these
flanging trajectories differ substantially, and make it clear that
clear that our simple single field algorithm will be unjustified in
some regions of the space of initial conditions. Even though the
transverse quantum jitter is characterized by very small $H/M_P$ with
width much smaller than the size of the curves in the figure, we
recognize that tiny changes during the initial period of evolution in
certain areas could produce big effects, an area for future
investigation.

\subsubsection{Number of e-folds $N$ and the wavenumber of perturbations} \label{sec:kNrelation}

Since our trajectory computations are in terms of $\ln a/a_{end}$,
whereas the observables probe $k \sim Ha$ in ${\rm Mpc}^{-1}$, we need
to connect the two. The CMB+LSS probe $k$ from $\sim 10^{-4}\, {\rm
Mpc}^{-1}$ to $\sim 1\, {\rm Mpc}^{-1}$, about 10 e-foldings.
According to general lore (see e.g. \cite{Podolsky:2005bw}) the number
of e-folds $N$ as a function of $k$ is given by \be
\label{efold}
N(k) = 62 - {\rm ln} \frac{k}{6.96\times 10^{-5} \, {\rm Mpc}^{-1}} +
\Delta ,
\ee
where $6.96\times 10^{-5} {\rm Mpc}^{-1}$ is the inverse size of the
present cosmological horizon and $\Delta$ is defined by the physics
after inflation:
\be
\Delta =-{\rm ln} \, \frac{10^{16} {\rm GeV}}{V_k^{1/4}}+
\frac{1}{4}{\rm ln} \frac{V_k}{V_{\rm end}}-
 \frac{1}{3} {\rm ln} \frac{V^{1/4}_{\rm end}}{\rho_{\rm reh}^{1/4}} \ ,
\label{DeltaDef}
\ee
where $\rho_{\rm reh}$ is the energy density at the end of reheating,
$V_k$ is the value of the inflaton potential at the moment when the
mode with the comoving wavenumber $k$ exits the horizon at inflation
and $V_{\rm end}$ is the value at the end of inflation.
We will not go into any details about this formula or its derivation, but 
merely motivate some numbers for the individual terms.

Starting with $\Delta$, we note that for its first term, $H\approx
3\times 10^{-10} M_{\mathrm{pl}}$ corresponds to $V_k^{1/4}\sim
10^{13}$ GeV. The second term can be neglected in our case, and the
last term in the expression for $\Delta$ depends on the details of
(p)reheating (which can be perturbative preheating, non-perturbative
reheating or reheating involving KK-modes) which we put in the range 
of $1 < \Delta < 10$.
Putting it all together we find that $\Delta \in [-17,-8]$.
Therefore the observable range is about $10$ e-folds inside the
interval [35,55], with the exact location of the former depending on
the details of reheating.  For the purpose of our discussion, we take
the observable $N\in [40,50]$, and indicate this range by the black
bar in Fig~\ref{fig:powerspectra}. We note that the required number 
of e-foldings is significantly lower than $65$.

\section{Stochastic Regime of Self-Reproduction}\label{sec:stoch}

Our potential is flattening exponentially rapidly as $\tau$ increases.
Therefore the velocity of the fields which are placed at large enough
$\tau$ will be rather small. This opens the possibility for stochastic
evolution of the fields due to their quantum fluctuations dominating
over classical slow-roll \cite{stoch,sb91}.  We now show that in the
model (\ref{pot}) there is region of $(\tau, \theta)$ space where the
regime of self-reproduction operates.  We distinguish this regime from
the regime of eternal inflation due to bubble nucleations between
different string vacua.

The criterion for self reproduction is actually that scalar
perturbations are at least weakly nonlinear and that perturbation
theory breaks down. The drift in each e-fold of the scalar $\phi^i$ is
$\Delta \phi^i = G^{ij}(P_j/a^3)/(2H)$ where $P_j/a^3$ is the
canonically-normalized field momentum. The corresponding drift in the
normalized inflaton $\psi$ is $\Delta \psi = \sqrt{\epsilon}$. The
{\it rms} diffusion due to stochastic kicks is $\delta \psi = [H/(2\pi
M_P)]/\sqrt{2}$, as given in \S~\ref{sec:stochastic}. The {\it rms}
kick beats the downward drift when ${\cal P}_s^{1/2} \approx [H/(2\pi
M_P)]/\sqrt{2\epsilon}$ exceeds unity, that is the fluctuations become
non-perturbative. With such a flat nearly de Sitter potential, this is
possible.

\begin{figure}
  \begin{center}
    \includegraphics[width=\mypicturewidth]{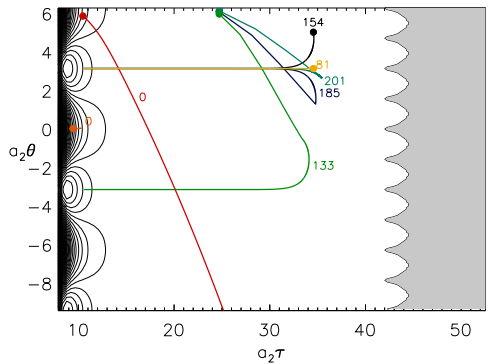}
  \end{center}
  \caption{The self-reproduction region in which stochastic kicks
    dominate over the classical downward drift is indicated by the
    grey shaded area superposed upon the contour plot of the potential
    and some of its inflationary trajectories for parameter set 1.
    The most interesting trajectories start from outside of the
    quantum region.  Those coming from the stochastic region will be
    attracted to the $\theta=\frac{2\pi l}{a_2}$ $\tau$-trough
    trajectories.}
\label{fig:eternal}
\end{figure}

We now wish to consider this boundary as a function of $\theta$ as
well as $\tau$. In general it is impossible to bring both of their
kinetic terms simultaneously into canonical form. However, when we
consider the fluctuations of fields, we can revert to the
approximation in which the K\"ahler metric stays approximately
constant over the time-scales of the fluctuations. The
result is the following condition for the region where quantum
fluctuations dominate over drift:
\begin{eqnarray}
  \frac{V}{12\pi^2} > \frac{1}{G_{\tau\tau}}\left(\frac{V_{,\tau}}{V}\right)^2 
&,&        \,\,\,      \frac{V}{12\pi^2} >\frac{1}{G_{\theta\theta}}\left(\frac{V_{,\theta}}{V}\right)^2.
\end{eqnarray}
This result is consistent with starting from
$\ddot{\phi}^i+3H\dot{\phi}^i+\Gamma^i_{jk}\dot{\phi}^j \dot{\phi}^k +
G^{ij}V_ {,\phi^j}=0$, using slow-roll to neglect terms of
$O(\dot{\phi}^2, \ddot{\phi})$ and using the diagonality of $G$.
In Fig.~\ref{fig:eternal} we replot potential contours with various
trajectories for parameter set 1, but with the drift/diffusion
boundary now shown. Starting trajectories with arbitrary $\theta$ at
the boundary should lead to large kicks in $\theta$ as well as $\tau$,
but ultimately a settling into a $\tau$-trough well before the
observable $N$ range is approached. In that case, the $\theta$
complexity of trajectories would not manifest itself. 

Will the universe truly be in a stochastic regime of self-reproduction
-- the scenario of eternal inflation \cite{Linde:1986fc} -- within
such a model? The picture is of the last hole jittering about in size
at $\tau$ far from the potential minimum. If the last hole can jitter
about far away from equilibrium, then the same phenomenon would be
expected for other holes that we took to be stabilized before the last
stages of $T_2$ settle-down. In that case, it is unclear how the
initial conditions for $T_2$ would be fed, but the most probable
source of trajectories would not necessarily be from the $T_2$
self-reproduction boundary. In all cases, it is unclear whether
further corrections to the potential way out there will uplift it to a
level in which drift steps exceed diffusion steps. Fortunately, we are
not dependent on such an asymptotically flat potential for the
inflation model explored here to work.

\section{Discussion and Summary} \label{sec:conc}

We have investigated inflation in the ``large volume''
compactification scheme in the type IIB string theory model of
\cite{BB,BBCQ,CQS}.  Dynamics in the model are driven
by K\"ahler moduli associated with four cycles and their axionic
partners, $T_i=\tau_i+\theta_i$.  We focused on the situation when
all K\"ahler moduli/axions and the compact volume are stabilized at
their minima except for a last modulus/axion pair $T_2=(\tau, \theta)$
which we identified with the ``observed'' early universe inflaton.  We
showed that $T_2$ dynamics does not perturb the global minimum of the
total potential if $T_2$ operates on a lower energy
scale than that associated with the earlier stabilizations. To do this, 
three or more moduli were needed. We explicitly demonstrated
stabilization with a three field example, $T_1, T_2, T_3$. We also
showed volume destabilization can occur even if it is initially
stabilized if there are only two K\"ahler moduli, $T_1, T_2$.

The two-dimensional potential $V(\tau, \theta )$ has a
rich ridge-trough, hill-valley structure.  We derived and solved the
equations of motion of the $(\tau, \theta )$ system appropriate to an
expanding universe using the Hamiltonian formalism of
\cite{supercosmo}. The non-canonical nature of the kinetic terms
played an important role in defining the evolution.

The ensemble of inflationary trajectories is a rich set. We first
studied inflation in the $\tau$ direction, along the valleys of the
potential surface, which are late-time attractors for most of the
general trajectories.  For example, trajectories originating in the
region of self-reproduction we identified in \S~\ref{sec:stochastic} ---
where stochastic diffusive kicks can beat classical field drift --- would
invariably finish in the $\tau$-troughs.  We calculated the number of
e-folds $N$, the time variation of the Hubble parameter as a function of
$\ln a$, $H(\ln a)$, the time variation of the acceleration history as
encoded in the ``first slow roll parameter'' $\epsilon(\ln a)$, and
the power spectrum for scalar ${\cal P}_s$ and tensor ${\cal P}_t$
fluctuations for many $\tau$-trajectories in the different
realizations of the potential we explored. We can use the tools of
single field inflation with confidence to compute the spectra if we
are only following evolution of the inflaton within the K\"ahler
modulus valley.  Our results support the conclusions of Conlon and
Quevedo \cite{CQ} who first suggested this model of inflation. In
particular, trajectories exist in the ``prior'' ensemble which satisfy
the CMB+LSS data.

Our main objective was to study inflation with general initial
conditions for $(\tau, \theta)$. We found that trajectories
originating in elevated parts of the potential (near its maxima or its
ridges) with moderate to large initial $\tau$ values can pass over
many valleys and hills in $\theta$ before settling into a final
$\tau$-valley approach to the end of inflation. These roulette wheel
trajectories have an enhanced numbers of e-foldings relative to pure
$\tau$ trajectories. No fine tuning of parameters was needed to get
plenty of long-lasting inflation trajectories. In this respect, the
K\"ahler modulus/axion inflation is superior to  ``better racetrack''
inflation \cite{Blanco-Pillado:2006he}. It is also preferable to
have 200 good trajectories with a single complex field rather than one
good trajectory with 200 complex fields as occurs in the string
version \cite{em} of ``assisted inflation'' \cite{Liddle:1998jc}. The
power spectra of scalar fluctuations for trajectories along a general
direction need to be calculated using tools appropriate to 
multiple-field inflation. In order to estimate the amplitude we used
expression (\ref{eq:ps}) for single field stochastic inflation.  We
found indications of weak running in the spectral index over the
observable range for some trajectories, and no running for others. To
the extent that these single-inflaton estimations are reasonable, we
again found plenty of roulette trajectories exist which satisfy the
CMB+LSS data. Including the influence of the isocurvature degree of
freedom on the observed power spectrum is needed to be completely
concrete about such conclusions.

Although there can be pure $\tau$-trajectories, we have not found pure
$\theta$ trajectories to be possible. The ``natural inflation'' model
explored in \cite{natural} had a radial and an angular field, in a
Mexican hat potential in which the Goldstone nature of the angular
direction was broken into a $\cos(a_2 \theta)$ type term only after
the radial motion settled into the $\theta$-trough with random
$\theta$ values. (The $a_2$ was also associated with non-perturbative
terms similar to those invoked here.) The radial motion could come
from either the small or large direction, the latter being like a
chaotic inflation scenario for a first (unobserved) stage of
inflation. The observed inflaton was to be identified with the angular
motion from near a $\theta$ maximum, where some fraction of the random
$\theta$ would reside, towards a $\theta$ minimum. One of the features of
the model was that near the maximum one could use the potential shape
to get small gravity waves and yet significant scalar tilts.

Why is our model so different? In a $\cos$ potential, inflation is
possible in the immediate neighbourhood of the maximum, and so it is
with our potentials if $\tau$ could be fixed. However we have found
that beginning in the $\theta$ heights of our potentials near
$\tau_{2,min}$ results in an outward radial flow, and, incidentally,
insufficient e-folds to be of interest. These low $\tau$ heights are
not populated from inward $\tau$-flows. This is because angular
dynamics are intimately tied to radial dynamics since the symmetry
breaking of $\theta$ and $\tau$ are due to the same process, so hills
and valleys in $\theta$ are at all radii, ultimately guiding $\theta$
trajectories to the $\tau$-valleys. And our approach to the minimum is
from a large radius flat potential rather than a growing chaotic
inflation one. In general, we believe that our two-field K\"ahler
modulus model gives a more natural inflation than natural inflation.

The inflaton potential at large $\tau$ is very shallow and in both
the K\"ahler modulus and axion directions the motion rolls very slowly,
even more so for larger $\tau$. In both $\tau$ and $\theta$ quantum
kicks in the (quasi) de Sitter geometry can dominate over the
classical slow roll drift. We identified this regime of the
self-reproducing universe in our models.

However, to have a self-reproduction regime depends on the immunity of
the model against perturbative corrections, as we discussed in
\S~\ref{sec:corr}.  Corrections may lead to polynomial terms in $\tau$
in $V(\tau, \theta )$ which may spoil the flatness, and possibly even
spoil inflation itself. This problem is specific to the models of
inflation associated with large $\tau$, as here. Models where
inflation is realized with small values of the K\"ahler moduli, like
``(better) racetrack inflation'', are not sensitive to the alteration
of the potential at large $\tau$. 

The issue is unclear because the exact form of the corrections is not
yet known, so we are left with exploring the worst and the best case
scenarios.  The best case scenario is that corrections are absent or
suppressed by a large volume factor $1/{\mc{V}}$. The worst case
scenario is where corrections generate significant terms for large
$\tau$. We note that inflation based on the potential $V(\tau,
\theta)$ with two variables is more protected from the corrections
than inflation based solely upon the potential $V(\tau)$.  Indeed, if
the potential $V(\tau, \theta)$ is altered, the runaway character
along the $\tau$-ridges may be changed to give a shallow minimum along
$\tau$, and the region around this minimum might turn out to provide
another suitable terrain for inflationary behaviour if there is slow
roll in the $\theta$ direction. Although asymptotic flatness and
self-reproduction may be destroyed by induced masses, the observable
window at significantly smaller $\tau$ may still allow inflations with
enough e-foldings. As often happens in the investigation of string
theory cosmology, assumptions such as the gentle nature of the uplift
imposed here have to be made. That a viable stringy inflation seems
feasible should motivate further work by the string theory community
on the corrections and their role in inflation model building.

In this paper, we did not treat the issues of reheating at the end of
inflation and the relation of the model to the observed particle
physics.  We just assumed that all the inflaton energy is transferred
to the energy of ultra relativistic particles of the Standard Model.
We require that there should be no overproduction of dangerous
particles such as other long-lived moduli or (non LSP)
gravitinos. Such dangerous relics may ``overclose the universe'' or
decay late enough to destroy the success of the Big Bang
Nucleosynthesis of the light elements. Details of reheating depend on
the K\"ahler moduli dynamics at the end of inflation and on the
interactions of the K\"ahler moduli with other fields, including the
Standard Model particles.  The dynamics of $\tau$ depends on the
character of the potential around the minimum. In our picture all the
other moduli except the inflaton $T_2$ are stabilized and stay at the
global minimum. If the value of $\tau_{2,min}$ exceeds the string
scale, the QFT description of the inflaton potential around the
minimum will be valid, with the inflaton beginning to oscillate after
slow roll ends.  The coupling to other degrees of freedom, in
particular those of the Standard Model, that appear in the reheating
process is unclear.  The canonically-normalized inflaton may have
interaction via the gravitational coupling and this interaction must
be sufficiently suppressed ({\it e.g.}, by the volume $1\slash
\mc{V}$) to avoid significant radiative corrections to the inflaton
potential. On the other hand, the oscillating inflaton should decay
into Standard Model particles fast enough (in $< 100$ sec) to preserve
successful Big Bang Nucleosynthesis. These requirements may put 
interesting constraints on the model for the couplings.

An alternative possibility is that $\tau_{2,min}$ is comparable to the
string scale so that stringy effects play a direct role at the end of
inflation. In this case one has to go beyond the QFT description of
the processes. We note that small $\tau_{2,min}$ does not constrain the
K\"ahler modulus/axion inflation, since the observed e-folds take place
at large $\tau$: only the approach to, and consummation of, preheating
would be affected, and would be quite different than in the QFT case.
One may envisage the following as a possible scenario: the K\"ahler
modulus, which corresponds to the geometrical size of the four-cycles,
shrinks to a point corresponding to the disappearance of the hole. The
energy of $\tau$ cascades first into the excited closed string
loops, then further into KK modes in the bulk which interact with the
Standard Model on the brane. Such a story is based on an analogy with
the string theory reheating in warped brane inflation due to
brane-antibrane annihilation investigated in \cite{Kofman:2005yz}.  In
either the QFT or stringy case, reheating in K\"ahler moduli/axion
inflation in the large volume stabilization model is an interesting
and important question worthy of further study.

Even within the context of the K\"ahler modulus/axion model explored
here, the statistical element of the theory prior probability in the
landscape of late stage moduli is unavoidable, so the terminology
``roulette inflation'' is quite appropriate --- quite aside from the
specific roulette trajectories we have identified that have a dominant
angular motion before settling into a $\tau$-trough on the way to the
minimum. Within the landscape, we may have to be content with error
bars on inflationary histories that have a very large ``cosmic
variance'' due to the broad range of theory models and trajectories
possible as well as the data errors due to cosmic microwave background
and large scale structure observational uncertainties.
 
\section*{Acknowledgments}
It is a pleasure to thank V.~Balasubramanian, J.~P.~Conlon,
K.~Dasgupta, K.~Hori, S.~Kachru, A.~Linde, F.~Quevedo, K.~Suruliz,
L.~Verde and S.~Watson and especially R.~Kallosh for valuable
discussions and suggestions.  This work was supported by NSERC and the
Canadian Institute for Advanced Research Cosmology and Gravity
Program. LK thanks SITP and KIPAC for hospitality at Stanford.

\end{document}